\documentclass[%
onecolumn,
superscriptaddress,
nofootinbib,
amsmath,amssymb,
aps,
10pt, 
]{revtex4-2}

\usepackage{hyperref,url}
\hypersetup{breaklinks,colorlinks=true,linkcolor=blue,citecolor=blue,urlcolor=blue,filecolor=blue}

\usepackage[svgnames,x11names]{xcolor}
\usepackage{graphicx}%

\usepackage{booktabs}
\usepackage{array}
\usepackage{multirow}
\usepackage{dcolumn}%

\usepackage{subcaption} %

\usepackage{physics}
\usepackage{amsmath}
\usepackage{bm}%
\usepackage{mathtools}
\usepackage{amsfonts}
\usepackage{amsthm} \theoremstyle{definition}

\usepackage{tikz}
\usetikzlibrary{quantikz2} %

\usepackage[ruled]{algorithm2e}

\usepackage[capitalise]{cleveref}%

\newcommand{\Google}{\affiliation{Google Quantum AI, Mountain View, CA, USA}}

\usepackage[english]{babel}

\makeatletter
\def\bbl@set@language#1{%
  \edef\languagename{%
    \ifnum\escapechar=\expandafter`\string#1\@empty
    \else\string#1\@empty\fi}%
  \@ifundefined{babel@language@alias@\languagename}{}{%
    \edef\languagename{\@nameuse{babel@language@alias@\languagename}}%
  }%
  \select@language{\languagename}%
  \expandafter\ifx\csname date\languagename\endcsname\relax\else
    \if@filesw
      \protected@write\@auxout{}{\string\select@language{\languagename}}%
      \bbl@for\bbl@tempa\BabelContentsFiles{%
        \addtocontents{\bbl@tempa}{\xstring\select@language{\languagename}}}%
      \bbl@usehooks{write}{}%
    \fi
  \fi}
\newcommand{\BibDeclareLanguageAlias}[2]{%
  \global\@namedef{babel@language@alias@#1}{#2}%
}
\makeatother

\BibDeclareLanguageAlias{en}{english}

\makeatletter
\DeclareFontFamily{OMX}{MnSymbolE}{}
\DeclareSymbolFont{MnLargeSymbols}{OMX}{MnSymbolE}{m}{n}
\SetSymbolFont{MnLargeSymbols}{bold}{OMX}{MnSymbolE}{b}{n}
\DeclareFontShape{OMX}{MnSymbolE}{m}{n}{
    <-6>  MnSymbolE5
   <6-7>  MnSymbolE6
   <7-8>  MnSymbolE7
   <8-9>  MnSymbolE8
   <9-10> MnSymbolE9
  <10-12> MnSymbolE10
  <12->   MnSymbolE12
}{}
\DeclareFontShape{OMX}{MnSymbolE}{b}{n}{
    <-6>  MnSymbolE-Bold5
   <6-7>  MnSymbolE-Bold6
   <7-8>  MnSymbolE-Bold7
   <8-9>  MnSymbolE-Bold8
   <9-10> MnSymbolE-Bold9
  <10-12> MnSymbolE-Bold10
  <12->   MnSymbolE-Bold12
}{}

\let\llangle\@undefined
\let\rrangle\@undefined
\DeclareMathDelimiter{\llangle}{\mathopen}%
                     {MnLargeSymbols}{'164}{MnLargeSymbols}{'164}
\DeclareMathDelimiter{\rrangle}{\mathclose}%
                     {MnLargeSymbols}{'171}{MnLargeSymbols}{'171}
\makeatother

\SetKwFor{UnaryFor}{multiplexing over register}{do}{end}
\SetKw{MiddleFor}{, for}

\crefname{algocf}{alg.}{algs.}
\Crefname{algocf}{Algorithm}{Algorithms}

\newcommand{\bigo}[1]{\mathcal{O}(#1)}

\usepackage{csquotes}
\usepackage{makecell}
\usepackage{siunitx}

\SetBlockEnvironment{italicquote}

\usepackage{caption}
\DeclareQuoteStyle{italic}[\itshape][\itshape]
{\textquotedblleft}{\textquotedblright}
{\textquoteleft}{\textquoteright}
\setquotestyle{italic}

\usetikzlibrary{arrows.meta} %
\usetikzlibrary{patterns}
\usepackage{siunitx}
\AtBeginDocument{\RenewCommandCopy\qty\SI}
\DeclareSIUnit{\us}{\micro\second}
\DeclareSIUnit{\s}{\second}

\usepackage{scrextend}

\newenvironment{rotatepage}%
    {\clearpage\pagebreak[4]\global\pdfpageattr\expandafter{\the\pdfpageattr/Rotate 90}}%
    {\clearpage\pagebreak[4]\global\pdfpageattr\expandafter{\the\pdfpageattr/Rotate 0}}%

\begin{document}

\title{The \textbf{FL}uid \textbf{A}llocation of \textbf{S}urface code \textbf{Q}ubits (FLASQ) cost model
	\\
	for early fault-tolerant quantum algorithms}

\author{William J.
	Huggins}\email{whuggins@google.com} \Google

\author{Tanuj Khattar} \Google

\author{Amanda Xu} \Google \affiliation{University of Wisconsin--Madison, Madison, WI, USA}

\author{Matthew Harrigan} \Google

\author{Christopher Kang} \Google \affiliation{Department of Computer Science, University of Chicago, Chicago, IL, USA}

\author{Guang Hao Low} \Google

\author{Austin Fowler} \Google

\author{Nicholas C.
	Rubin} \Google

\author{Ryan Babbush} \Google

\date{\today}

\begin{abstract}
	Holistic resource estimates are essential for guiding the development of fault-tolerant quantum algorithms and the computers they will run on.
	This is particularly true when we focus on highly-constrained early fault-tolerant devices.
	Many attempts to optimize algorithms for early fault-tolerance focus on simple metrics, such as the circuit depth or T-count.
	These metrics fail to capture critical overheads, such as the spacetime cost of Clifford operations and routing, or miss they key optimizations.
	We propose the \textbf{FL}uid \textbf{A}llocation of \textbf{S}urface code \textbf{Q}ubits (FLASQ) cost model, tailored for architectures that use a two-dimensional lattice of locally connected qubits to implement the two-dimensional surface code.
	FLASQ abstracts away the complexity of routing by assuming that ancilla space and time can be fluidly rearranged, allowing for the tractable estimation of spacetime volume while still capturing important details neglected by simpler approaches.
	At the same time, it enforces constraints imposed by the circuit's measurement depth and the processor's reaction time.
	We apply FLASQ to analyze the cost of a standard two-dimensional lattice model simulation, finding that modern advances (such as magic state cultivation and the combination of quantum error correction and mitigation) reduce both the time and space required for this task by an order of magnitude compared with previous estimates.
	We also analyze the Hamming weight phasing approach to synthesizing parallel rotations, revealing that despite its low T-count, the overhead from imposing a 2D layout and from its use of additional ancilla qubits will make it challenging to benefit from in early fault-tolerance.
	We hope that the FLASQ cost model will help to better align early fault-tolerant algorithmic design with actual hardware realization costs without demanding excessive knowledge of quantum error correction from quantum algorithmists.
\end{abstract}

\maketitle

\section{Introduction}
Quantum error correction (QEC) promises to eventually enable fault-tolerant quantum computation, assembling a large error-free quantum computer out of noisy physical components.
In the nearer term, error-corrected devices will remain highly resource-constrained.
These early fault-tolerant quantum computers (EFT) will be severely limited in the number of physical qubits available to perform QEC, and they may have higher-than-desired physical error rates as well.
The algorithms we run on such machines will ideally be space-efficient, tolerant of non-negligible error rates, and highly optimized for specific architectures.

In order to develop algorithms and explore architectural design choices, it is useful to have simple abstractions that help us answer questions such as ``How long will it take to execute my algorithm?'' or ``Which implementation of this subroutine should I use?''
``Cost models" aim to abstract away irrelevant details while providing insight into important limiting factors.   
Good cost models must balance simplicity with accuracy:
Abstract away too few details, and the cost model may not be useful because it requires manual intervention or intractable computation.
Abstract away too many details, and the cost model may fail to capture the true constraints of the quantum hardware and lead to incorrect conclusions.

Accurate cost models can focus attention on important challenges and drive progress.
For example, the focus on accounting for the number of non-Clifford T or Toffoli gates has led to reductions by many orders of magnitude in the number of T gates required for standard benchmarks and algorithmic primitives~\cite{maslov2016optimal, Low2018-uu, von-Burg2021-yq, Lee2021-jp, Low2025-ei}.
This cost model was motivated by the high cost of implementing non-Clifford gates via traditional magic state distillation~\cite{Bravyi2005-vi,Fowler2012-li}.
Recent advances, such as magic state cultivation~\cite{Craig2024-kv, Wan2025-dx, Wan2025-du, Chen2025-cu, Vaknin2025-yt, Hirano2025-rl}, have dramatically reduced the overhead of implementing T gates.
Consequently, costs previously ignored by the T-count model, such as the spacetime volume required for Clifford operations (via lattice surgery), can have a significant impact on the feasibility of particular algorithms.
Furthermore, architectural constraints can impose additional overheads; in two-dimensional platforms only supporting local connectivity, the overhead associated with routing and managing logical qubits is significant and should not be abstracted away by assuming free rearrangement via Clifford operations~\cite{von-Burg2021-yq, Lee2021-jp, Low2025-ei}.
For example, applying a CNOT or SWAP gate between a pair of qubits on the opposite sides of a square lattice with $N$ qubits requires using $\Omega(\sqrt{N})$ qubits to mediate the gate.
At the same time, simple constructive strategies that account for the Clifford operations at the expense of serializing the non-Clifford gates are suboptimal compared to approaches that allow for parallelization~\cite{litinski_game_2019,Beverland2022-dh}.

As we move towards an era of early fault-tolerant quantum computers, new cost models are needed that better reflect these devices' constraints and the factors that will drive the cost of algorithms.
Various works have proposed tools for early fault-tolerant algorithms, allowing for reduced circuit depth~\cite{Lin2022-zw,Wang2019-wm,Wan2022-ro,Wang2022-ac,Dong2022-ye,Ding2023-zt,Dutkiewicz2024-hl,Ni2023-mp,Liang2024-yl,Wang2025-jj} or logical qubit counts~\cite{Lin2022-zw,Wan2022-ro,Dong2022-ye,Ding2023-ff,Ding2023-zt,Ni2023-mp,Dutkiewicz2024-hl,Liang2024-yl,Wang2025-jj}, frequently at the expense of larger constant factors or more incoherent repetitions of the circuit~\cite{Lin2022-zw,Wang2019-wm,Wang2022-ac,Dong2022-ye,Dutkiewicz2024-hl,Liang2024-yl,Wang2025-jj}.
However, it can be unclear when these optimizations are useful in isolation.
Actual early fault-tolerant applications will be characterized by a careful balancing of various interrelated tradeoffs; a quantitative picture of the relative costs of various choices will be essential for both algorithm developers and architecture designers.

In this work, we introduce the \textbf{FL}uid \textbf{A}llocation of \textbf{S}urface code \textbf{Q}ubits (FLASQ) model, a new cost model tailored to algorithms implemented on early fault-tolerant quantum devices.
We specifically focus on architectures implementing the two-dimensional surface code on a square lattice of locally-connected physical qubits, operating in a regime where physical error rates are well below the threshold, but the number of qubits is limited.
Our goal is to provide easy-to-calculate estimates of the time and space required to implement a given quantum circuit in this particular early fault-tolerant architecture.
Rather than generate rigorous upper or lower bounds, we aim to provide tools that researchers can use to generate heuristic forecasts of the performance achievable by a future optimized compilation stack.

The FLASQ model provides a holistic cost estimate of a program, including the costs of mapping a computation to surface code primitives and laying them out in space and time.
The actual routing and compilation problem is not solved in FLASQ; instead, FLASQ treats available ancilla space as a ``fluid'' resource that can be flexibly used.
This abstraction assumes that the (extra) spacetime volume required for operations in the surface code is conserved across various implementations and can be efficiently rearranged. 
This assumption is motivated by techniques like the walking surface codes~\cite{McEwen2023-gk} and selective gate teleportation~\cite{Fowler2012-ti}, which allow for a great deal of freedom in arranging surface code computations in space and time.
This idealization enables the tractable estimation of the total spacetime volume required by a program, though it necessarily ignores the inefficiencies that we would expect in a real compilation.
We balance the optimism of this abstraction by being conservative in our estimates of the volume required for individual gates, and by imposing a hard constraint based on the circuit’s ``measurement depth'' and the processor’s ``reaction time.''
This guarantees that the FLASQ model respects the serial nature of certain (mostly non-Clifford) operations and the physical limitations imposed by classical control latency.

The FLASQ model enables the exploration of design tradeoffs in early fault-tolerance.
For example, previous work has proposed the combination of quantum error correction and quantum error mitigation (QEM)~\cite{Piveteau2021-nm, Lostaglio2021-rh, Suzuki2022-wq, Akahoshi2024-uk, Dutkiewicz2024-hl}, but many of those proposals have focused on a setting where error mitigation is used only to correct errors arising from non-Clifford gate operations.
Because FLASQ provides a quantitative estimate of the spacetime volume required for a particular application, we can predict the overhead incurred by, e.g., using a lower code distance and addressing the residual errors using error mitigation.
We can also use FLASQ to quantify the benefits and drawbacks of trading off between basic algorithmic resources, such as decreasing the circuit depth at the expense of increasing the number of repetitions required, or using more ancilla qubits to reduce the number of non-Clifford gates required.

We explore these applications of the FLASQ model in two case studies.
First, \Cref{sec:background} details our architectural assumptions and reviews the surface code, introducing key concepts like lattice surgery, magic state cultivation, and reaction time.
Then, in \Cref{sec:flasq_model}, we formally introduce the FLASQ model and its core assumptions.
With this foundation laid, \Cref{sec:ising_model} analyzes the 2D Ising model, studying the resources required to implement a classically challenging simulation task.
We then substantiate the model's predictive power by comparing its estimates with a hand-optimized compilation of this same task.
In that section, we also explore the crossover between a near-term intermediate-scale (NISQ) mode of operation for a quantum processor and a fault-tolerant one, closing by comparing FLASQ's estimates with previously published resource estimates.
In \Cref{sec:hwp_example}, we consider the Hamming weight phasing approach to cheaply synthesizing parallel rotations, highlighting why it may be misleading to focus on optimizing algorithms solely by reducing their T-count.
We conclude in \Cref{sec:discussion}.

We emphasize that FLASQ is not designed to replace the detailed analysis provided by careful hand-compilation, nor can it perfectly predict the performance of a future, optimized compiler. 
Instead, it is designed to serve as a tractable and significantly better proxy for the true implementation cost than simpler metrics such as T-count or circuit depth. 
By providing a more comprehensive view of the resource landscape, we aim to help align algorithmic development with the realities of early fault-tolerant superconducting hardware.

\section{Background and architectural assumptions}
\label{sec:background}

In this section, we explain the hardware constraints we assume and review some aspects of surface code quantum error correction, establishing the foundation and terminology used throughout the paper.

\subsection{Target architecture}

The FLASQ model is designed to estimate resources for a specific class of quantum architectures relevant to early fault-tolerance: a rectangular grid of two-dimensional logical surface code qubits realized by a two-dimensional grid of locally-coupled physical qubits.
This is a commonly-pursued architecture for superconducting qubit platforms~\cite{Corcoles2015-zi,Google-Quantum-AI2023-mp,Google-Quantum-AI-and-Collaborators2025-oz}.
We specifically assume a square lattice of physical qubits that can only interact with their nearest neighbors.
This locality constraint is motivated by the experimental challenges of implementing long-range couplers without introducing significant crosstalk or complexity~\cite{Bialczak2011-tr}.
While other connectivities are possible (e.g., hexagonal lattices~\cite{McEwen2023-gk}), we restrict the scope of this work to the square lattice due to its simplicity as well as its prevalence in theoretical and experimental studies.
Specifically, we consider one monolithic rectangle of physical qubits, although the actual design of such a large-scale quantum computer may involve a more complicated modular architecture~\cite{Mollenhauer2025-dd, Field2023-wd}.

The speed of the underlying physical operations dictates the timescale of error correction.
The fundamental unit of time is the ``surface code cycle'' ($t_{cyc}$), which is the time required to execute the sequence of gates and measurements for one round of stabilizer measurements.
For most of the concrete estimates in this paper, we assume $t_{cyc} = \qty{1}{\us}$ (see \Cref{app:standard_assumptions_details}), consistent with current superconducting platforms.
Another fundamental time scale is the ``reaction time,'' the time it takes to implement a measurement, process the results classically, and communicate the action conditioned on the outcome of that processing to the quantum processor.
We discuss the role of this delay below in more detail, but for the concrete estimates in the paper, we assume a reaction time of \(\qty{10}{\us}\) unless otherwise stated.

\subsection{The surface code}
\label{sec:surface_code}

To protect quantum information within this locally connected architecture, the FLASQ model assumes the use of the two-dimensional rotated surface code~\cite{Fowler2012-li,Horsman2012-vd,Tomita2014-zk}.
This realization of the surface code uses a \(d \times d\) patch of physical data qubits to encode a single logical qubit with a code distance of \(d\).
The code is defined by its stabilizers, products of X or Z operators on adjacent data qubits, which are repeatedly measured using \(d^2 - 1\) additional physical measure qubits.
The outcomes of these measurements, the syndrome data, are used by the classical control software to detect and correct errors.

The performance of the surface code under realistic noise depends on the decoding strategy~\cite{Dennis2002-pp, Tomita2014-zk,iOlius2023-jf}.
Empirical data~\cite{Google-Quantum-AI2023-mp,Google-Quantum-AI-and-Collaborators2025-oz} and analytic arguments~\cite{Dennis2002-pp} suggest that we can approximate the logical error rate (per cycle per logical qubit) using the expression:
\begin{equation}
	p_{cyc} = c_{cyc} \Lambda^{-(d + 1) / 2},
\end{equation}
where \(c_{cyc} \approx .03\) is a constant~\cite{Fowler2012-li}, and \(\Lambda\) is the error suppression factor, given by:
\begin{equation}
	\Lambda = \frac{p_{th}}{p_{phys}}.
\end{equation}
Here, \(p_{phys}\) is a single-parameter abstraction of the physical error rate, and \(p_{th}\) is the surface code threshold, often estimated around \(.01\)~\cite{Raussendorf2007-sa, Wang2011-ud, Fowler2012-li, Stephens2013-xb}.

To ensure fault tolerance against measurement errors, stabilizers must be measured repeatedly.
In this work, we assume that this requirement necessitates performing \(d\) surface code cycles between (or as part of) most logical operations~\cite{Horsman2012-vd}, although the number may be proportional to \(d\) but slightly smaller in practice~\cite{Gidney2024-fy}.
This defines one of the fundamental units of the FLASQ model, the ``logical timestep,'' which we take to be the time required to perform \(d\) surface code cycles.
We define another unit, the ``block,'' to be the spacetime volume corresponding to a single logical qubit operating for one logical timestep.
This volume consists of \(2\left( d + 1 \right)^2\) physical qubits, multiplied by \(d\) cycles, which accounts for the data and measure qubits, plus a small overhead to account for tiling logical qubit patches in the plane.

\subsection{Computation with lattice surgery}

The surface code belongs to a family of codes known as Calderbank-Shor-Steane (CSS) codes.
Computation can be achieved in CSS codes through a variety of mechanisms, including homological measurement, transversal gates, and automorphisms.%
We assume the use of lattice surgery to perform (most) computation in a 2D surface code architecture because it can be performed efficiently and fault-tolerantly even with a restriction to nearest-neighbor interactions between the physical qubits~\cite{Horsman2012-vd,Fowler2018-he}, and with less overhead than braiding defects~\cite{Horsman2012-vd}.
When lattice surgery was first introduced, it was cast in terms of the ability to implement Clifford gates by dynamically deforming logical qubit patches through a set of primitives: growing, shrinking, merging, and splitting.

Growing and shrinking operate as their names suggest, allowing us to change the number of physical qubits that support a logical qubit.
Merge operations execute fault-tolerant multi-qubit Pauli measurements.
For example, a merge of two qubits A and B can measure \(X_A \otimes X_B\) or \(Z_A \otimes Z_B\), leaving behind a single merged qubit whose logical operators are related to the unmeasured logical operators of the original qubits~\cite{Horsman2012-vd}.
Split operations reverse this process.
A merge followed immediately by a split implements a nondestructive Pauli measurement between the two patches.
These basic operations (and a few more specialized ones~\cite{Litinski2019-nu,Gidney2024-fy}) form building blocks that allow for the implementation of arbitrary multi-qubit Pauli measurements and arbitrary Clifford circuits~\cite{Horsman2012-vd,Fowler2018-he}. 
In this formalism, Pauli gates can be implemented virtually using Pauli frame tracking~\cite{Knill2005-ah, Fowler2012-li}, or, more generally, by updating the classical control software's interpretation of measurement outcomes.

This description of lattice surgery is still accurate, but modern practitioners design lattice surgery implementations using the correspondence between lattice surgery and the ZX calculus formalism rather than working with these basic operations directly~\cite{de-Beaudrap2017-yk, Gidney2019-qi, Craig2024-kv, Khattar2025-yr}.
ZX calculus is a diagrammatic language that extends standard tensor network notation~\cite{Bridgeman2017-kg}.
Tensor network notation represents $k$-index tensors as shapes with $k$ protruding legs, and denotes summation over a shared index by connecting the legs of two tensors.
ZX calculus extends this notation, enabling the contents of these tensors (in many cases) to be indicated by color or other simple annotations.
Crucially, ZX calculus includes a series of rules for simplifying and manipulating the diagrams.
By converting a logical circuit into a ZX diagram, simplifying it, and then realizing it directly as a lattice surgery construction, it is often straightforward to derive compact implementations that would otherwise be challenging to discover.
In some of our explicit constructions, we show circuits encoded as ZX graphs as well as their realizations as lattice surgery pipe diagrams.

Implementing Clifford gates via lattice surgery can require a significant amount of ancilla space (which we sometimes refer to as the ``footprint'' of a logical operation).
This is particularly true when the logical qubits are separated in space and a connecting path of logical ancilla qubits is required.
Because a logical timestep's worth of repeated measurements is generally required after initializing, merging, or splitting logical qubits, this large space requirement leads to a high cost in terms of spacetime volume.
Cost models that only count the number of T gates either neglect these costs or (as we discuss in \Cref{app:ising_gosc_csc_comparison}) avoid implementing Clifford gates at the expense of serializing the non-Clifford gates~\cite{litinski_game_2019}.

\subsection{Beyond Clifford gates}

To achieve universal quantum computation, Clifford gates are supplemented by a non-Clifford gate, typically the T gate or Toffoli gate.
In the architecture we consider, T gates are implemented via gate teleportation, consuming a ``magic state'' such as \(\ket{T} = T \ket{+}\)~\cite{Gottesman1999-ky}.
This protocol uses a Clifford circuit to perform a measurement between a data qubit and the magic state, teleporting the action of the T gate onto the data qubit.
This teleportation step randomly introduces a byproduct operator, performing the T gate up to a possible Clifford correction, which can be implemented using an \(S\) gate (and possibly some additional virtual Pauli corrections).

In addition to the spacetime volume used to consume the $T$ state, it is also necessary to account for the cost of magic state preparation.
Traditional approaches rely on magic state distillation~\cite{Bravyi2005-vi,Reichardt2006-cy,Litinski2019-ek}, which (especially in its earlier forms) incurs a substantial spacetime overhead.
In this work, we assume the use of the recently developed magic state cultivation technique~\cite{Craig2024-kv, Wan2025-dx, Wan2025-du, Chen2025-cu, Vaknin2025-yt}.
Cultivation offers significantly improved efficiency (lower spacetime volume for a target fidelity) compared to traditional distillation when targeting modest magic state infidelities useful for EFT applications.\footnote{Cultivation also reduces the resources required for magic state distillation by providing a better input to the process, but in this work we focus on the regime where cultivation on its own is sufficient.}
Additionally, cultivation has a much smaller spatial footprint than distillation, reducing the challenge of routing by allowing for magic states to be generated nearby when and where they are needed.

Magic state cultivation complicates the task of estimating resource requirements for early fault-tolerance.
Prior estimates using distilled T states overwhelmingly predicted that magic state usage would be the primary driver of costs for quantum algorithms.
Because cultivation allows for the preparation of magic states with a cost comparable to Clifford operations, it shifts the overall computational bottleneck away from the implementation of non-Clifford gates and towards the cost of routing and Clifford operations.
For the early fault-tolerant algorithms we consider here, the error rate achieved by cultivation alone is likely low enough to be sufficient.
However, cultivated T states can also be used as high-quality inputs to magic state distillation, reducing the resource requirements even when lower error rates are required.

\subsection{Spacetime tradeoffs}

The execution of a fault-tolerant algorithm involves complex tradeoffs between space and time across multiple levels of the quantum computing stack.
There is a range of algorithmic techniques for trading off between depth, gate count, and space~\cite{Cleve2000-xt, Svore2013-tl, Low2018-uu, zakablukov2017asymptotic1, zakablukov2017asymptotic2, gidney2015using, bravyi2021efficient}.
Furthermore, compilers will be able to parallelize many instructions by using more space, using, e.g., gate teleportation or optimization at the phase polynomial and ZX calculus level~\cite{Fowler2012-ti, amy_polynomial-time_2014,Litinski2022-yw,Gidney2025-bn}.
However, this flexibility is not absolute: both latency and routing overheads constrain the parallelizability of quantum computation. 

Many operations require ``fixups,'' Clifford or Pauli corrections that can only be determined after the classical control infrastructure processes some measurement and syndrome data.
For example, when implementing a non-Clifford gate via teleportation (e.g., a $T$ gate), a Clifford correction may be required, depending on the measurement outcome from the teleportation protocol~\cite{Gottesman1999-ky,Bravyi2005-vi}.
Identifying needed corrections requires the classical control software to process the preceding syndrome data and determine the correct interpretation of the measurement.
These corrections can be converted into a choice of measurement settings for one or more ancilla qubits using techniques such as selective gate teleportation~\cite{Fowler2012-ti}, but the result is still ultimately a serial process.
We call the time required for this classical feedback loop the ``reaction time'' ($t_{react}$) of the quantum processor.

We refer to the minimum number of times this classical feedback loop must be performed as the ``reaction depth'' or ``measurement depth'' of a computation.
To constructively explain this notion of measurement depth, we say that one measurement depends on another when the previous measurement outcome must be known before performing the subsequent measurement.
This usually shows up in one of two ways: either 1) the choice of measurement basis for one measurement depends on a previous measurement outcome, or 2) the choice of whether or not to apply some Clifford correction depends on a previous measurement outcome and this Clifford operation influences a subsequent measurement in a way that cannot be accounted for in classical postprocessing.
The measurement depth of a computation is equal to the length of the critical path (i.e., the longest sequence) of such dependent measurements.
Given a circuit, determining how to realize it with an optimal measurement depth may be challenging.
However, we can efficiently bound the measurement depth from the circuit, as we describe below in \Cref{sec:flasq_model}.

The other constraint that hampers smoothly trading off between space and time is the cost of routing.
The architecture modeled by FLASQ assumes a 2D plane of logical qubits, and performing long-range operations with lattice surgery requires a connecting path of logical ancilla space (technically a connecting path of ancilla spacetime is sufficient, since standard lattice surgery operations can be freely routed backwards and forwards in ``time'').
The extra overhead required to free up these connecting paths could be burdensome when using standard lattice surgery techniques.
If there is a high opportunity cost to providing extra space where it is needed, then it might be challenging to avoid serializing operations that could, in principle, be parallelized.

\begin{figure*}
    \centering
    \tikzset{
        occupied/.style={fill=gray!20, draw=black, thick},
        free/.style={fill=white, draw=black, thick},
        target/.style={preaction={fill, gray!20}, pattern=north west lines, pattern color=yellow!50!orange, draw=black, thick},
        destination/.style={pattern=north east lines, pattern color=blue!60, draw=black, thick},
        movementarrow/.style={red, very thick, -{Stealth[length=3mm]}}
    }
    
    \def\figscale{0.7}

    \begin{subfigure}[t]{0.48\textwidth}
        \centering
        \begin{tikzpicture}[scale=\figscale]
            
            \foreach \x in {0,...,10} {
                \draw[occupied] (\x,0) rectangle (\x+1,1);
            }

            \foreach \x in {0,...,9} {
                \draw[occupied] (\x,1) rectangle (\x+1,2);
            }
            
            \draw[free] (10,1) rectangle (11,2);
            
            \foreach \x in {1,...,10} {
                \draw[movementarrow] (\x , 1.5) -- (\x + .6, 1.5);
            }
            
        \end{tikzpicture}
        \caption{Initial state. All patches are occupied with logical qubits (grey background) except for the top right, which is available for use as an ancilla (white background). Red arrows indicate the desired movement.}
        \label{fig:simplified_move_a}
    \end{subfigure}
    \hfill
    \begin{subfigure}[t]{0.48\textwidth}
        \centering
        \begin{tikzpicture}[scale=\figscale]

            \foreach \x in {0,...,10} {
                \draw[occupied] (\x,0) rectangle (\x+1,1);
            }
            
            \draw[free] (0,1) rectangle (1,2);
            
            \foreach \x in {1,...,10} {
                \draw[occupied] (\x,1) rectangle (\x+1,2);
            }
            
        \end{tikzpicture}
        \caption{Final state. Each of the qubits in the top row has been moved one patch to the right, freeing the space at the top left (white background).}
        \label{fig:simplified_move_b}
    \end{subfigure}
    \caption{A visualization of a simple situation where rearranging ancilla space is much faster using walking surface codes.
	The squares represent logical qubits arranged in a \(2 \times 11\) rectangle.
	All qubits but the top right are in use and we wish to move the ancilla space to the top left by shifting the top row over to the right.
	Using standard lattice surgery techniques, each qubit would have to be moved sequentially. 
	These movement operations can be performed in one logical timestep, so moving the entire row of qubits would take \(10\) logical timesteps.
	Walking surface codes allow for all of the qubits to be shifted together in \(2\) logical timesteps.}
    \label{fig:simplified_move}
\end{figure*}
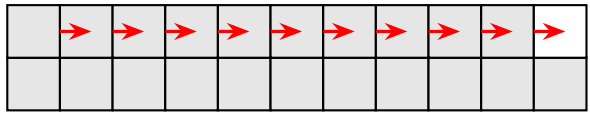
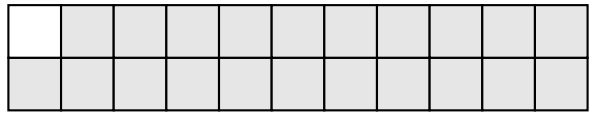

To address this, we assume the use of ``walking surface codes'' (a type of dynamic surface code)~\cite{McEwen2023-gk,Eickbusch2025-xl,Gidney2024-fy}.
Rather than treating logical qubits as static objects and relying on lattice surgery to connect or move distant logical qubits, walking codes allow for the collective movement of logical qubits across the physical lattice.
They do so by dynamically exchanging the roles of data and measure qubits in a pattern that causes the logical qubit(s) to drift in space at a rate of \(1\) logical qubit patch's worth of distance per \(2d\) surface code cycles~\cite{McEwen2023-gk}.
This motion can be vertical, horizontal, or diagonal, and there is some indication from prior works that faster walking speeds are possible as well~\cite{Gidney2024-fy}.
We show a simple example that illustrates how walking surface codes can help efficiently rearrange ancilla space in \Cref{fig:simplified_move}.

The freedom to rearrange logical qubits offered by these walking surface code constructions is fundamental to the FLASQ model.
The efficiency of walking codes provides the physical justification for abstracting away the fine details of routing and treating the available ancilla space as a ``fluid'' resource that can be efficiently reallocated where needed.

\section{The FLASQ model}
\label{sec:flasq_model}

The purpose of our model is to estimate the outcome of a careful compilation process that takes a quantum circuit defined on a two-dimensional grid of qubits and outputs a valid set of surface code instructions.
In particular, we would like to predict i) the amount of time it takes to execute the circuit and ii) the total spacetime volume used in its execution.
From these, we can easily estimate other meaningful quantities, such as the failure probability or the total runtime including the overhead from error mitigation.
We summarize our algorithm for estimating these properties below in \Cref{alg:simple_flasq_estimate}.
We explain the two constraints that are used to generate this estimate below, while also highlighting the assumptions and simplifications that we make.

\begin{algorithm}[t]
	\caption{Applying the FLASQ model to a quantum circuit}
	\label{alg:spacetime_estimation}
	\KwIn{
		\\
		\(\mathcal{C}\), a quantum circuit expressed in terms of primitive gates acting on logical qubits arranged in a two-dimensional grid\;
		\(N_{tot}\), the total number of logical qubits on the device, including the qubits not used in the circuit but available for use as (fluid) ancilla\;
		\(t_{react}\), the reaction time of the processor, in units of logical timesteps\;
		\(v_{cult}\), the spacetime volume required to cultivate a T state, in units of blocks (logical qubits \(\times\) timesteps)\;
	}
	\BlankLine
	\KwOut{
		\\
		\(L\), the total number of logical timesteps\;
		\(S\), the total spacetime volume\;
		\(M\), the total number of magic states consumed\;
	}

	\BlankLine
	\BlankLine

	\tcp{The ancilla volume and the T-count are calculated by simple sums over each gate.}
	\(V \leftarrow \sum_{g \in \mathcal{C}} \textsc{Ancilla Volume}\left( g \right)\)\;
	\(M \leftarrow \sum_{g \in \mathcal{C}} \textsc{T Count}\left( g \right)\)\;

	\BlankLine

	\tcp{An upper bound on the measurement depth is efficiently computable from the compute graph of the circuit.}
	\(D \leftarrow \textsc{Measurement Depth Bound}\left( \mathcal{C} \right)\)\;
	\BlankLine
    
	\tcp{We use a heuristic to calculate the maximum number of qubits simultaneously used by the circuit.}
	\(Q \leftarrow \textsc{Maximum Qubit Usage}\left( \mathcal{C} \right)\)\;
	\tcp{We determine the number of qubits available for use as fluid ancilla in our FLASQ calculations conservatively (assuming that the \(Q\) qubits may be in use continuously).}
	\(A \leftarrow N_{tot} - Q\)\;

	\BlankLine
    \tcp{We demand that the circuit has enough logical timesteps to accommodate both constraints.
	}
	\(L \leftarrow \max\left( V/A, t_{react}
	D\right)\)\;
	\BlankLine

    \tcp{The total spacetime volume is the sum of the fluid ancilla volume and the volume contributed by qubits explicitly used in the circuit.}
	\(S \leftarrow L Q + V \)\;
	\tcp{If we are spacetime limited then \(V = LA\) and we have \(S = L N_{tot}\).}
	\BlankLine

	\Return{\(L, S, M\)}
	\label{alg:simple_flasq_estimate}
\end{algorithm}

The primary constraint we expect to limit the operation of an early fault-tolerant quantum computer is the spacetime constraint.
We observe that nearly every logical operation requires some extra ancilla space and time, but that there is a great deal of flexibility in how this ancilla spacetime is provided.
For example, we can frequently trade off between space and time~\cite{Fowler2018-he,Litinski2019-nu}, even using teleportation to effectively vary the order of certain operations~\cite{Gottesman1999-ky, Fowler2012-ti}.
We can also use walking surface codes to efficiently move the ancilla space we do have to the locations where it is needed~\cite{McEwen2023-gk}.
Motivated by these examples, we make the simplifying assumption that ancilla space is fluid; i.e., we neglect the problem of routing ancilla patches and simply assume that it is available where and when it is needed.\footnote{We sketch an example of this assumption in action in \Cref{app:move_gate_details}, when discussing the ``Move'' operation.}

\begin{figure}
	\centering
	\includegraphics[width=.6\textwidth]{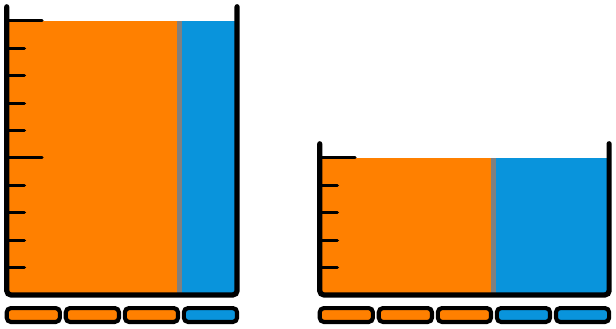}
	\caption{A cartoon that shows how the FLASQ model functions when we are spacetime-limited.
		We visualize space horizontally and time vertically.
		Left and right are two cartoons representing the same computation with varying amounts of fluid ancilla space.
		The orange rounded rectangles represent algorithmic qubits and the blue rounded rectangles represent fluid ancilla qubits.
		The fluid ancilla spacetime volume (blue fill) is conserved between the two scenarios, although the overall spacetime volume is not (sum of blue fill and orange fill).
	}
	\label{fig:flasq_space_diagram}
\end{figure}

We translate this assumption into a concrete model by assigning a ``fluid ancilla volume'' to each operation in our circuit and demand that enough time and extra ``fluid ancilla'' qubits are allocated for the computation to account for the total required volume.
We expect this abstraction to be most useful when there is a limited amount of extra space available and an optimizing compiler has many opportunities to make use of efficient tradeoffs between space and time.
\Cref{fig:flasq_space_diagram} illustrates how varying the number of fluid ancilla affects the depth and the overall spacetime volume when we are ``spacetime limited'' (when this constraint is dominant).

In \Cref{tab:gate_costs_small}, we provide an estimate of the ancilla volume required to implement a basic set of simple gates.
We describe our methodology for constructing these estimates in \Cref{app:flasq_volume_details} and also include a more comprehensive table (\Cref{tab:gate_costs_large}).
The ancilla volume required for multi-qubit gates depends on the length of a path that connects the qubits (denoted by \(p(q_1, q_2)\) for two qubits).
We assume that non-Clifford gates are compiled to a Clifford + T gate set, with the T gates being implemented using magic state cultivation~\cite{Craig2024-kv}.
This leads to an ancilla volume that depends on the physical error rate \(p_{phys}\) as well as the desired logical error rate \(p_{cult}\), as we explain in \Cref{app:cultivation_details}.

\begin{table}
	\centering
	\setlength{\tabcolsep}{10pt}
	\resizebox{1\textwidth}{!}{%
		\linespread{1.0}\selectfont%
		\renewcommand{\arraystretch}{2.5}
		\begin{tabular}
			{@{}llll@{}}
			\toprule
			Basic gates           & FLASQ ancilla volume                          & Measurement depth & Notes
			\\
			\midrule
			\(X \;/\;Y\;/\;Z\)    & \(0\)                                         & 0                 & Implemented in software
			\\
			\makecell{\(X\;/\;Z\) basis measurement
			\\
			(or initialization)}  & \(0\)                                         & 0                 &
			\\
			\(H\)                 & \(7\)                                         & 0                 & Includes the cost of a patch rotation
			\\
			\(S \;/\; S^\dagger\) & \(5.5\)                                       & 0                 &
			\\
			\(T \;/\; T^\dagger\) & \(1.5 v(p_{phys}, p_{cult}) + t_{react} + 6\) & 1                 & \makecell[l]{
				Depends on physical error rate, \(p_{phys}\),
			\\
				and target logical error rate, \(p_{cult}\)
			}
			\\
			Move                  & \(5 p(q_1, q_2)\)                             & 0                 & Moves a qubit to an empty patch
			\\
			\(CNOT \;/\; CZ\)     & \(5 p(q_1, q_2)\)                             & 0                 &
			\\
			\bottomrule
		\end{tabular}
	}
	\caption{
		Estimates of the fluid ancilla (spacetime) volume required to implement various gates (in units of ``blocks,'' i.e., \(d^3\)), as well as the measurement depth we assume (in units of ``logical timesteps,'' i.e., \(d\)).
		The volume is in addition to the volume required to store the qubit(s) the gate acts on and includes a conservative estimate of the cost of using walking surface codes to rearrange the necessary ancilla space.
		The function \(p(q_1, q_2)\) denotes the Manhattan distance between two qubits.
		The volume required to cultivate a magic state (\(v(p_{phys}, p_{cult})\)) depends on the underlying physical error rate (\(p_{phys}\)) and the target logical error rate (\(p_{cult}\)).
		We implicitly multiply the reaction time (\(t_{react}\)) by one logical qubit to obtain the correct units.
		We discuss the methodologies for generating these estimates in \Cref{app:flasq_volume_details}, but they are generally derived by considering canonical implementations of the gates and (in some cases) extrapolating how the cost should scale.
	}
	\label{tab:gate_costs_small}
\end{table}

In talking about the FLASQ model, it is sometimes helpful to refer to three different kinds of logical qubits.
The ``fluid ancilla'' qubits are the ones accounted for implicitly by our model, the ones that mediate surface code operations.
When we are analyzing a simple quantum circuit that consists of only unitary quantum gates, we refer to all of the qubits that the circuit acts on as ``data qubits.''
However, many quantum algorithms make temporary use of ancilla qubits at the algorithmic level.
We use the phrase ``algorithmic ancilla'' to refer to these qubits that are defined at the circuit level and occupy explicit positions on a 2D grid.
For example, optimized algorithms for loading classical data make temporary use of additional qubits~\cite{Low2018-uu, Huggins2025-pa}.

Our estimates of the time (and total spacetime volume) depend heavily on the number of fluid ancilla qubits available to mediate a computation.
Given a quantum circuit (\(\mathcal{C}\)) and a total number of logical qubits available on the device (\(N_{tot}\)), it is helpful to automatically determine this number.
At present, we do so by using a heuristic provided by the Qualtran software library to calculate the maximum simultaneous qubit usage for a circuit~\cite{Harrigan2024-rj}.
We describe this heuristic in \Cref{app:max_qubit_heuristic}, but the basic idea is that we order the operations in the circuit sequentially.
We then proceed through the operations, keeping track of the maximum number of data qubits and algorithmic ancilla that are simultaneously in use.
We call this number \(Q\), and assume that \(A = N_{tot} - Q\) qubits are available for use as fluid ancilla.
This heuristic risks both overestimating and underestimating the amount of space available to mediate computation, and we discuss possible extensions of the model in \Cref{sec:discussion}.
For simple circuits that use few algorithmic ancilla (or a consistent number throughout their execution) these issues will not arise.

The second constraint, the measurement depth constraint, is based on the fact that certain measurements in the surface code must be performed serially.
As we discussed in \Cref{sec:background}, we refer to the number of such sequential steps required to implement the circuit as the ``measurement depth,'' and the time required between sequential measurements as the ``reaction time.''
To account for this limitation, we always allocate a total time for executing a circuit that is at least as long as the product of the measurement depth and the reaction time.
We call computations where this constraint is dominant ``reaction limited.''
We expect that most early fault-tolerant computations will not be reaction limited, since the amount of space will be highly constrained.
Nevertheless, we still enforce this constraint in our model.

Assuming that we know the measurement depth of every operation in a circuit (which is true for all of the primitive operations we consider), we can efficiently upper bound the measurement depth of the entire circuit.
We do this by treating the circuit as a directed acyclic graph whose nodes are the primitive operations and whose edges are the qubits acted on by those operations.
We assign a weight to each node equal to the measurement depth of that operation.
Then we use a standard longest-path algorithm to calculate the length of the critical path, which yields an upper bound to the true measurement depth of the circuit.\footnote{This is not a tight bound because some operations with non-zero measurement depths may act on an overlapping set of qubits without actually increasing the measurement depth.
	As a trivial example, multiple T gates (which each have measurement depth \(1\)) targeting the same qubit can be implemented in parallel using only additional Clifford operations, implying that they have measurement depth \(1\).
	Of course, they could also be simplified to a product of Clifford gates and \(0\) or \(1\) T gates.
}

The FLASQ model makes several approximations in order to efficiently generate reasonable cost estimates without actually solving the compilation problem.
Its core simplification is that it assumes that ancilla space can be fluidly moved throughout the computation so that it can be efficiently provided when and where it is needed.
This avoids the need to explicitly handle routing data or packing operations together in space and time.
Instead, the model only requires the calculation of the total volume of ancilla spacetime required, neglecting the impact of any routing congestion or packing inefficiency that might arise in an explicit compilation.

Its second major simplification is that this spacetime constraint is treated independently from the measurement depth constraint, when in reality the problem of laying out the computation would be coupled with the need to manage inherently serial measurement and feedback steps.
This decoupling may cause the model to underestimate costs when both constraints are simultaneously relevant because the measurement depth constraints may make it challenging to pack operations in space and time.
Finally, the strategy for determining the amount of fluid ancilla space relies on a heuristic scheduling proxy (see \Cref{app:max_qubit_heuristic}) that may over- or under-estimate the amount of available spacetime for mediating computations that also make use of algorithmic ancilla.

The optimism of these simplifications is counterbalanced by several guardrails.
First, the model's application of the measurement depth constraint provides a hard floor on runtime, preventing pathological behavior (in most cases) when the amount of space available is very large.
More generally, the regime we are most focused on is the space-limited regime where we do not expect such pathological behavior (or the measurement depth constraint itself) to be relevant.
Second, the costs that the model assigns to individual gates are intentionally conservative (see \Cref{app:flasq_volume_details}).
Third, the notion of treating the gates individually is itself conservative, neglecting powerful optimizations that come from using ZX-calculus operations to fuse multiple gates together.
The model's core simplifications ensure its tractability, while the balance of optimistic and conservative assumptions are tailored to give it predictive power.
In \Cref{sec:hand_compiled_ising} and \Cref{app:validation_details}, we provide numerical evidence that this balance of optimism and pessimism yields reasonable estimates, especially when we are concerned with estimating the relative costs between related circuits.

\section{Case study: Dynamical simulation of the two-dimensional quantum Ising model}
\label{sec:ising_model}

One of the earliest scientifically-interesting applications of fault-tolerant quantum computing will be the dynamical simulation of model Hamiltonians.
Because of its simplicity and its status as a canonical example in quantum many-body physics, simulating the time evolution of the 2D transverse-field Ising model (TFIM) is a standard benchmark for classical methods~\cite{Begusic2025-pk,Pavesic2025-co, Krinitsin2025-wf,Mandra2025-sa}.
It has also been studied as an early target for quantum advantage~\cite{Beverland2022-dh,Kim2023-nm,Haghshenas2025-kd,Mandra2025-sa}.
The Hamiltonian for this model is given by
\begin{equation}
	H = -J \sum_{\langle i, j \rangle} Z_i Z_j + g \sum_i X_i,
\end{equation}
where \(J\) is the coupling strength between neighboring spins, \(g\) is the strength of the transverse field, and the first sum ranges over pairs of adjacent sites.

Given a sufficiently large system size and long evolution time, this system becomes challenging to classically simulate with known methods for certain choices of Hamiltonian parameters.
While the exact boundary for classical simulation of the 2D TFIM is uncertain~\cite{Haghshenas2025-kd,Begusic2025-pk,Pavesic2025-co, Tindall2024-oy, Krinitsin2025-wf, Mandra2025-sa}, we focus on parameter regimes that have so far resisted accurate simulation by classical methods.
The authors of \cite{Haghshenas2025-kd} argued that a 56-qubit quantum processor simulating a specific high-entanglement quench of the 2D TFIM showed a potential advantage over a suite of classical simulation techniques.
This work focused on the task of estimating the expectation value of the observable \(Z_{tot}^2 = \frac{1}{N^2} \sum_{j,k} Z_j Z_k\) as a function of time, using up to \(20\) second-order Trotter steps together with a finely-tuned choice of initial state and other Hamiltonian parameters. 
In \cite{Mandra2025-sa}, the authors studied the same simulation task for system sizes up to \(11 \times 11\) using a method that combines matrix product state (MPS) simulations with an empirically-tuned rescaling parameter.
Their numerical experiments suggest that their method is accurate out to longer times and larger sizes than the approaches considered in \cite{Haghshenas2025-kd}, but that it struggles to achieve high precision at a system size of \(11 \times 11\).

We apply the FLASQ model to this problem and relate the cost estimates we obtain with prior art.
Specifically, we provide FLASQ estimates for the resources required to estimate the expectation value of \(Z_{tot}^2\) (or any other diagonal observable with unit norm) after applying a Trotterized approximation to the time-evolution operator.
We consider the model defined on a \(\sqrt{N} \times \sqrt{N}\) square lattice with periodic boundary conditions (except in \Cref{sec:nisq_ft_crossover}).
We take \(N = 121\) (except in \Cref{sec:ising_beverland_comparison}, where we take \(N = 100\) to compare with \cite{Beverland2022-dh}).
We target an overall error from rotation synthesis of \(0.001\) (except where noted in \Cref{sec:ising_beverland_comparison}).
Our FLASQ estimates are produced by optimizing over the code distance and magic state cultivation postselection probability (see \Cref{app:standard_assumptions_details}).
Except for one of the comparisons in \Cref{sec:ising_beverland_comparison}, we assume the use of probabilistic error cancellation to mitigate the residual errors after error correction (\cite{Temme2017-hj}, see \Cref{app:pec_details}), targeting a standard deviation of \(0.0045\).

We Trotterize the time evolution by dividing the Hamiltonian into two parts, \(H = A + B\), where \(A\) contains all of the \(X_i\) terms and \(B\) contains all of the \(Z_i Z_j\) terms.
We use second-order Trotter formulas except in \Cref{sec:ising_beverland_comparison} (where we use a fourth-order formula for the sake of comparison).
We detail the circuits and Trotter formulas in more detail in \Cref{app:trotter_formulas}, but their basic structure is as follows: 
Time evolution by the \(X_i\) terms is performed by implementing all \(N\) single-qubit rotations in parallel.
To implement the \(Z_i Z_j\) terms, we divide them into four groups, interacting each qubit with the one above, below, to the right, and to the left in turn.
Each \(Z_i Z_j\) interaction is implemented by a single-qubit \(R_Z\) rotation conjugated by two CNOT gates.

We present several related FLASQ estimates and comparisons.
We begin in \Cref{sec:ising_classical_benchmark_comparison} by presenting FLASQ estimates for Ising model simulation in a parameter regime informed by a recent classical benchmarking study~\cite{Mandra2025-sa}.
In \Cref{sec:hand_compiled_ising}, we present a hand-optimized compilation of the same simulation task.
We summarize the results of this comparison in \Cref{tab:ising_hand_compiled_comparison}.
We then examine the crossover point between two strategies for operating a large quantum processor in \Cref{sec:nisq_ft_crossover}: one which uses quantum error correction and one which performs many noisy simulations in parallel.
We close by setting our FLASQ estimates side-by-side with the fault-tolerant resource estimates for Ising simulation presented in \cite{Beverland2022-dh}.
More details regarding the methodologies we used to generate these comparisons are provided in \Cref{app:ising_details}, and we present additional comparisons with other compilation frameworks in \Cref{app:ising_gosc_csc_comparison}, including a new automated compiler that generates a constructive upper bound to the computation time.

These comparisons suggest that the crossover to fault-tolerant quantum advantage may occur sooner than previously anticipated, although we emphasize that the particular examples we consider are not definitively beyond the reach of classical simulation.
This conclusion is driven partly by the architectural assumptions embedded in the FLASQ model and partly by ongoing advances in quantum error correction and algorithms.

\begin{table*}[]
	\centering
	\renewcommand{\arraystretch}{1.5}
	\setlength{\tabcolsep}{6pt}

	\resizebox{\textwidth}{!}{%
		\begin{tabular}{
				@{} %
				l r r r r r r r
				@{} %
			}
			\toprule
			\textbf{Example}      &
			{\makecell{\textbf{Logical}
			\\
			\textbf{qubits}}}     &
			{\makecell{\textbf{Code}
			\\
			\textbf{distance}}}   &
			{\makecell{\textbf{Physical}
			\\
			\textbf{qubits}}}     &
			{\makecell{\textbf{Number of}
			\\
			\textbf{rotations}}}  &
			{\makecell{\textbf{Logical timesteps}
			\\
			\textbf{(FLASQ)}}}    &
			{\makecell{\textbf{Logical timesteps}
			\\
			\textbf{(by hand)}}}  &
			{\makecell{\textbf{Ratio}
			\\
						\textbf{(FLASQ / hand)}}}
			\\
			\midrule

			{\makecell[l]{11x11 Ising Model
			\\
			20 steps, 2nd order}} &
			160                   &
			13                    &
			\num{62720}           &
			\num{7381}            &
			\num{55995}           &
			\num{73810}           &
			0.76
			\\

			\addlinespace

			{\makecell[l]{10x10 Ising Model
			\\
			20 steps, 4th order}} &
			140                   &
			14                    &
			\num{63000}           &
			\num{30100}           &
			\num{235284}          &
			\num{270900}          &
			0.87
			\\

			\bottomrule
		\end{tabular}
	}
	\caption{
		The number of logical timesteps required to execute the two Ising model simulation circuits we consider in this section (see \Cref{sec:ising_classical_benchmark_comparison} and \Cref{sec:ising_beverland_comparison}).
		We show the FLASQ estimates and the estimates from a hand-optimized compilation of each task.
		The approach taken for the hand-compiled estimates is described for the \(11 \times 11\) example of \Cref{sec:ising_classical_benchmark_comparison} in \Cref{sec:hand_compiled_ising}.
		The details for the \(10 \times 10\) example of \Cref{sec:ising_beverland_comparison} along with more details about both cases is presented in \Cref{app:hand_compiled_ising_details}.
		We find that the FLASQ model obtains good qualitative agreement with the hand-compiled estimates, but it reports a somewhat lower depth in both cases.
	}
	\label{tab:ising_hand_compiled_comparison}
\end{table*}

\subsection{FLASQ estimates for a classically challenging regime} \label{sec:ising_classical_benchmark_comparison}

To the best of our knowledge, the work of \cite{Mandra2025-sa} represents the most advanced simulations of Ising model dynamics in the classically challenging regime.
That work presented results up to a lattice size of \(11 \times 11\), although uncontrolled errors make it difficult to determine the accuracy of their method at larger system sizes.
We characterize the high-accuracy regime for this \(11 \times 11\) simulation task as ``classically challenging'' (but not necessarily intractable) due to the difficulty of fully converging classical calculations to a high precision and bounding their error.
This makes it an appealing example to consider for our resource estimates, although we expect that improved classical methods will be developed over time.

\begin{figure}
	\includegraphics[width=.75\textwidth]{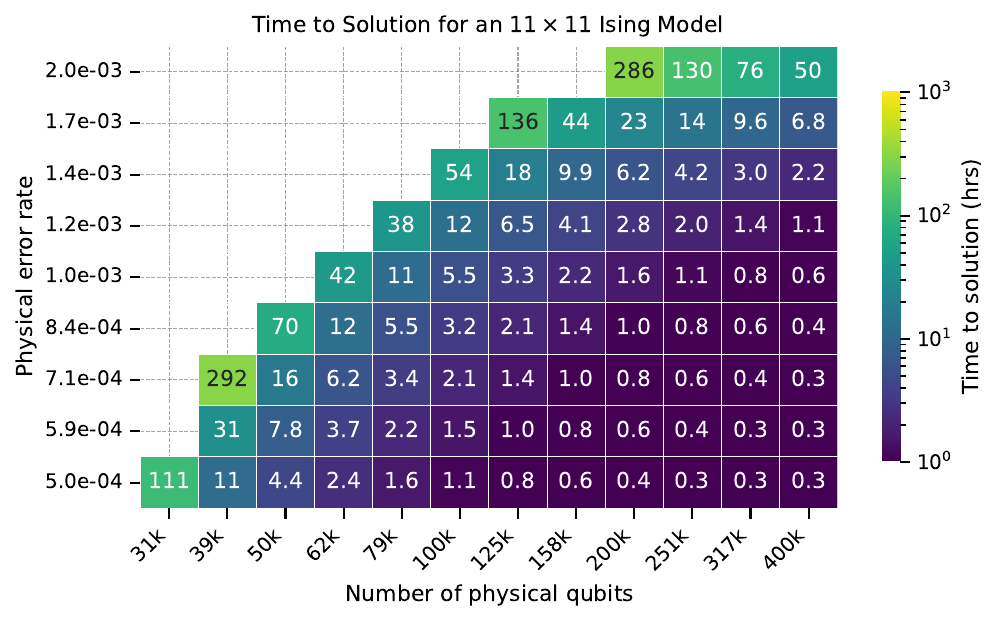}
	\caption{
		FLASQ estimates of the total time required to estimate a diagonal observable with unit norm to within \(.01\) after \(20\) second-order Trotter steps for an \(11\times11\) two-dimensional Ising model as a function of the physical error rate and the number of physical qubits.
		This task is challenging for existing classical methods in certain parameter regimes~\cite{Haghshenas2025-kd, Mandra2025-sa}.
		These estimates assume a surface code cycle time of \(\qty{1}{\us}\) and a reaction time of \(\qty{10}{\us}\).
		Each square of the heatmap is colored and annotated according to the total amount of time required (in hours).
	}.
	\label{fig:ising_heatmap}
\end{figure}

We focus on providing FLASQ estimates for the deepest circuits considered in \cite{Mandra2025-sa}, showing the cost for estimating \(\ev{Z_{tot}^2}\) after \(20\) second-order Trotter steps.
We target an overall error in the expectation value of \(Z_{tot}^2\) that is \(\leq .01\) with \(> 95\%\) probability.
We present these estimates in \Cref{fig:ising_heatmap} and \Cref{tab:ising_hand_compiled_comparison}.
FLASQ predicts that as few as \(62{,000}\) physical qubits are sufficient to complete this task at a physical error rate of \(10^{-3}\).
We conjecture that the resources required to demonstrate quantum advantage in quantum simulation will fall below this threshold, even accounting for the development of improved classical heuristics, due to the continued improvement of quantum algorithms and error correction.

\subsection{Lattice Surgery Compilation for the 2D Ising Model circuits}
\label{sec:hand_compiled_ising}

In this section, we present a hand-optimized lattice surgery compilation for the 2D Ising model simulation.
This serves two purposes: to validate the estimates from our FLASQ model and to provide an updated, concrete benchmark for this task.
Here we focus on the second-order \(11 \times 11\) example considered in \Cref{sec:ising_classical_benchmark_comparison} and we assume an error suppression factor of \(\Lambda = 10\).
We outline the general compilation strategy, including the physical layout of logical qubits and the implementation of rotation synthesis here, deferring some additional details (and a second example) to \Cref{app:hand_compiled_ising_details}.
The key parameters and results of this compilation are summarized in \Cref{tab:ising_hand_compiled_comparison}.

\begin{figure}
	\centering
	\begin{minipage}[]{0.68\textwidth}
		\centering
		\begin{subfigure}[]{\textwidth}
			\includegraphics[width=.8\textwidth]{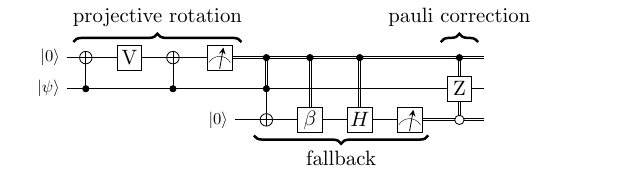}
			\caption{
				A circuit diagram showing a modified implementation of the mixed fallback rotation synthesis of \cite{Kliuchnikov2023-vm}, which is detailed in \Cref{app:rotation_synthesis_details}.
				The single-qubit operations $V$ and $\beta$ are both instances of diagonal unitary approximations and can be expressed as alternating sequences of $T^{\pm1}_X$ and $T^{\pm1}_Z$ gates, together with an arbitrary single-qubit Clifford gate~\cite{Gidney2025-bn}.
				Synthesizing a single qubit rotation with precision $\epsilon$ requires $n_T \approx 0.53 \log_2 \epsilon^{-1} + 4.86$ T gates.
				The fallback rotation (\(\beta\)) is only necessary when the first qubit is measured in the \(\ket{1}\) state, which happens less than \(1\%\) of the time by construction~\cite{Kliuchnikov2023-vm}.
				We assume that \(\beta\) is implemented with sufficiently high precision that we can neglect any error.
				The Pauli correction corrects the \(Z\) error from the initial failed rotation as well as the possible \(Z\) error from the measurement-based uncomputation of the second ancilla.
			}
			\label{fig:mixed_fallback_circuit}
		\end{subfigure}
		\\
		\begin{subfigure}[]{\textwidth}
			\hfill
			\includegraphics[width=0.15\textwidth]{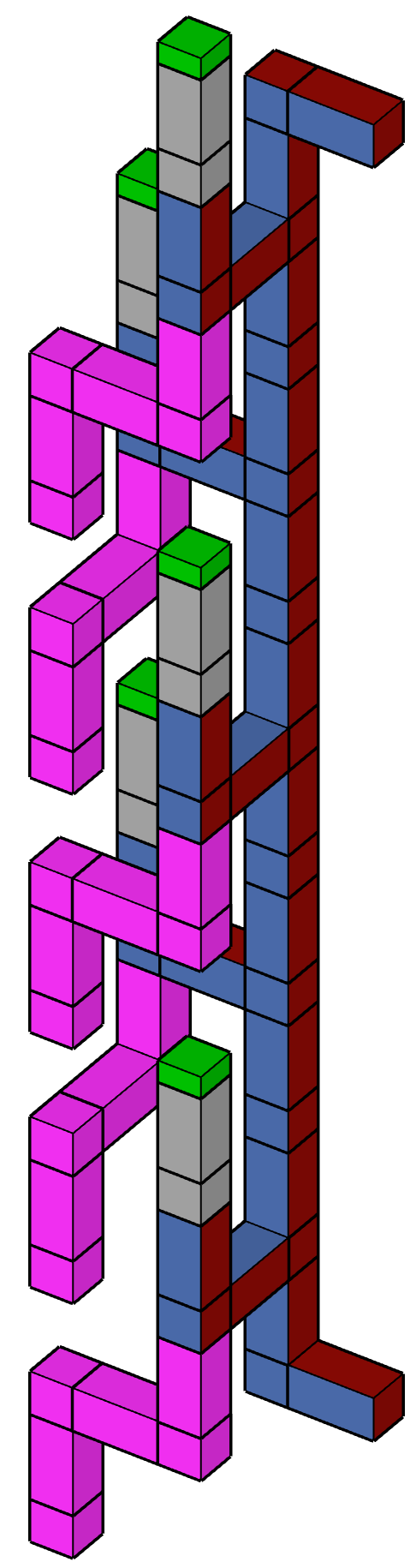}
			\hfill
			\includegraphics[width=0.27\textwidth]{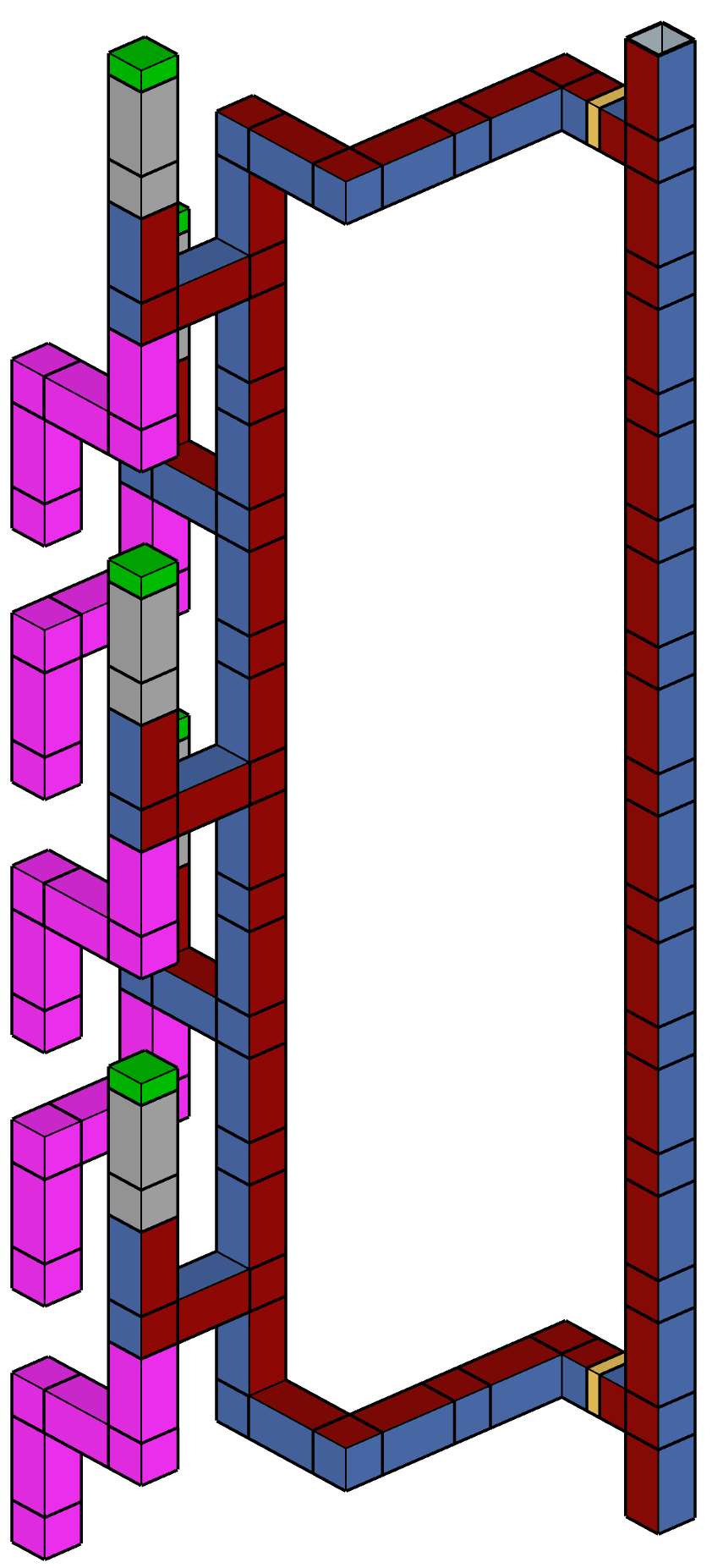}
			\hfill
			\includegraphics[width=0.27\textwidth]{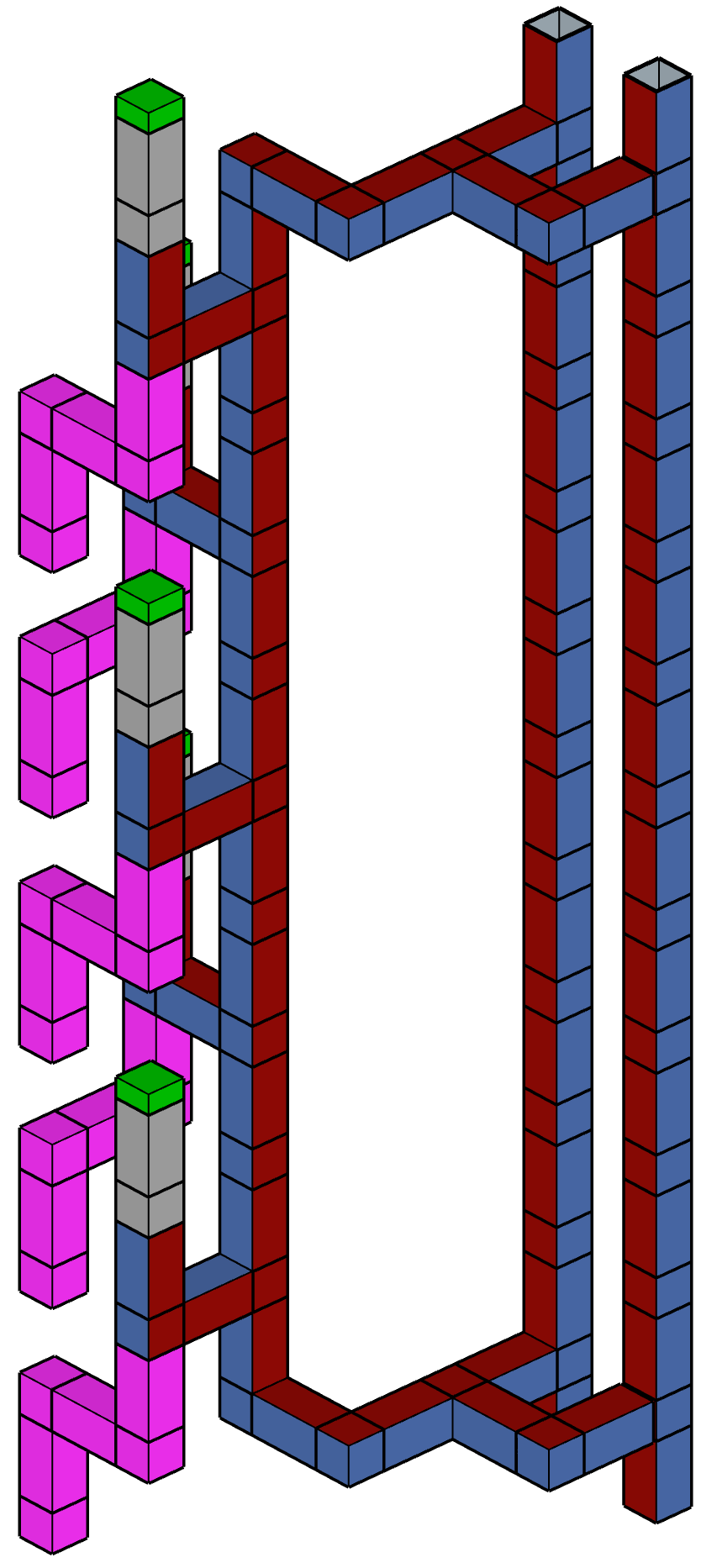}
			\hfill
			\caption{
				Lattice surgery compilation of the projective rotation circuit from \cref{fig:mixed_fallback_circuit}.
				Left: A $2\times 2$ rotation synthesis unit which implements the single qubit rotation $V$ as an alternating sequence of $T^{\pm 1}_X$ and $T^{\pm 1}_Z$ gates.
				Pink boxes correspond to spacetime used for magic state cultivation \cite{Craig2024-kv} in the X or Z basis.
				Grey boxes correspond to waiting time needed for the classical control system to process the results of previous measurements.
				Green boxes represent a choice of \(X\) or \(Y\) basis measurements used to apply the phase correction to the injected T state if necessary~\cite{Gidney2024-fy}.
				The rotation synthesis unit consumes a \(T\) state every \(2d\) surface code cycles.
				If the decomposition has $n_T$ $\ket{T}$ states, then a height of $2\times n_T + 2$ units is sufficient to also incorporate a single qubit Clifford at the end of the decomposition (not shown in this figure).
				The horizontal poles on the bottom and top correspond to the CNOT gates in \cref{fig:mixed_fallback_circuit}, and should be connected to the system qubits on which the single qubit rotation should be applied.
				Middle: An \(R_X\) gate applied on the system qubit, represented using the vertical pole on the right, using the $2\times 2$ rotation synthesis unit on the left.
				Right: An \(R_{ZZ}\) gate applied on the two system qubits, represented using the two vertical poles on the right, using the $2\times 2$ rotation synthesis unit on the left.
			}
			\label{fig:mixed_fallback_pipes}
		\end{subfigure}
	\end{minipage}
	\hfill
	\begin{minipage}[]{0.28\textwidth}
		\centering
		\begin{subfigure}[]{\textwidth}
			\includegraphics[width=0.9\textwidth]{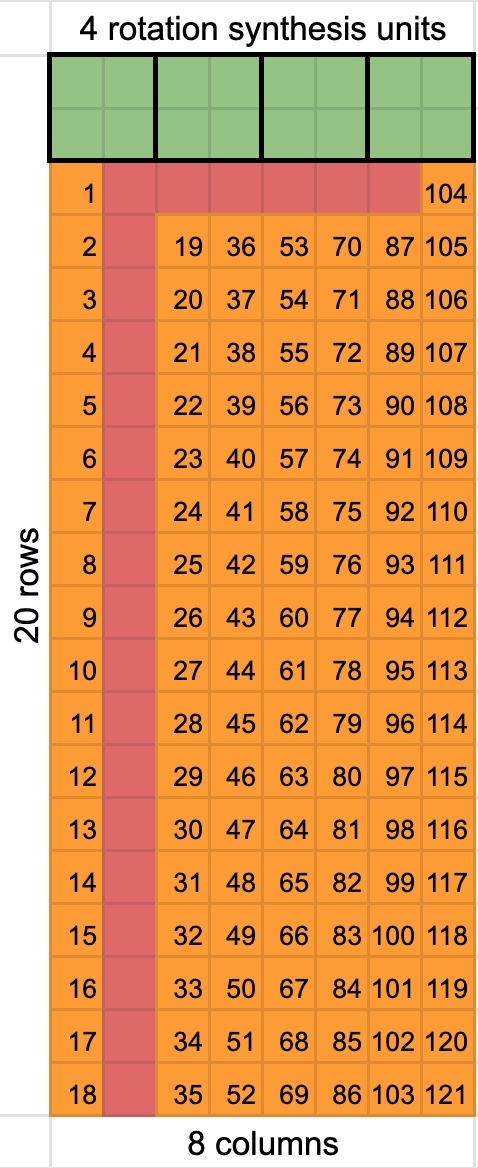}
			\caption{
				Mockup of the physical layout for the \(11 \times 11\) 2D Ising model simulation, realized using a \(20 \times 8\) rectangular grid of logical qubits.
				Orange patches correspond to system qubits and are numbered according to their ordering (top to bottom, then left to right) in the \(11 \times 11\) Ising model lattice.
				Green boxes denote single qubit rotation synthesis units from \cref{fig:mixed_fallback_pipes}.
				Red patches are ancilla qubits used as routing space for connecting the orange system qubits with green boxes.
				The red ancilla region will sweep from left to right and right to left as the computation proceeds.
				This sweeping can be performed between timesteps when the rotation synthesis gadgets have to be connected to the system qubits in almost all cases.
				Most of the two qubit interactions happen between pairs of orange patches that lie within adjacent columns.
			}
			\label{fig:11_x_11_ising_model_layout}
		\end{subfigure}
	\end{minipage}
	\caption{
		An overview of our compilation strategy for implementing time evolution by the \(2D\) Ising model Hamiltonian using the mixed fallback protocol of \cite{Kliuchnikov2023-vm} for rotation synthesis (see also \Cref{app:rotation_synthesis_details}).
	}
	\label{fig:constructive_compilation_overview}
\end{figure}

Our compilation strategy is built around a rectangular grid of logical qubits, with dedicated zones for performing arbitrary rotations.
The system qubits for the Ising model simulation are laid out in a zig-zag pattern to keep nearest-neighbor interactions in the simulation relatively local on the 2D grid of logical qubits.
The remaining space is populated with routing ancillas and several parallel rotation synthesis units, as illustrated in \Cref{fig:11_x_11_ising_model_layout}.
Both \(R_X\) and \(R_{ZZ}\) rotations are implemented using the same underlying rotation synthesis gadget, which is based on the mixed fallback protocol (see \Cref{fig:mixed_fallback_circuit}, \Cref{app:rotation_synthesis_details}, and \cite{Kliuchnikov2023-vm}).
As shown in the pipe diagrams in \Cref{fig:mixed_fallback_pipes}, these gadgets occupy a \(2 \times 2\) block of logical qubits and can synthesize a rotation in \(2m+2\) logical timesteps, where \(m\) is the number of T gates required.
With several of these units operating in parallel, we can pipeline the synthesis of the many rotations required by the algorithm.
For example, in the \(11 \times 11\) case, we use four parallel units.
Including a buffer of 4 logical timesteps for routing a system qubit to and from a unit, we estimate a new rotation can be completed every \((36+4)/4 = 10\) logical timesteps.

We detail both the physical resources required and the calculation itself in \Cref{app:hand_compiled_ising_details}, where they are summarized in \Cref{tbl:ising_model_physical_estimates}.
The estimates are derived from the compilation strategy described above, using the specific hardware and error model assumptions listed in the table.
For example, the \(20\) second-order Trotter steps for the \(11 \times 11\) model require \(7{,}381\) total arbitrary rotations.
Using the mixed fallback rotation synthesis method, an average of \(17\) T gates per rotation is sufficient to achieve a cumulative error from rotation synthesis of less than \(.001\) in the diamond norm.
The single-qubit rotation synthesis gadgets we illustrate in \Cref{fig:mixed_fallback_pipes} require \(2 m + 2 = 36\) logical timesteps to implement a rotation.
With four working in parallel and a buffer of \(4\) additional logical timesteps for routing, we assume a rotation can be implemented every \(10\) logical timesteps.
The whole simulation therefore requires \(73{,}810\) logical timesteps.
We estimate in \Cref{app:hand_compiled_ising_details} that a code distance of \(13\) is sufficient to achieve a reasonably high fidelity of \(\approx 0.6\), so the total number of physical qubits is \(62{,}720\).

As we presented in \Cref{tab:ising_hand_compiled_comparison}, FLASQ reports an estimate of the logical depth that is approximately \(24\%\) lower than we estimate using this hand-optimized compilation for the \(11 \times 11\) case.
It is unclear how much of this difference is due to FLASQ underestimating the true minimum cost, and how much is due to inefficiencies in the hand-optimized compilation.
For example, the FLASQ estimates implicitly assume that T gates are cultivated and rotations are synthesized close to the qubits where they are being applied, but the layout presented in \Cref{fig:11_x_11_ising_model_layout} uses rotation synthesis zones that are fixed at the top of the layout.
It is possible that the cost could be reduced by changing how the rotations are implemented, but achieving a more sophisticated layout by hand would be challenging and involve error-prone manual optimization.
Overall, this comparison shows that FLASQ achieves good qualitative accuracy while also highlighting that its estimates are subject to considerable uncertainty.

\subsection{Probing the crossover between NISQ and fault-tolerance}
\label{sec:nisq_ft_crossover}

In this section, we illustrate how FLASQ can enable us to explore the landscape of early fault-tolerance by analyzing the tradeoff between two strategies for using an early fault-tolerant scale quantum processor.
The first strategy uses the device as a large, noisy (NISQ) processor to run many error-mitigated simulations in parallel.
The time and space costs of each simulation are much lower in this approach, but the exponential overhead of error mitigation sets in as the error rate increases.
The second uses the same device in a fault-tolerant (FT) mode, performing a single error-corrected calculation at a time (and mitigating the residual errors).
The way we model errors and error mitigation for both modes of operation (but particularly the NISQ mode) is an idealization designed to show the qualitative features of the two approaches.
We leave a more quantitatively accurate comparison to a future work, but we expect that FLASQ will prove useful for this task as well.

We study the same Ising model simulation task as in \Cref{sec:ising_classical_benchmark_comparison} (and \Cref{fig:ising_heatmap}), except that we consider open boundary conditions rather than periodic boundary conditions.
This is because while the FLASQ cost estimates are relatively insensitive to the boundary conditions, it would significantly increase the depth to compile the interactions that wrap around the lattice when restricted to 2D connectivity on the NISQ machine.
We expect that open boundary conditions will make the classical simulation substantially easier, but we choose to use them because they simplify the comparison.

To estimate the error mitigation overhead, we assume the use of probabilistic error cancellation (PEC) in both modes.
While costly, PEC provides an unbiased estimator given a known noise model, and it is tractable to analyze its overhead.
We also adopt simplified single-parameter error models for both the NISQ and FT simulations and conflate their noise strength parameter.
For the NISQ simulations, we assume that each qubit experiences single-qubit depolarizing noise with a fixed strength after each layer of two-qubit gates, and that there are no additional coherent or incoherent errors.
The estimates for the error mitigation overhead in the fault-tolerant context use the FLASQ estimate of the spacetime volume, combining error correction with PEC while assuming the noise model outlined in \Cref{app:pec_details}.
Conflating these two error models implicitly assumes that the threshold for the fault-tolerant device under the NISQ noise model would be \(\approx .01\). 
We describe our specific calculations in more detail in \Cref{app:flasq_ising_nisq_details}.

\begin{figure}
	\centering
	\includegraphics[width=.75\textwidth]{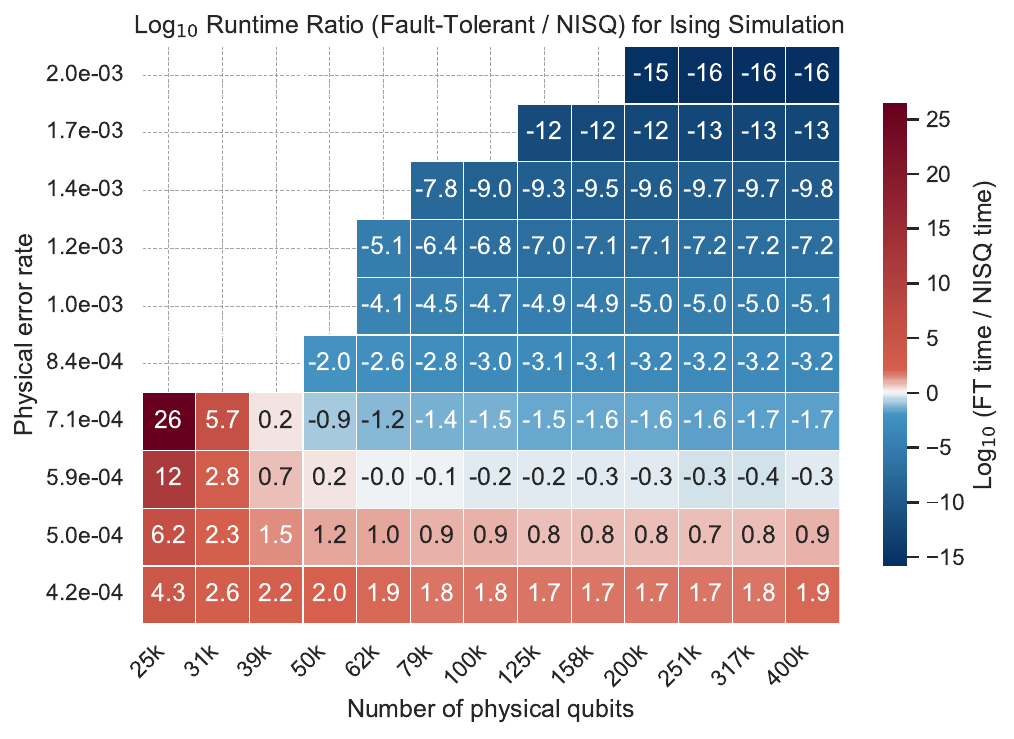}
	\caption{
		A heatmap showing the logarithm of the runtime ratio between an optimized fault-tolerant (FT) approach and a noisy, intermediate-scale (NISQ) approach to a specific Ising model simulation task as a function of the number of physical qubits (x axis) and the noise strength (y axis).
		The NISQ approach applies error mitigation (specifically, probabilistic error cancellation) to many parallel simulations to address noise.
		The FT approach applies error mitigation to a single error-corrected simulation.
		In both cases, the goal is to measure a diagonal observable with unit norm after performing \(20\) second-order Trotter steps of an \(11 \times 11\) Ising model Hamiltonian with open boundary conditions.
		Blue regions (negative values) indicate where the FT approach is faster, while red regions (positive values) indicate where the NISQ approach is faster.
		The white area marks the crossover frontier where performance is comparable.
		For the NISQ simulations, the error rate is the strength of the single-qubit depolarizing noise applied after each layer of two-qubit gates, whereas for the FT simulations, we use a simple phenomenological error model detailed in \Cref{app:resource_estimates}.
		A more comprehensive analysis would likely shift the crossover.
	}
	\label{fig:nisq_ft_crossover}
\end{figure}

We plot the logarithm of the ratio of the fault-tolerant runtime to the NISQ runtime in \Cref{fig:nisq_ft_crossover}.
The transition from blue (FT advantage) to red (NISQ advantage) illustrates the fundamental tension between the high initial overhead of QEC and the exponential cost of using error mitigation alone.
While the exact location of the crossover (white area) is dependent on our idealized assumptions, the figure clearly delineates how improvements in error rate or increases in machine size have the potential to shift the balance.
This suggests that careful analysis will be required to determine when it is preferable to pursue an early fault-tolerant approach.
We expect the exact numerical details to vary based on exact assumptions about the error models, mitigation strategies, and other factors.

\begin{figure}
	\centering
	\includegraphics[width=.75\textwidth]{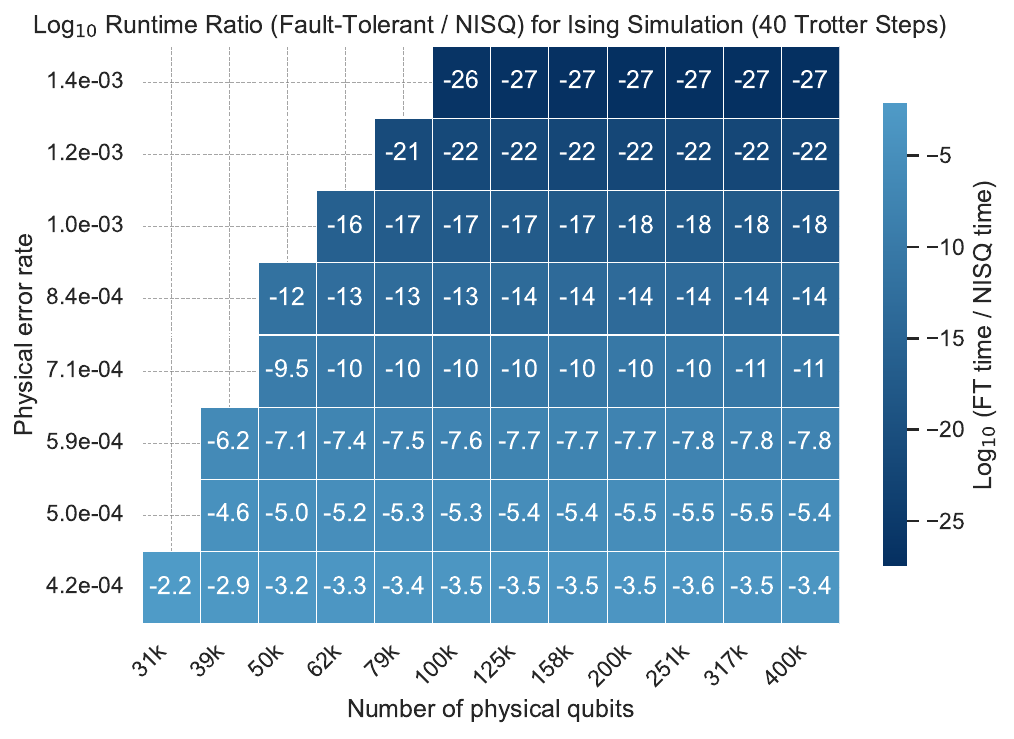}
	\caption{
		A heatmap showing the logarithm of the runtime ratio between an optimized fault-tolerant (FT) approach and a noisy, intermediate-scale (NISQ) approach to a specific Ising model simulation task as a function of the number of physical qubits (x axis) and the noise strength (y axis).
		Here we consider \(40\) second-order Trotter steps of an \(11 \times 11\) Ising model with open boundary conditions instead of the \(20\) in \Cref{fig:nisq_ft_crossover}.
		The crossover between the NISQ and FT modes of operating shifts significantly at this higher depth, with only a small region of NISQ advantage visible on the plot.
	}
	\label{fig:nisq_ft_crossover_40}
\end{figure}

Additionally, we expect that the location of this crossover regime will depend strongly on the particular computational task.
In \Cref{fig:nisq_ft_crossover_40}, we perform the same analysis for the same simulation problem (estimating a diagonal observable to within an error of \(.01\) after time-evolving an \(11 \times 11\) Ising model) using \(40\) second-order Trotter steps instead of \(20\).
Doubling the number of Trotter steps decreases the ratio (shifting the balance towards the early fault-tolerant mode of operation) by more than five orders of magnitude at every combination of error rate and number of physical qubits that we consider.
This highlights the `exponential wall' inherent in NISQ error mitigation.
Fundamentally, the NISQ mode is overwhelmed by the exponential scaling of error mitigation, while the fault-tolerant mode offers a scalable path forward.

This comparison highlights that the transition to the fault-tolerant era will not occur at a single point in time.
Instead, it will gradually occur at different rates for different problems and architectures. Providing a quantitative estimate of the crossover in this case is possible because the FLASQ model lets us estimate the spacetime volume required for the fault-tolerant quantum computation.
A more sophisticated analysis could take into account the subtleties of error mitigation for the NISQ calculations, including the challenge of characterizing the noise accurately, the impact of coherent errors, and the possibility of using more efficient error mitigation strategies~\cite{Tran2023-qr, Eddins2024-ij}. %
In particular, the simple one-parameter error model we use here may not accurately capture the way that various components of a quantum computer have significantly different noise strengths.

\subsection{Reduced resource estimates compared with previous work}
\label{sec:ising_beverland_comparison}

We compare our results with the resource estimation framework presented in \cite{Beverland2022-dh}.
This framework models a 2D surface code architecture using a compilation strategy termed Parallel Synthesis Sequential Pauli Computation (PSSPC).
Similar to the Game of Surface Codes framework that we review in \Cref{app:ising_gosc_csc_comparison}, PSSPC entirely eliminates the need to perform Clifford unitaries by commuting them through the circuit.
Naively, this approach results in replacing the T gates in the circuit with an equivalent number of many-qubit Pauli rotations that must be implemented serially (because, in general, each rotation may act on all of the qubits).
The PSSPC approach to compilation improves the situation by recompiling the circuit to implement the rotations on a separate set of synthesis qubits.
As a result, the number of serial operations scales with the number of rotations, rather than the number of T gates, allowing for a substantial amount of parallelism.

Besides this alternative compilation framework, the analysis of \cite{Beverland2022-dh} differs from the standard assumptions we make with the FLASQ model in two other ways: It relies on conventional magic state distillation rather than cultivation, and it sets a stringent error budget of \(0.001\), which it achieves using QEC alone and not error mitigation.
To perform a more direct comparison, we study the same circuit they consider in their quantum dynamics benchmark using the FLASQ model with the same error budget, although we still eschew the use of magic state distillation in favor of cultivation.
Their benchmark focuses on the simulation of a \(10 \times 10\) Ising model simulation using a 4th-order Trotter formula (see \Cref{app:trotter_formulas} for details).
We analyze a circuit consisting of 20 Trotter steps, noting that while the main text of \cite{Beverland2022-dh} mentions 10 steps, their appendix and reported rotation counts are consistent with 20 steps.
We adopt their ``nanosecond'' qubit profiles, considering two specific physical error rates ($p_{phys}=10^{-3}$ and $10^{-4}$) and a surface code cycle time of $t_{cyc}= \qty{400}{\ns}$.
They neglect the reaction time entirely, but we choose to set one that aligns with their general optimism, assuming a reaction time of 10 cycles (\(\qty{4}{\us}\))\footnote{Assuming a finite reaction time provides a guardrail in the FLASQ model, which might otherwise estimate an unphysical level of parallelism}.

\begin{figure}
	\centering
	\includegraphics[width=0.75\textwidth]{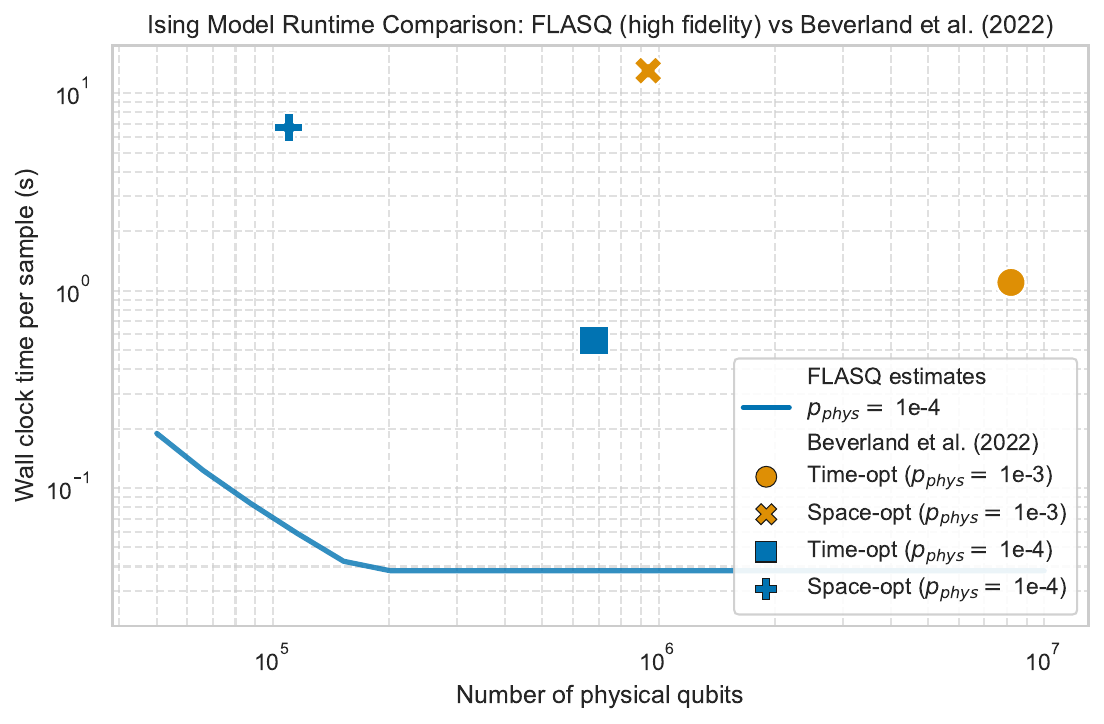}
	\caption{
	The time to execute \(20\) \(4\)th order Trotter steps of a \(10 \times 10\) Ising model.
	The lines represent FLASQ estimates generated by optimizing over the code distance and magic state cultivation error probabilities to minimize the time.
	The points represent the Time-Optimal and Space-Optimal estimates from \cite{Beverland2022-dh}.
	Both models target an overall infidelity of at most \(.001\), with the error budget spread equally between rotation synthesis, logical errors, and magic state errors.
	There is no curve for $p_{phys}=10^{-3}$ (orange) because we are unable to satisfy the desired error budget using FLASQ (specifically using magic state cultivation).
	At $p_{phys}=10^{-4}$ (blue), FLASQ predicts a substantial reduction in resources compared to the estimates of \cite{Beverland2022-dh}.
		}
	\label{fig:flasq_vs_beverland_loglog_high_fidelity}
\end{figure}

In \Cref{fig:flasq_vs_beverland_loglog_high_fidelity}, we compare the timings predicted by FLASQ with those from \cite{Beverland2022-dh}.
Here we plot the time required to generate a single high-fidelity sample as a function of the total number of qubits.
For both approaches, the total error budget is set to \(.001\), which is divided equally between rotation synthesis error, magic state error, and errors in the remaining logical operations.
At $p_{phys}=10^{-4}$, the FLASQ model predicts a substantial reduction in resources compared to the PSSPC estimates.
However, at $p_{phys}=10^{-3}$, we provide no FLASQ estimates because we are unable to satisfy the desired error budget.
Specifically, magic state cultivation (as modeled here) cannot achieve the fidelity required to support the $\approx 571{,}900$ T gates within the \(\frac{.001}{3}\) error budget at this physical noise level.
An approach that uses only quantum error correction to suppress errors in this regime would likely necessitate the use of higher-overhead magic state distillation, illustrating the challenges of achieving very high fidelities for early fault-tolerant quantum computations.

\begin{figure}
	\centering
	\includegraphics[width=0.75\textwidth]{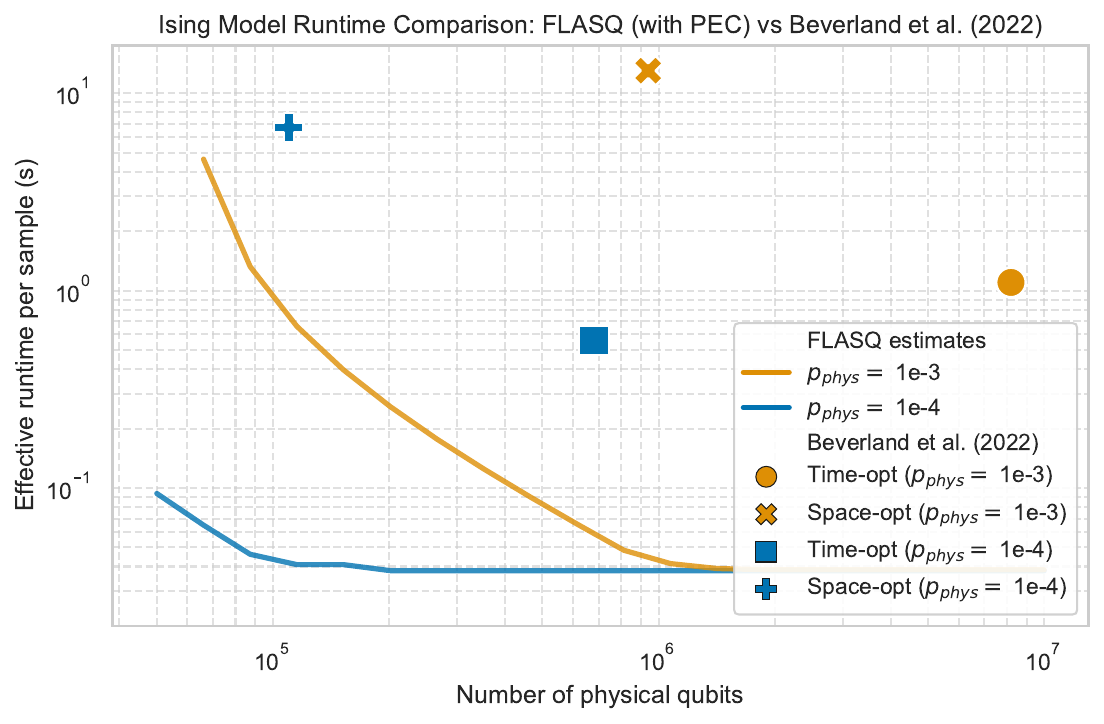}
	\caption{
	The expected time required to generate data equivalent to a single high-precision sample from a circuit consisting of \(20\) \(4\)th order Trotter steps of a \(10 \times 10\) Ising model.
	For the FLASQ estimates (lines), we plot the average time required to achieve the same variance as one would achieve with a noiseless device, allowing for the use of PEC and a rotation synthesis error of \(.001\).
	For the data from \cite{Beverland2022-dh}, we report the time required to generate a single high-fidelity sample with an overall error budget of \(.001\) (using QEC alone).
	The combination of QEC and QEM enables a significant resource reduction for the FLASQ estimates and makes the $p_{phys}=10^{-3}$ (orange) regime accessible.
			Across all regimes we consider, the FLASQ estimates predict a considerable reduction in the resource requirements compared with the analysis of \cite{Beverland2022-dh}.
		}
	\label{fig:flasq_vs_beverland_loglog_PEC}
\end{figure}

The challenge we face in meeting the stringent error budget at realistic physical error rates for an early fault-tolerant machine highlights the need for a hybrid strategy that combines QEC and quantum error mitigation (QEM).
In \Cref{fig:flasq_vs_beverland_loglog_PEC}, we compare the same data from \cite{Beverland2022-dh} with FLASQ estimates generated using this hybrid strategy.
As outlined in \Cref{app:pec_details}, we use FLASQ to optimize over the choice of code distance and the magic state cultivation error, assuming the use of probabilistic error cancellation (PEC) to mitigate the remaining errors.
In addition to slightly relaxing the error budget for rotation synthesis (since all \(.001\) of the allowed error can be allocated to this component of the error), the use of error mitigation allows us to tolerate significantly higher error rates without biasing the expectation values of any observables we might measure.
The metric we plot is now the ``effective runtime per sample,'' which incorporates the PEC sampling overhead, showing the time (in expectation) required to achieve the same variance as a single noiseless sample.
For the estimates of \cite{Beverland2022-dh}, we report the same numbers as above.

The model that combines FLASQ estimates with QEM using PEC predicts runtimes that are at least two orders of magnitude less than \cite{Beverland2022-dh} when considering equivalent error rates and numbers of physical qubits.
Alternatively, one could say that the FLASQ model predicts that the same runtimes could be achieved with a combination of higher error rates and lower numbers of physical qubits.
Crucially, this strategy enables the use of magic state cultivation in the $p_{phys}=10^{-3}$ (orange) regime that was inaccessible for the FLASQ model when using QEC alone.
Because PEC handles the residual cultivation errors, we no longer require the extremely high fidelities that prevented a solution previously.
Overall, the comparisons suggest that the resources required for fault-tolerant quantum simulation may be significantly lower than previously estimated.

\section{Case study: Hamming weight phasing}
\label{sec:hwp_example}

In this section, we demonstrate how FLASQ can shed light on algorithmic design decisions in early fault-tolerance by analyzing the Hamming weight phasing (HWP) approach to parallel rotation synthesis.
HWP is a strategy to reduce the cost of implementing multiple parallel copies of an \(R_Z(\theta)\) rotation~\cite{Gidney2018-xg, Kivlichan2020-am}.
Instead of implementing \(N\) parallel rotations independently, one first calculates the Hamming weight of the \(N\) target qubits into an ancilla register of size \(\approx \log_2 (N)\).
It then suffices to apply scaled rotations to this ancilla register: \(R_Z (\theta)\) to the least significant bit, \(R_Z(2 \theta)\) to the next bit, \(R_Z (4 \theta)\) to the bit after that, and so on.
After uncomputing the Hamming weight register, the overall result is the same as applying \(N\) parallel \(R_Z(\theta)\) operations.

When implemented with the optimized adder circuits of \cite{Gidney2018-xg}, this approach significantly reduces the T gate complexity required to implement the rotations to within an error \(\epsilon\), from \(N \cdot \mathcal{O}\left( \log \epsilon^{-1} \right)\) to \(4 N + \mathcal{O}\left( \log N \cdot \log \epsilon^{-1} \right)\).
However, this theoretical advantage comes at a significant cost.
The approach requires not just the \(\approx \log_2 N\) ancilla qubits to store the Hamming weight, but also additional ancilla used to manage the computation.
The overall \textit{algorithmic} ancilla cost for the implementation analyzed here is \(\approx N\).
While this approach results in the best-known scaling of the T-count, the opportunity cost of using so many ancilla to calculate the Hamming weight is large, as these qubits cannot be used to cultivate T states or mediate the computation in other ways.
Furthermore, while the number of T gates is lower, the number of Clifford gates and the complexity of the routing and layout for the ancilla qubits is significantly higher.

We use the FLASQ model to quantitatively analyze the tradeoff between the T-count reduction of this implementation of Hamming weight phasing and the opportunity cost of its ancilla usage (together with the spacetime cost of its Clifford operations).
We compare a naive strategy that implements \(N\) \(R_Z\) rotations in parallel against Hamming weight phasing, as implemented in the Qualtran software package~\cite{Harrigan2024-rj}.
We focus on a regime relevant to early fault-tolerance, targeting an error per rotation of \(10^{-7}\) from rotation synthesis and \(10^{-5}\) from magic state cultivation.\footnote{For example, these parameters would allow one to implement a circuit with \(10{,}000\) rotations with an overall rotation synthesis error of \(.001\), and an overall error from magic state cultivation of \(0.1\) (which could easily be addressed using probabilistic error cancellation).}
We use the default conservative gate costs outlined in \Cref{tab:gate_costs_large} and assume a surface code distance of \(14\), with a cycle time of \(\qty{1}{\us}\).
HWP as implemented in Qualtran is not optimized to minimize the measurement depth, so we focus on exploring the spacetime-limited regime and neglect the measurement depth constraints.

\begin{figure}
	\centering
	\includegraphics[width=.75\textwidth]{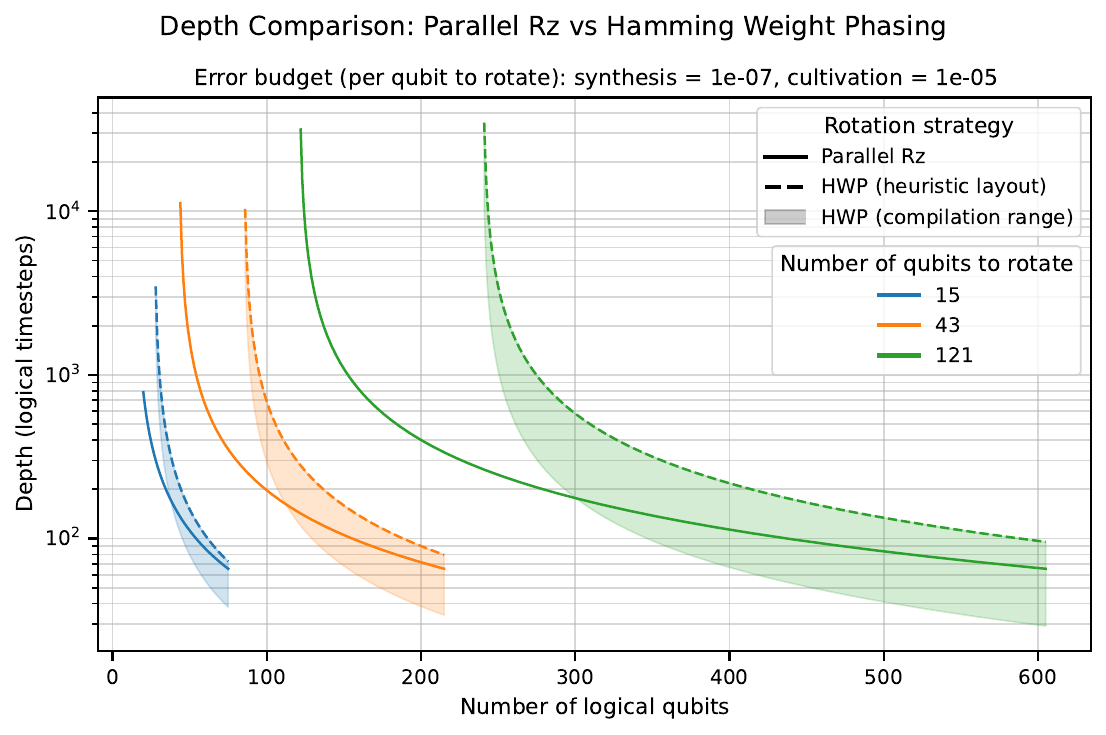}
	\caption{
		The number of logical timesteps required to perform \(N\) parallel \(R_Z\) rotations, targeting an error from rotation synthesis of \(N \times 10^{-7}\) and an error from magic state cultivation of \(N \times 10^{-5}\).
		Solid lines represent the naive Parallel \(R_Z\) approach.
		For Hamming weight phasing (HWP), the shaded region represents a range of potential performance bounded by using a heuristic to lay out the qubits on a 2D grid before applying the FLASQ model (dashed upper line) or neglecting the distance-dependent components of the cost (bottom edge of the shaded region).
		All curves begin only when enough total logical qubits are available and terminate when the total number of logical qubits is \(5N\) (in order to focus on the spacetime-limited regime).
		The FLASQ estimates reveal that realizing even a part of the advantage offered by Hamming weight phasing's lower T-count will require carefully optimizing its implementation, layout, and usage of ancilla qubits.	}
	\label{fig:HWP_big_plot}
\end{figure}

HWP involves structured arithmetic circuits whose cost (when including Clifford gates) is highly sensitive to how they are compiled and routed on a 2D lattice.
Because the available implementation does not compile the operation down to an optimized 2D layout, we analyze the performance by providing a range of estimates for the total number of timesteps required.
Specifically, we generate a pessimistic upper bound (to the optimal FLASQ estimate) by fully accounting for the distance-dependent costs using a heuristically-generated layout of the qubits on a 2D lattice (see \Cref{app:hwp_details} for more details).
We also generate a lower estimate by neglecting the distance-dependent components of the cost entirely, which represents a possibly unattainable idealized layout where these costs are negligible.
In \Cref{fig:HWP_big_plot}, we plot the resulting number of logical timesteps as a function of the total number of available logical qubits for several choices of \(N\) (the number of parallel rotations).

Several considerations are apparent from the figure.
First, the requirement for a significant number of algorithmic ancilla means HWP demands a minimum footprint of roughly \(2N\) logical qubits, restricting its use when the availability of extra space is low (e.g., the \(N=121\) variant requires \(7\) algorithmic ancilla to store the Hamming weight plus another \(116\) to calculate it using this approach). %
Second, the wide shaded region demonstrates that HWP's cost is sensitive to the quality of the 2D compilation.
The FLASQ estimates of the depth for the heuristic layout (dashed upper bound) are consistently larger than those for the parallel \(R_Z\) strategy across the analyzed ranges.
Regardless of the distance-dependent costs, we find that HWP only shows significant reductions in depth (or overall spacetime volume, since we are in the spacetime-limited regime) when using a sufficiently large number of qubits.

These findings contrast with the conclusion one might draw purely from considering the T gate count, presented in \Cref{tab:hwp_comparison_T_count_table}.
At this precision, the T-count analysis suggests that HWP is a factor of two to nearly four times cheaper than parallel rotation synthesis.
This comparison highlights the importance of a cost model based on spacetime volume.
Accounting for the cost of all operations using the FLASQ model reveals that the total ancilla volume is at most \(3\) times smaller for HWP than for parallel rotation synthesis and that the advantage may even disappear if our heuristic layout cannot be improved upon substantially.
Additionally, the opportunity cost of HWP's high algorithmic ancilla usage (in this formulation) further reduces the potential advantage by limiting the amount of space to mediate the computation (i.e., the number of fluid ancilla).
We present additional data substantiating these conclusions in \Cref{app:hwp_details}, where we examine their sensitivity to several factors and also explain our methodology in more detail.
In particular, we show that the qualitative conclusions presented in this section are not very sensitive to the exact estimates for the ancilla volumes of specific gates by presenting an alternative version of \Cref{fig:HWP_big_plot} using a more optimistic set of volume estimates (detailed in \Cref{app:flasq_volume_details})

\begin{table}[]
	\centering
	\setlength{\tabcolsep}{8pt}
	\resizebox{\textwidth}{!}{%
		\linespread{1.0}\selectfont%
		\renewcommand{\arraystretch}{1.2}
		\begin{tabular}{lcccccc}
			\toprule
			\makecell{\textbf{Qubits to}
			\\
			\textbf{Rotate}}               & \makecell{\textbf{HWP}
			\\
			\textbf{T-Count}}              & \makecell{\textbf{Parallel \(R_Z\)}
			\\
			\textbf{T-Count}}              & \makecell{\textbf{T-Count Ratio}
			\\
			\textbf{(\(R_Z\)/HWP)}}        & \makecell{\textbf{HWP Fluid}
			\\
			\textbf{Ancilla Volume}}       & \makecell{\textbf{Parallel \(R_Z\)}
			\\
			\textbf{Fluid Ancilla Volume}} & \makecell{\textbf{Ancilla Volume Ratio}
			\\
				\textbf{(\(R_Z\)/HWP)}}
			\\
			\midrule
			15                             & \(\approx 112\)                     & \(\approx 270\)   & \(\approx 2.4\times\) & \(1{,}829 \text{ to } 3{,}474\)   & \(3{,}920\)  & \(\approx 1.1 \text{ to } 2.1\times\)
			\\
			43                             & \(\approx 252\)                     & \(\approx 774\)   & \(\approx 3.1\times\) & \(4{,}427 \text{ to } 10{,}337\)  & \(11{,}238\) & \(\approx 1.1 \text{ to } 2.5\times\)
			\\
			121                            & \(\approx 576\)                     & \(\approx 2{,}178\) & \(\approx 3.8\times\) & \(10{,}663 \text{ to } 34{,}813\) & \(31{,}624\) & \(\approx 0.9 \text{ to } 3.0\times\)
			\\
			\bottomrule
		\end{tabular}
	}
	\caption{
		A comparison of the T gate counts and FLASQ ancilla volume for Hamming weight phasing (HWP) versus a naive parallel implementation for \(N\) rotations, targeting a rotation synthesis error of \(10^{-5}\).
		Two numbers are presented for the HWP ancilla volume, the smaller one neglects the distance-dependent costs, while the larger one uses our explicit heuristic layout.
		Despite the significant T-count advantage for HWP, the FLASQ analysis indicates that it will be challenging to obtain a similar reduction in ancilla volume once the distance-dependent costs are included.
		Furthermore, the potential advantage is also affected by the fact that HWP uses a large number of ancilla qubits, reducing the number of fluid ancilla available to provide the needed volume.
	}
	\label{tab:hwp_comparison_T_count_table}
\end{table}

\section{Discussion}
\label{sec:discussion}

The transition to early fault-tolerant quantum computing demands cost models that move beyond focusing on a single parameter, such as the T gate count, in isolation.
The \textbf{FL}uid \textbf{A}llocation of \textbf{S}urface code \textbf{Q}ubits (FLASQ) model that we have introduced addresses this necessity by also incorporating several important factors, including the cost of Clifford operations and the ability to make a range of tradeoffs between space and time.
FLASQ is intended to predict the resources required to implement quantum algorithms for a specific surface code architecture incorporating a variety of modern advances.
By assigning quantitative estimates to the impact of various design choices, FLASQ enables the systematic exploration of tradeoffs for early fault-tolerant algorithms.

Two case studies illustrate this capability.
The first case study analyzed the simulation of time dynamics in the 2D transverse field Ising model (TFIM) in \Cref{sec:ising_model}.
We applied FLASQ to estimate the resources required for simulating the 2D TFIM in a classically challenging (but perhaps not intractable) regime as a function of the physical error rate and the number of physical qubits available.
We found that the FLASQ model's predictions agreed with a hand-optimized compilation of this task, predicting the circuit depth to within a factor of \(\approx 25\%\).
We also used the model to estimate the relative costs of NISQ (many parallel noisy simulations) and fault-tolerant (one single error-corrected simulation) approaches to the same simulation task.
Finally, FLASQ estimates indicate that both the space and the time required for simulating the 2D TFIM are more than an order of magnitude lower than previously reported.
This suggests that the timeline to early fault-tolerant quantum advantage can be accelerated by optimally combining recent advancements, including magic state cultivation, walking surface codes, and a hybrid approach that uses a mixture of quantum error correction and mitigation.

As a second case study, we used FLASQ to analyze the Hamming weight phasing (HWP) approach to applying multiple copies of the same single-qubit rotation in parallel in \Cref{sec:hwp_example}.
Hamming weight phasing is designed to reduce T gate counts.
However, this reduction comes at the expense of increased implementation complexity, higher routing overheads, and additional ancilla qubits.
Recent advances, particularly the development of magic state cultivation~\cite{Craig2024-kv}, have reduced the cost of implementing high-fidelity T gates and called into question the single-minded focus on minimizing the T-count.
Applying FLASQ to estimate the spacetime volume of Hamming weight phasing suggests that the benefit of a lower T-count can easily be overwhelmed by other factors.
This is especially true when the difficulty of laying out the arithmetic circuits on a 2D lattice is considered, or when extra space is limited.
While our analysis leaves open the possibility of benefiting from HWP, it demonstrates that its optimization -- and algorithmic development in general -- must move beyond reducing the number of T gates and account for the interrelated factors that determine an algorithm's true cost.

The FLASQ model is designed to produce heuristic estimates rather than upper or lower bounds.
To understand its predictive power and its limitations, we compared its predictions against hand-compiled examples for both the Ising simulation (\Cref{sec:ising_model}) and arithmetic circuits (\Cref{app:validation_details}).
We find that the absolute accuracy of the FLASQ model is context-dependent.
The default (conservative) model parameters align well with the Ising estimates (within 13-24\%).
However, the conservative model overestimates the cost of dense, hand-optimized arithmetic circuits by more than a factor of 2 (see \Cref{app:validation_details}).

This discrepancy highlights a key difference between computational regimes.
The conservative model assumes a high overhead for allocating and restoring ancilla space for every operation.
Optimized arithmetic circuits, in contrast, reuse the same workspace repeatedly, achieving a high efficiency by amortizing this cost.
For such cases, the optimistic model parameters (introduced in \Cref{app:flasq_volume_details}) are significantly more predictive.
Despite this variation in absolute numbers, we observe that FLASQ's relative accuracy when comparing similar circuits is consistently high, regardless of parameter choice and context.
This suggests that FLASQ is particularly valuable as a tool for choosing between various approaches to the same task.

We also estimated the uncertainty in the FLASQ model's predictions by analyzing its sensitivity to various input parameters.
In the context of the  HWP case study, we generated two different FLASQ estimates to determine the sensitivity of our conclusions to the routing costs incurred by laying HWP out on a 2D grid.
The relatively high sensitivity we found highlights that minimizing the FLASQ cost requires thinking carefully about the layout of the input circuit.
We believe that this observation is not an artifact of the FLASQ model and will also apply to practical fault-tolerant compilation frameworks.
We perform a similar sensitivity analysis in \Cref{app:flasq_ising_classical_benchmark}, where we determine the minimal resources required to perform the Ising model simulation task considered in \Cref{sec:ising_classical_benchmark_comparison} using the more optimistic FLASQ parameters.
We find that even substantial reductions in the spacetime volumes assigned to individual gates result in only a modest shift in the number of physical qubits required to perform the simulation in a reasonable amount of time.
This indicates that the FLASQ's predictions about the onset of early fault-tolerance are robust to the specific assumptions about the cost of individual gates.

The uncertainties and limitations of the FLASQ model arise from the same core assumptions that make it tractable to calculate.
The model's main abstraction is treating ancilla space as a ``fluid'' resource.
This is motivated by efficient techniques for making spacetime tradeoffs (such as selective gate teleportation~\cite{Fowler2012-ti}) and efficiently rearranging logical qubits (using walking surface codes~\cite{McEwen2023-gk}).
This avoids the need to solve the problem of explicitly routing data and packing by assuming that the required spacetime volume is efficiently rearrangeable.
However, it is an idealization.
This idealization may break down and lead to underestimation of the true cost in scenarios where there is a high degree of routing congestion, or where the ancilla space is naturally fragmented and difficult to use.
Additionally, the model makes the simplifying assumption that we can decouple the spacetime volume constraint from the reaction time constraint, simply taking the maximum of the two.
This should yield reasonable behavior in the limit where one constraint dominates (and we expect the spacetime constraint to dominate for the earliest days of the early fault-tolerant era).
However, neglecting the interactions between these two limiting factors may result in an underestimation in the regime where both constraints are relevant.
Finally, our heuristic for determining the maximum simultaneous qubit usage is oversimplified, and algorithms with highly fluctuating space requirements might realize better performance than FLASQ currently predicts.

The FLASQ model we have presented here represents a first attempt at a holistic cost model for this particular early fault-tolerant architecture, and we expect that it will be built on over time.
One key next step is to continue calibrating and validating the model against a wider variety of hand-compiled examples, or against tools that output constructive upper bounds (such as an improved version of the CSC compiler we discuss in \Cref{app:ising_gosc_csc_comparison}).
Besides improving the FLASQ model, this calibration may also yield more rigorous insights about how an expert researcher can adjust the parameters of the model to adapt to the structure of the algorithm or subroutine being analyzed.
The gap between what is achievable by currently available automated compilation tools and what FLASQ predicts emphasizes the urgent need to develop new approaches to compilation.
We also anticipate improvements to the FLASQ model, including strategies for incorporating hand-compiled subroutines, better tools for estimating the amount of space available for use as fluid ancilla, and refined estimates of its uncertainty.
It would be interesting to extend our resource estimation framework to treat other architectures, although the use of long-range connections~\cite{Bluvstein2024-wc, Litinski2022-yw,Yoder2025-av}, other types of error correcting codes~\cite{Yoder2025-av, Breuckmann2021-qw, Bravyi2024-nb}, or more exotic architectural features~\cite{Jintae2024-uf} may necessitate other approaches to resource estimation.

Developing algorithms and applications in the early fault-tolerant era will require a quantitative understanding of the many tradeoffs and constraints offered by different architectures.
The FLASQ model takes an important step for early fault-tolerant resource estimation, going beyond simply counting the number of T gates or the circuit depth.
As the Hamming weight phasing case study reveals, optimizing EFT algorithms requires considering other factors, including the overhead of Clifford operations and the opportunity cost of using more ancilla.
The Ising model analysis highlights the dramatic impact that a combination of modern techniques can make in reducing the resources required for performing classically challenging tasks.
By providing a tractable and holistic framework for early fault-tolerant resource estimation, FLASQ enables the systematic exploration of this complex design landscape.
While FLASQ provides a heuristic estimate and does not replace the need for optimized hand-compilation, it offers a significantly better proxy for true implementation costs than metrics that ignore architectural constraints and the spacetime volume of Clifford operations.
We believe this comprehensive approach is essential for accelerating progress in the early fault-tolerant era.

\section*{Code Availability}
A beta version of the code implementing the FLASQ model is available upon request, as is the code used to generate the specific plots in this work.
We intend to integrate the FLASQ model into the Qualtran software package in a future release.

\begin{acknowledgments}
We thank Sergio Boixo, Dmitri Maslov, Craig Gidney, Oscar Higgot, Salvatore Mandra, Zlatko Minev, and Tom O'Brien for helpful discussions.
\end{acknowledgments}

%

\appendix

\section{Ancilla spacetime volume estimates in the FLASQ model}
\label{app:flasq_volume_details}

In this appendix, we explain our philosophy and methodology for estimating the ancilla spacetime volume.
The overall goal of the FLASQ model is to predict the resources required to implement quantum algorithms after they have been optimized by a (currently non-existent) compilation stack.
As compilers for classical computers today do, quantum compilers will take a high-level description of a quantum program as input and perform a series of complex optimizations (e.g., at the ZX-calculus level, or directly using SAT solvers like in \cite{Tan2024-cn}) that are difficult to model without actually implementing them.
We aim to produce estimates that are more predictive than simply counting the number of T gates or one- and two-qubit gates by implicitly accounting for some of the complex optimizations available to such a compiler.
To do so, we make use of the observation (and assumption) that it is often possible to flexibly trade between space and time in how we implement parts of a computation, and that these tradeoffs tend to conserve the extra ``ancilla'' spacetime volume required.

Based on this principle, we build our model by estimating the ancilla volume on a gate-by-gate basis.
This approach is conservative in some respects, as a true compiler would likely fuse certain gates together, reducing their overall cost.
To account for this, we design the model to be extensible; its accuracy can be improved by incorporating larger, pre-compiled operations as new primitives.\footnote{In the future it may be necessary to extend the FLASQ model with additional constraints to account for the properties of these large pre-compiled operations, but we leave this to a future work.}
This is why, while \Cref{tab:gate_costs_small} in the main text already contains costs for a universal set of gates, we present an augmented table that includes additional operations in \Cref{tab:gate_costs_large}.
In this section, we first discuss the principles and methodology used to arrive at these cost estimates before proceeding to explain how we arrived at the estimates of the ancilla spacetime volumes for each of the gates in the larger table.

Our approach focuses on estimating the cost of each primitive gate by studying one or two typical lattice surgery implementations.
For multi-qubit gates, we specifically analyze the gate acting on adjacent qubits.
This process involves concretely counting the amount of ancilla space and time required to implement the gate, often by examining its visual representation as a ``pipe diagram.''
This sometimes requires qualitative judgments, such as discounting small, isolated "stranded ancilla" regions that are unlikely to be usable by a real compiler.
Performing this step yields a concrete estimate of the ancilla spacetime volume for each primitive gate.

However, a single number is an oversimplification, as the actual ancilla volume required will vary based on the layout of the input and output qubits, the specific routing and packing challenges at particular points in the circuit, as well as the probabilistic nature of certain operations.
To help account for this, we actually propose two costs for each gate, one based on a default ``conservative'' model and a more ``optimistic'' one.
The difference between these two estimates of the ancilla volume represents an estimate of the scale of our uncertainty and allows us to test the sensitivity of our final results to the underlying assumptions.

The two models differ in several respects.
The conservative model assumes a larger cost for routing, roughly corresponding to the expected cost to free up ancilla space using walking surface codes and then restore the configuration of the qubits after performing the gate.
It also assumes the worst case in terms of required volume for operations whose cost is uncertain (with the exception of arbitrary rotations, which are discussed in \Cref{app:rotation_synthesis_details}).
The optimistic model assumes that one must pay a cost to free up ancilla spacetime volume, but that the qubits do not need to be restored to a canonical position after the gate is performed.
In this model, we instead imagine a sea of qubits constantly shuffling slightly to allocate ancilla for performing specific gates.
It also assumes a faster speed for the walking surface movement, based on results alluded to in \cite{Gidney2024-fy} but not published or carefully analyzed elsewhere.
The optimistic model also uses expected values rather than worst-case values for costs that are uncertain (but could in principle be determined at compile time; we do not discount costs that will necessarily remain uncertain until partway through the execution of an algorithm).
We present both the conservative and optimistic estimates for a range of primitive gates in the expanded \Cref{tab:gate_costs_large}.

The choice between the conservative and optimistic models should be guided by the structure of the algorithm or subroutine being analyzed and by the purpose of the analysis. 
The conservative model is designed to provide a robust and relatively safe estimate for general algorithms, particularly those involving significant long-range interactions and complex routing challenges (such as the Ising simulation in \Cref{sec:ising_model}). 
However, as shown in \Cref{app:validation_details}, it may be overly pessimistic for dense, localized computations (such as optimized arithmetic circuits) where ancilla space can be efficiently reused and the cost of providing it amortized. 
For such cases, the optimistic model may provide a more accurate estimate of the achievable performance.
When it is useful to establish a plausible range of performance, then we recommend considering the variation between two models as a rough estimate of uncertainty.
When it is more important to avoid underestimating the cost, then defaulting to the conservative model would be a safer choice, although its estimates could be improved by hand-compiling specific primitives.

The estimates of ancilla spacetime volume we have discussed so far are for single-qubit gates or for multi-qubit gates acting on adjacent qubits.
However, it is straightforward to perform lattice surgery operations between distant qubits, provided that a connecting corridor of unused ancilla qubits is available.
In the past, this has motivated resource estimates using layouts that ensure that all of the qubits are accessible with such routing corridors (also referred to as ``hallways'').
However, the walking surface code techniques of Refs.~\citenum{McEwen2023-gk,Eickbusch2025-xl} allow for us to collectively shift whole blocks of qubits, making it efficient to open up these corridors on demand.
The implications of this for routing are elaborated on in Appendix C of \cite{McEwen2023-gk}, and we highlight them with our first example of a long-range gate in \Cref{app:move_gate_details}.

Essentially, we can usually expect to open up a corridor of qubits using a cost that depends only on the length of the path (in the Manhattan distance), multiplied by some constant factor.
For the conservative model, we take this factor to be \(4\), using the speed estimates from \cite{McEwen2023-gk} (a block of logical qubits can be moved over one space in \(2\) logical timesteps), together with the conservative assumption that one should pay the same cost to restore the qubits to their previous position after performing the operation.
For the optimistic model, we take this factor to be \(1\).
We justify this decision partly by the assumption that each operation can cause the qubits to ``jiggle'' a bit and therefore we don't need to pay the cost to return displaced qubits to their previous positions.
Additionally, \cite{Gidney2024-fy} alluded to the possibility of walking surface code constructions that move at faster speeds, although more research is needed to explore if this is truly possible without increasing error rates.
Besides these points, we also expect it to frequently be the case that several operations will be able to make use of the same ancilla space, implying that the cost we assign to any individual operation should actually be a smaller amortized cost that accounts for this.

In order to support multi-qubit gates between distant qubits in our model, we use this requirement for a connecting sequence of qubits to determine how the geometric component of the cost ``should scale.''
We add the cost of opening (and possibly closing) the corridor to the cost of actually performing any lattice surgery steps with a distance dependent ancilla volume requirement.
We then derive cost formulas by combining a base cost, determined from the canonical adjacent-qubit example, with a term that captures the geometric scaling.
For example, for two-qubit gates, the cost scales linearly with some multiple of the Manhattan distance between the qubits, \(p(q_1, q_2)\).

For three qubits, it can be shown that the minimal length of a connecting path between the qubits is given by the formula
\begin{equation}
	p(q_1, q_2, q_3) = \frac{1}{2} \left( p(q_1, q_2) + p(q_1, q_3) + p(q_2, q_3) \right).
\end{equation}
The problem of finding the shortest connecting path in the general case is known as the rectilinear Steiner tree problem, and it is known to be NP hard~\cite{Hwang1992-uj}.
However, it is possible to efficiently approximate the best path by finding the rectilinear minimum spanning tree.
Such an approximation is at most \(50\%\) larger than the shortest path~\cite{Hwang1976-yv}.
We therefore let \(p(q_1, \dots, q_k)\) denote such an approximation for \(k \geq 4\), and we use this as the basis of our costs for multi-qubit gates with four or more qubits.

One key simplification the FLASQ model makes is that we separately enforce constraints based on the ancilla spacetime volume and the measurement depth required by a computation.
We simply calculate the minimum number of logical timesteps required for each and take the maximum.
While we generally follow this principle of decoupling these two components of the cost, for certain primitive gates (e.g., a Toffoli gate) the ancilla volume itself has a small component that necessarily scales with the reaction time (\(t_{react}\)).
Where they can be determined, we include these direct costs by making the assumption that we are not in the reaction-limited regime (and therefore the delays for the classical control software can be computed gate-by-gate in isolation).
We make this choice in order to make the model as predictive as possible while keeping it tractable, although it means that we ignore the more complex coupling between the required spacetime volume and the measurement depth that may be important in some regimes.

Having explained the broad principles we use to generate our estimates of the ancilla volume, we now proceed entry by entry through the gates contained in \Cref{tab:gate_costs_large}.

\begin{rotatepage}
	\begin{turnpage}
		\begin{table*}[]
			\centering
			\setlength{\tabcolsep}{10pt}
			\resizebox{1\textheight}{!}{%
				\linespread{1.0}\selectfont%
				\renewcommand{\arraystretch}{2.2}
				\begin{tabular}
					{@{}lllll@{}}
					\toprule
					Basic gates                                                                                                             & Conservative estimate of ancilla volume                                                                & Optimistic estimate of ancilla volume                                                     & Measurement depth & Notes
					\\
					\midrule
					\(X \;/\;Y\;/\;Z\)                                                                                                      & \(0\)                                                                                                  & \(0\)                                                                                     & 0                 & Implemented in software
					\\
					\makecell{\(X\;/\;Z\) basis measurement
					\\
					or initialization}                                                                                                      & \(0\)                                                                                                  & \(0\)                                                                                     & 0                 &
					\\
					\(H\)                                                                                                                   & \(7\)                                                                                                  & \(1.5\)                                                                                     & 0                 & Including the cost of patch rotations
					\\
					\(S \;/\; S^\dagger\)                                                                                                   & \(5.5\)                                                                                                & \(1.5\)                                                                                  & 0                 &
					\\
					Cultivate                                                                                                               & \(1.5 v(p_{phys}, p_{cult})\)                                                                          & \(v(p_{phys}, p_{cult})\)                                                                 & 0                 & \makecell[l]{
						Depends on physical error rate, \(p_{phys}\),
					\\
						and target logical error rate, \(p_{cult}\)
					}
					\\
					\(T \;/\; T^\dagger\)                                                                                                   & \(\textsc{Vol}\left( \textrm{Cultivate} \right) + t_{react} + 6\)                                      & \(\textsc{Vol}\left( \textrm{Cultivate} \right) + t_{react} + 2.5\)                       & 1                 & Includes freeing space for cultivation
					\\
					Move                                                                                                                    & \(5 p(q_1, q_2)\)                                                                                      & \(2 p(q_1, q_2)\)                                                                         & 0                 & Moves a qubit to an empty patch
					\\
					\(CNOT \;/\; CZ\)                                                                                                       & \(5 p(q_1, q_2)\)                                                                                      & \(2 p(q_1, q_2)\)                                                                         & 0                 &
					\\
					\midrule
					Additional primitive gates                                                                                              & Conservative estimate of ancilla volume                                                                & Optimistic estimate of ancilla volume                                                     & Measurement depth & Notes
					\\
					\midrule
					\(T_X = H T H \;/\; T_X^\dagger\)                                                                                       & \(\textsc{Vol}\left( T \right)\)                                                                       & \(\textsc{Vol}\left( T \right)\)                                                          & 1                 & Basically the same as \(T\) and \(T^\dagger\)
					\\
					\(R_Z(\theta) / R_X(\theta)\)                                                                                           & \makecell[l]{\( (0.53 \log_2(\epsilon^{-1}) + 4.86) \left( 2 + \textsc{Vol}\left( T \right) \right) \)
					\\
					\( + 2 \textsc{Vol}\left( S \right) + 2 \textsc{Vol}(H) + 2 \textsc{Vol}\left( CNOT \right) + 10 \)}                    & \makecell[l]{\( (0.53 \log_2(\epsilon^{-1}) + 4.86) \left( 1 + \textsc{Vol}\left( T \right) \right) \)
					\\
					\( + \frac{7}{6} \textsc{Vol}\left( S \right) + \frac{5}{6} \textsc{Vol}(H) + 2 \textsc{Vol}\left( CNOT \right) + 5 \)} & \(0.53 \log_2(\epsilon^{-1}) + 4.86\)                                                                                                    & Depends on rotation synthesis error, \(\epsilon\)
					\\
					SWAP                                                                                                                    & \(6 p(q_1, q_2)\)                                                                                      & \(3 p(q_1, q_2)\)                                                                         & 0                 & Swaps the state of two qubits
					\\
					AND                                                                                                                     & \(4 \textsc{Vol}\left( \text{Cultivate} \right) + 2 t_{react} + 5 p(q_1, q_2, q_3) + 64\)              & \(4 \textsc{Vol}\left( \text{Cultivate} \right) + 2 t_{react} + 2 p(q_1, q_2, q_3) + 36\) & 1                 &
					\\
					AND\(^\dagger\)                                                                                                         & \(\textsc{Vol}\left( CZ(q_1, q_2) \right)\)                                                            & \(\textsc{Vol}\left( CZ (q_1, q_2) \right)\)                                              & 1                 & Uncomputes the AND of \(q_1\), \(q_2\)
					\\
					Toffoli                                                                                                                 & \(4 \textsc{Vol}\left( \text{Cultivate} \right) + 5 t_{react} + 5 p(q_1, q_2, q_3) + 68\)              & \(4 \textsc{Vol}\left( \text{Cultivate} \right) + 5 t_{react} + 2 p(q_1, q_2, q_3) + 39\) & 2                 & 
					\\
					\makecell{\(Y\) basis measurement
					\\
					or initialization}                                                                                                      & \(1\)                                                                                                  & \(.5\)                                                                                    & 0                 &
					\\
					\bottomrule
				\end{tabular}
			} \caption{ 
				Estimates of the fluid ancilla (spacetime) volume required to implement various gates (in units of ``blocks,'' i.e., \(d^3\)), as well as the required measurement depth (in units of ``logical timesteps,'' i.e., \(d\)).
				Note that this volume is in addition to the volume required to store the qubit(s) the gate acts on.
				Informally, \(p(q_1, \cdots, q_k)\) denotes the minimum length (measured in the manhattan distance) required to connect the specified qubits.
				The volume required to cultivate a magic state (\(v(p_{phys}, p_{cult})\)) depends on the underlying physical error rate (\(p_{phys}\)) and the target logical error rate (\(p_{cult}\)).
				The table contains two columns for the estimated volume, one that errs on the conservative side and one that may be overly optimistic except in special cases.
			}
			\label{tab:gate_costs_large}
		\end{table*}
	\end{turnpage}
\end{rotatepage}

\subsection{Pauli gates (\(X, Y, Z\))}

Pauli gates can be implemented by updating the state of the classical control software, so they require zero ancilla spacetime.

This implementation is realized through Pauli frame tracking~\cite{Knill2005-ah,Fowler2012-li}.
As discussed in \Cref{sec:background} (specifically, the subsection on Computation via Lattice Surgery), this technique allows the classical control software to virtually track the accumulation of Pauli operators—whether they arise from intentional gates, error correction byproducts, or probabilistic gate teleportation.
By updating the interpretation of future measurement outcomes, physical Pauli operations on the quantum device are avoided entirely, incurring zero spacetime overhead.

\subsection{X/Z basis measurement and initialization}

In the rotated surface code, initialization and measurement in the X or Z basis are performed transversally.
To initialize a logical state (e.g., \(\ket{0}_L\)), all physical data qubits constituting the patch are simultaneously prepared in the corresponding physical state (e.g., \(\ket{0}\)).
Likewise, logical measurement is achieved by transversally measuring all data qubits in the desired basis.

While the physical initialization or measurement occurs in a single surface code cycle (independent of the code distance \(d\)), fault tolerance requires that these operations are accompanied by \(d\) cycles of stabilizer measurements to detect potential errors during the process before additional operations are performed on the qubit~\cite{Fowler2012-li,Horsman2012-vd}.
However, these operations require no additional \textit{ancilla spacetime volume} in the FLASQ model and so we assign them a cost of \(0\).
This corresponds roughly to the assumption that we can always compile things such that this delay is a negligible overhead that we would have incurred anyway while waiting to perform other operations.

\subsection{Hadamard gate (H)}

Applying a Hadamard gate to each of the data qubits performs a transversal Hadamard, except that it effectively rotates the logical qubit by \(90\) degrees (by swapping the X and Z boundaries of the surface code patch).
We can restore the original orientation by performing a patch rotation using a single ancilla qubit for three logical time steps~\cite{Litinski2019-nu}.
The conservative estimate comes from allocating a single ancilla qubit (\(4\) units of volume) and assuming that we must restore the patch to its original orientation and location after every Hadamard gate (\(3\)).
The optimistic estimate of \(1.5\) comes from assuming that we only need to perform the patch rotation with \(50\%\) probability and that we do not need to restore the exact position of the qubit (together with the cheaper cost of the ancilla qubit allocation).
In this case, the ancilla volume is effectively \(0\) half of the time because a transversal Hadamard is sufficient and \(3\) the rest of the time (\(1\) to allocate the space and \(2\) for a patch rotation that does not move the qubit back to its original location).

\subsection{Phase gate (S, S\(^\dagger\))}

Using the techniques of \cite{Gidney2024-fy}, we can implement an S gate on a stationary logical qubit using one ancilla qubit for \(1.5\) logical time steps.
If the qubit is in motion (i.e., the S gate is combined with a move operation for some other purpose), then we can perform the \(S\) gate using \(1\) logical qubit acting for \(.5\) time steps.
In our conservative estimates we use the larger static volume estimate, plus we account for the volume required to allocate and deallocate an ancilla qubit's worth of space (\(4\)).
In our optimistic estimates we assume that the S gate can be applied in motion \(50\%\) of the time.
When the gate is applied in applied in motion, we only account for the additional \(.5\) blocks of ancilla spacetime, since ancilla would already be in allocated for the operation that results in the motion.
The static cost in the optimistic model is \(1.5\) for the operation plus another \(1\) to free the space.
The average of these two costs is \(1.5\).

\subsection{Cultivation} \label{app:cultivation_details}

One of the modern advancements motivating the FLASQ model is the development of magic state cultivation~\cite{Craig2024-kv}, which we expect to dramatically reduce the cost of non-Clifford gates in the early fault-tolerant regime.
Even with this dramatic reduction, we expect the cost of non-Clifford gates to be a significant factor in many applications, and so the accuracy of our cost estimates will depend on having a good understanding of the spacetime volume required for cultivation, $v(p_{phys}, p_{cult})$.
This section details the methodology used to derive these costs, which closely follows the techniques and code from \cite{Craig2024-kv}.
We also briefly explain the underlying assumptions that we use to interpret these costs and include them in the FLASQ model.
We anticipate further work developing magic state cultivation and analyzing it in a variety of noise models (in fact, there are already some followup works~\cite{Wan2025-dx, Wan2025-du, Chen2025-cu, Vaknin2025-yt}).
As this line of research evolves, it will likely make sense to adjust the specific numbers used for the FLASQ model.

Magic state cultivation is a technique for generating high-fidelity magic states.
It works by injecting a noisy magic state into a small color code, performing checks to help ensure it was injected correctly (using techniques similar to those used in magic state distillation), and quickly growing the distance of the code.
Throughout this process, it relies heavily on post-selection, aborting any attempts where the decoder detects a significant failure probability.
\cite{Craig2024-kv} designed and validated this protocol with extensive numerical simulations, studying the performance at several different physical error rates and choices of the final postselection threshold.
Because we are interested in understanding the overall cost of algorithms as a function of the underlying physical error rate, we perform similar simulations at a larger variety of error rates.

To be precise, we use the code released alongside \cite{Craig2024-kv} to simulate the cultivation of the stabilizer state \(S \ket{+}\) under a variety of error rates.
Following their work, we use a uniform depolarizing noise channel, essentially using the same error rate to apply bit-flip errors upon measurement, single-qubit depolarizing noise after each physical single-qubit gate or idling step, and two-qubit depolarizing noise after each two-qubit gate.
We use the ``desaturation'' decoder specified in Section 2.5 of that work and make the same choice for the final code distance (\(15\)) and the number of rounds spent idling in the final code \(5\) as in their Figure 14.
We perform \(5 \times 10^9\) simulations at each physical error rate for the distance \(3\) cultivation protocol and \(6 \times 10^10\) simulations at each physical error rate for the distance \(5\) cultivation protocol.
For each physical error rate and choice of code distance for the cultivation protocol (\(3\) or \(5\)), we map out the achievable logical error rates by varying the threshold for the final postselection step.
This gives us an understanding of the number of cultivation attempts required as a function of the physical error rate and the target logical error rate.
For each physical error rate, given a final postselection threshold we take the logical error rate to be equal to the maximum likelihood estimate given the simulated data.
We note that there is significant uncertainty in the estimated logical error rates in the low logical error rate regime, and further numerical and experimental studies will be necessary to understand the exact performance of cultivation~\cite{Craig2024-kv}.

\begin{figure}
	\centering
	\includegraphics[width=.75\textwidth]{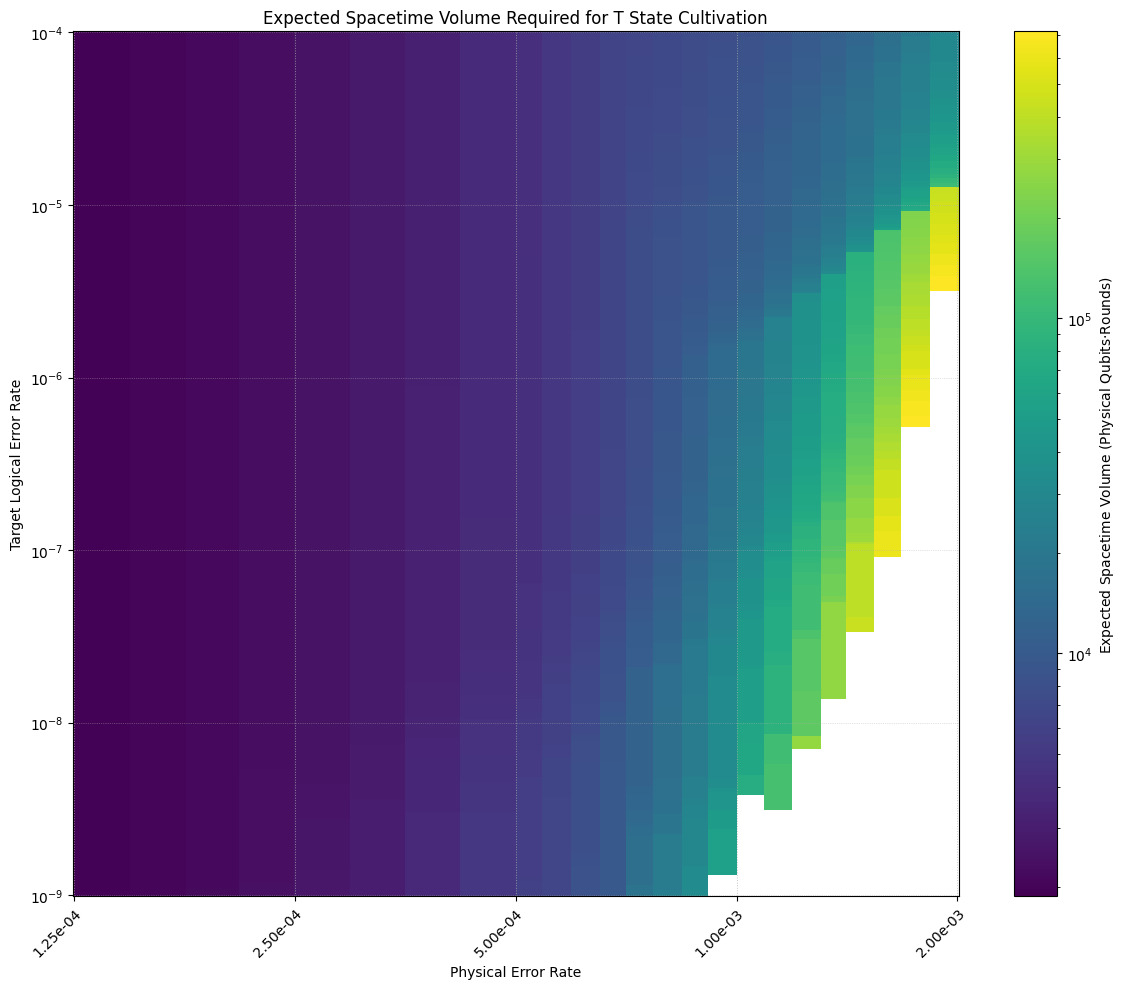}
	\caption{
		The expected spacetime volume of magic state cultivation as a function of the physical error rate and the target logical error rate.
		For each data point, we then estimate the expected spacetime volume taking into account attempts where we abort early, yielding an expected spacetime volume for each data point.
		To generate the heatmap, we report the volume associated with the closest data point by rounding the physical error rate up and the target logical error rate down (so that we produce an overestimate of the volume rather than interpolating).
		When there is a choice between using distance \(3\) and distance \(5\) cultivation, we choose whichever yields the lowest spacetime volume.
	}
	\label{fig:cultivation_expected_cost}
\end{figure}

Understanding the expected number of retries is useful to understanding the expected spacetime volume of the cultivation procedure, but we also need to account for the fact that a significant fraction of cultivation attempts may be aborted early in the process.
To do so, we again follow the methodology of \cite{Craig2024-kv} and estimate the expected spacetime volume in units physical qubits \(\times\) surface code cycles.
We follow the approach used to generate Figure 15 in that work, using simulated data to estimate the number of qubits in use after each cycle and the fraction of attempts that persist after each cycle.
Integrating the product of these two functions yields the expected spacetime volume, although this ignores packing constraints.
We perform this calculation at a variety of physical error rates and choices of the final postselection threshold, using the same uniform depolarizing noise model together with \(10\) rounds of idling in the final distance \(15\) code, as in the simulations of \cite{Craig2024-kv}.
We follow \cite{Craig2024-kv} and estimate that the logical error rate of a cultivated \(T\) state is twice that of our simulated \(S\ket{+}\) states, although this assumption may break down at low error rates (where it is not substantiated by the simulations in that previous work).
We plot this data in \Cref{fig:cultivation_expected_cost}, where we present an estimate of the spacetime cost of cultivation as a function of the physical error rate and the target logical error rate.

Following closely the methodology of \cite{Craig2024-kv} allows us to generate estimates of the spacetime volume in terms of physical qubits \(\times\) surface code cycles, which we convert into units of logical blocks (the conversion factor is distance dependent).
The values we calculate represent the total expected spacetime volume for the cultivation procedure, including the overhead from postselection and aborted attempts, but it does not account for the additional overhead required to pack aborted attempts into space and time.
The optimistic volume estimates for cultivation are taken directly from these simulation results, and therefore likely represent an underestimate of the true ancilla volume required (since it does not account for the practical overheads from packing).
For the conservative estimates, we use \(1.5 \times v(p_{phys}, p_{cult})\).
This \(1.5\times\) multiplicative factor serves as a heuristic buffer to account for the overhead from imperfectly ``packing'' the cultivation attempts in space and time~\footnote{This factor of \(1.5\) was suggested in a private communication with Gidney.}.
We expect that future work will address the challenges of optimizing magic state cultivation in practice, but it is beyond the scope of this present work.
Unlike other operations in our model, we do not explicitly account for the volume required to allocate an ancilla qubit for cultivation; this is because because we always make use of this volume estimate in the context of assigning a cost to other operations that account for the allocation of ancilla space.

As we explained, our estimates of the spacetime volume required for cultivation use a specific uniform depolarizing noise model (UDM) outlined in \cite{Craig2024-kv}.
This noise model is parameterized by a single noise strength, but it differs from the abstract physical error rate $p_{phys}$ used elsewhere in the FLASQ model to determine the surface code suppression factor $\Lambda$.
Both of these models are only rough approximations to the true noise realized by physical hardware.
However, the UDM model is closely related to the various noise models frequently used in analyzing surface code performance and thresholds~\cite{Fowler2012-li, Raussendorf2007-sa, Wang2011-ud}.
For the purposes of our analysis here, we conflate the noise strength of the UDM model from \cite{Craig2024-kv} and the abstract paramater \(p_{phys}\), assuming that we can treat treat them equivalently.
A slightly more careful study would analyze the surface code threshold using the same noise model as the cultivation protocol, and an even more careful study would use the most realistic noise models available for both simulations.
Nevertheless, for our purposes, we believe that equating the abstract parameter \(p_{phys}\) with the depolarizing noise strength for the UDM model is sufficient.

\subsection{T gate (\(T, T^\dagger\))}

The T gate is implemented via gate teleportation using a magic state prepared by cultivation.
This process involves preparing a magic state in an adjacent ancilla location, followed by a joint \(ZZ\) parity measurement between the magic state and the data qubit~\cite{Gidney2025-bn}.
This has a 50\% chance of applying the desired T or \(T^\dagger\) gate and a \(50\%\) chance of applying its conjugate (which necessitates a Clifford correction).
This correction can be implemented with a Y-basis measurement, but the decision on whether or not to perform this measurement requires waiting for the results of the parity measurement to be decided by the decoder.
As a result, implementing the T gate has a measurement depth of one.
The total ancilla volume is the sum of five components: the cultivation itself, the volume required to allocate a single ancilla qubit's worth of space, the volume of the parity measurement, the volume of waiting to resolve the outcome of the parity measurement, and the potential correction.

For the conservative estimate, the ancilla volume is given by
\begin{align*}
	\textsc{Vol} & = \textsc{Vol}\left( \textsc{Cultivate} \right) + \textsc{Vol}\left( \textsc{Alloc} \right) + \textsc{Vol}\left( \textsc{Parity} \right) + \textsc{Vol}\left( \textsc{Wait} \right) + \textsc{Vol}\left( \textsc{Correct} \right)
	\\
	             & = \textsc{Vol}\left( \textsc{Cultivate} \right) + 4 + 1 + t_{react} + \textsc{Vol}\left( Y_{meas} \right)
	\\
	             & = \textsc{Vol}\left( \text{Cultivate} \right) + 4 + 1 + t_{react} + 1 = \textsc{Vol}\left( \text{Cultivate} \right) + t_{react} + 6.
\end{align*}

For the optimistic estimate, the numbers differ because of two factors: First, we assign a cost of \(1\) to allocate an ancilla qubit rather than \(4\), and second, we do not round up the cost of the \(Y\) basis measurement to a full block.
For this model, the ancilla volume is given by
\begin{align*}
	\textsc{Vol} & = \textsc{Vol}\left( \textsc{Cultivate} \right) + \textsc{Vol}\left( \textsc{Alloc} \right) + \textsc{Vol}\left( \textsc{Parity} \right) + \textsc{Vol}\left( \textsc{Wait} \right) + \textsc{Vol}\left( \textsc{Correct} \right)
	\\
	             & = \textsc{Vol}\left( \textsc{Cultivate} \right) + 1 + 1 + t_{react} + \textsc{Vol}\left( Y_{meas} \right)
	\\
	             & = \textsc{Vol}\left( \text{Cultivate} \right) + 1 + 1 + t_{react} + .5 = \textsc{Vol}\left( \text{Cultivate} \right) + t_{react} + 2.5.
\end{align*}

\subsection{Move ``gate''} \label{app:move_gate_details}

The `Move` operation serves as a simple example of a long-range operation implemented via lattice surgery.
Physically, this is realized by `growing' the surface code patch to encompass the destination, waiting for a timestep (to ensure fault tolerance) and then `shrinking' the extended logical qubit away from the origin.
To perform this over a distance, a corridor of ancilla qubits must be cleared by moving neighboring data qubits out of the way.

\begin{figure}
	\centering

	\begin{subfigure}[t]{0.48\textwidth}
		\centering
		\begin{tikzpicture}[scale=0.7]
			\foreach \x in {0,...,9} { \foreach \y in {1,...,3} { \fill[gray!20] (\x,\y) rectangle (\x+1,\y+1); } }
			\fill[yellow!50!orange] (1,2) rectangle (2,3);
			\fill[white] (8,2) rectangle (9,3);
			\fill[pattern=north east lines, pattern color=blue!60] (8,2) rectangle (9,3);
			\foreach \x in {0,...,10} { \foreach \y in {0,...,3} { \draw (\x,\y) rectangle (\x+1,\y+1); } }
		\end{tikzpicture}
		\caption{\linespread{1.0}\selectfont Initial layout at \(t=0\).
			The yellow qubit is to be moved to the hatched location.
		}
	\end{subfigure}
	\hfill
	\begin{subfigure}[t]{0.48\textwidth}
		\centering
		\begin{tikzpicture}[scale=0.7]
			\foreach \x in {0,...,9} { \foreach \y in {1,...,3} { \fill[gray!20] (\x,\y) rectangle (\x+1,\y+1); } }
			\fill[yellow!50!orange] (1,2) rectangle (2,3);
			\fill[white] (8,2) rectangle (9,3);
			\fill[pattern=north east lines, pattern color=blue!60] (8,2) rectangle (9,3);
			\fill[red, opacity=0.3] (2,1) rectangle (8,3);
			\foreach \x in {2,...,7} { \draw[-{Stealth[length=3mm]}, very thick, red] (\x.5, 1) -- (\x.5, 0.4); }
			\foreach \x in {0,...,10} { \foreach \y in {0,...,3} { \draw (\x,\y) rectangle (\x+1,\y+1); } }
		\end{tikzpicture}
		\caption{\linespread{1.0}\selectfont Twelve data qubits (shaded red) move down to create a corridor.
			Ancilla Volume: 6 blocks, 2 logical timesteps (\(12\)).
		}
	\end{subfigure}

	\vspace{1em}

	\begin{subfigure}[t]{0.48\textwidth}
		\centering
		\begin{tikzpicture}[scale=0.7]
			\fill[gray!20] (0,1) rectangle (10,2); \fill[gray!20] (0,3) rectangle (10,4);
			\foreach \x in {0,9} { \fill[gray!20] (\x,2) rectangle (\x+1,3); }
			\fill[gray!20] (2,0) rectangle (8,1);
			\fill[yellow!50!orange] (1,2) rectangle (2,3);
			\fill[white] (8,2) rectangle (9,3);
			\fill[pattern=north east lines, pattern color=blue!60] (8,2) rectangle (9,3);
			\foreach \x in {0,...,10} { \foreach \y in {0,...,3} { \draw (\x,\y) rectangle (\x+1,\y+1); } }
		\end{tikzpicture}
		\caption{\linespread{1.0}\selectfont Layout at \(t=2\), after creating the corridor.}
	\end{subfigure}
	\hfill
	\begin{subfigure}[t]{0.48\textwidth}
		\centering
		\begin{tikzpicture}[scale=0.7]
			\fill[gray!20] (0,1) rectangle (10,2); \fill[gray!20] (0,3) rectangle (10,4);
			\foreach \x in {0,9} { \fill[gray!20] (\x,2) rectangle (\x+1,3); }
			\fill[gray!20] (2,0) rectangle (8,1);
			\fill[yellow!50!orange] (1,2) rectangle (8,3);
			\fill[yellow!50!orange] (8,2) rectangle (9,3);
			\fill[pattern=north east lines, pattern color=blue!60] (8,2) rectangle (9,3);
			\foreach \x in {0,...,10} { \foreach \y in {0,...,3} { \draw (\x,\y) rectangle (\x+1,\y+1); } }
		\end{tikzpicture}
		\caption{\linespread{1.0}\selectfont The logical qubit patch grows to the target location via lattice surgery.
			Ancilla Volume: 6 blocks, 1 logical timestep (\(6\)).
		}
	\end{subfigure}

	\vspace{1em}

	\begin{subfigure}[t]{0.48\textwidth}
		\centering
		\begin{tikzpicture}[scale=0.7]
			\fill[gray!20] (0,1) rectangle (10,2); \fill[gray!20] (0,3) rectangle (10,4);
			\foreach \x in {0,9} { \fill[gray!20] (\x,2) rectangle (\x+1,3); }
			\fill[gray!20] (2,0) rectangle (8,1);
			\fill[yellow!50!orange] (8,2) rectangle (9,3);
			\foreach \x in {0,...,10} { \foreach \y in {0,...,3} { \draw (\x,\y) rectangle (\x+1,\y+1); } }
		\end{tikzpicture}
		\caption{\linespread{1.0}\selectfont Layout at \(t=3\), after the patch shrinks from the original location, completing the move.}
	\end{subfigure}
	\hfill
	\begin{subfigure}[t]{0.48\textwidth}
		\centering
		\begin{tikzpicture}[scale=0.7]
			\fill[gray!20] (0,1) rectangle (10,2); \fill[gray!20] (0,3) rectangle (10,4);
			\foreach \x in {0,9} { \fill[gray!20] (\x,2) rectangle (\x+1,3); }
			\fill[gray!20] (2,0) rectangle (8,1);
			\fill[yellow!50!orange] (8,2) rectangle (9,3);
			\fill[red, opacity=0.3] (1,1) rectangle (2,2);
			\fill[red, opacity=0.3] (2,0) rectangle (8,2);
			\foreach \x in {1,...,7} { \draw[-{Stealth[length=3mm]}, very thick, red] (\x.5, 2) -- (\x.5, 2.6); }
			\foreach \x in {0,...,10} { \foreach \y in {0,...,3} { \draw (\x,\y) rectangle (\x+1,\y+1); } }
		\end{tikzpicture}
		\caption{\linespread{1.0}\selectfont Thirteen data qubits (shaded red) move back to restore the initial layout.
			Ancilla Volume: 7 blocks, 2 logical timesteps (\(14\)).
		}
	\end{subfigure}

	\vspace{1em}

	\begin{subfigure}[t]{0.48\textwidth}
		\centering
		\begin{tikzpicture}[scale=0.7]
			\foreach \x in {0,...,9} { \foreach \y in {0,...,3} { \fill[gray!20] (\x,\y) rectangle (\x+1,\y+1); } }
			\fill[white] (0,0) rectangle (10,1);
			\fill[white] (1,1) rectangle (2,2);
			\fill[yellow!50!orange] (8,2) rectangle (9,3);
			\foreach \x in {0,...,10} { \foreach \y in {0,...,3} { \draw (\x,\y) rectangle (\x+1,\y+1); } }
		\end{tikzpicture}
		\caption{\linespread{1.0}\selectfont The final layout at \(t=3\).}
	\end{subfigure}

	\caption{\linespread{1.0}\selectfont An example of a `Move` operation with a conservative estimate of the ancilla volume. (a) A logical qubit (yellow) is to be moved to the hatched location. (b) To create a corridor, surrounding data qubits are moved out of the way. (c-e) The logical qubit is moved by first 'growing' the patch to the destination and then 'shrinking' it from the origin. (f-g) The displaced data qubits are then moved back to restore the original, compact layout.
		The total ancilla volume of \(32 \approx 5 p(q_1, q_2) = 35\) accounts for creating the corridor, performing the move, and restoring the layout.
	}
	\label{fig:move_conservative}
\end{figure}
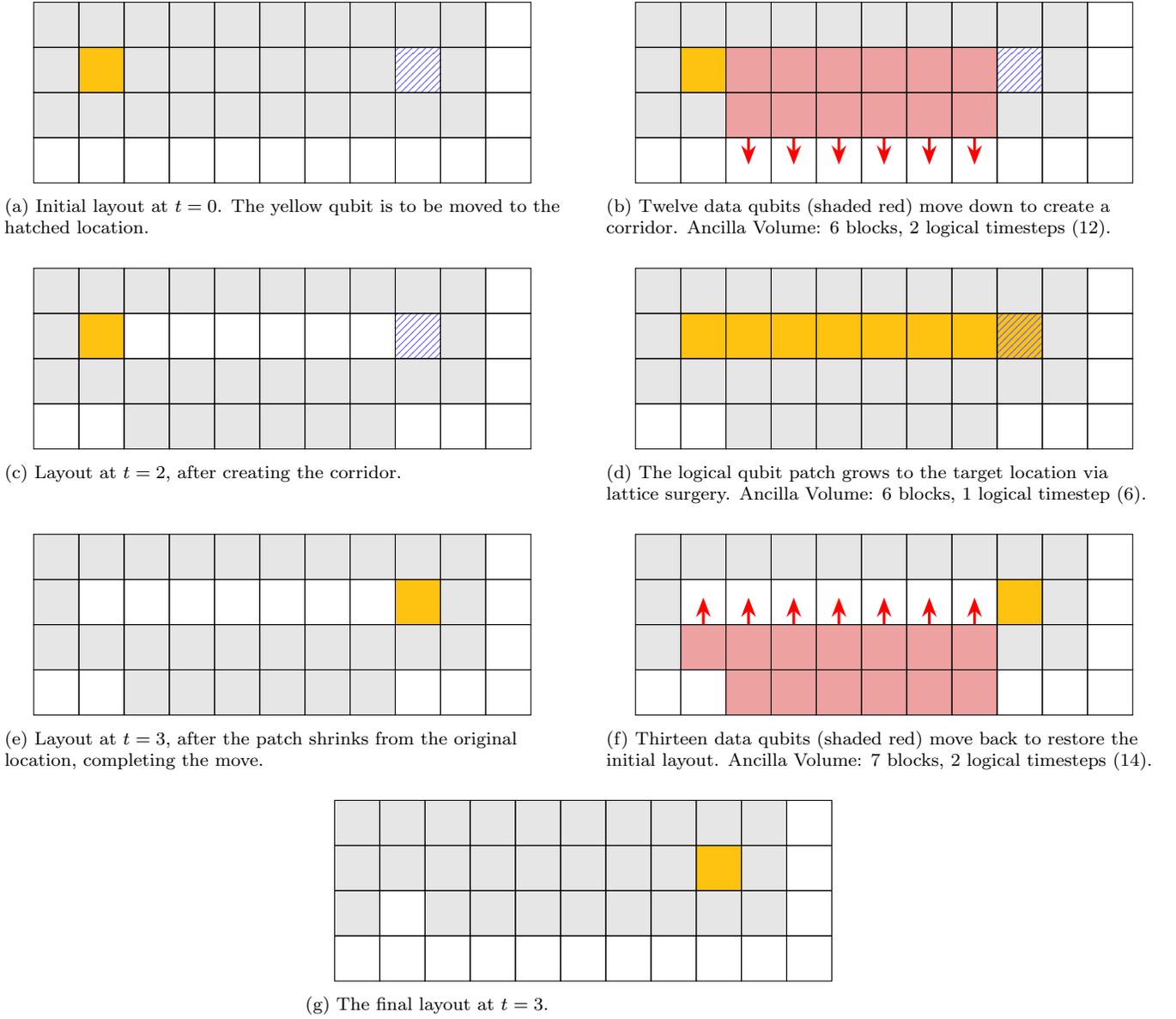

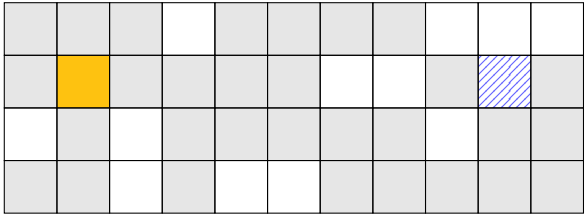
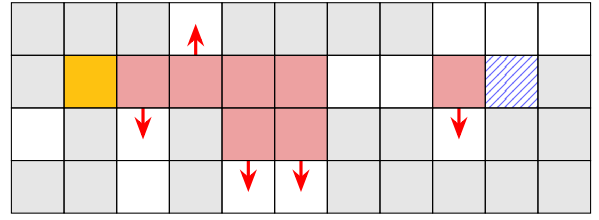
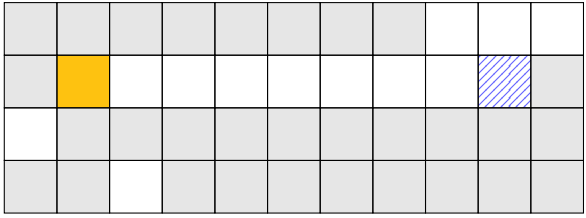
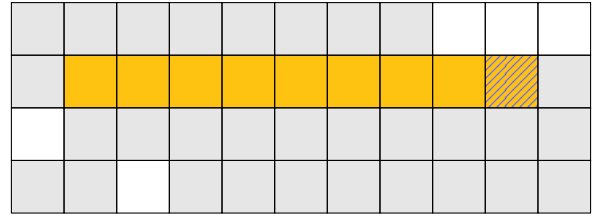
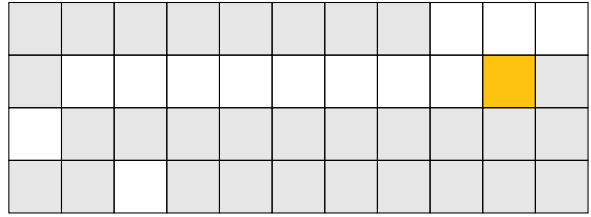
\begin{figure}
	\centering

	\begin{subfigure}[t]{0.48\textwidth}
		\centering
		\begin{tikzpicture}[scale=0.7]
			\foreach \x in {0,...,10} { \foreach \y in {0,...,3} { \fill[gray!20] (\x,\y) rectangle (\x+1,\y+1); } }

			\fill[white] (8,3) rectangle (11,4); %
			\fill[white] (0,1) rectangle (1,2);
			\fill[white] (2,0) rectangle (3,1);
			\fill[white] (2,1) rectangle (3,2);
			\fill[white] (3,3) rectangle (4,4);
			\fill[white] (3,0) rectangle (4,1);
			\fill[white] (6,2) rectangle (7,3);
			\fill[white] (7,2) rectangle (8,3);
			\fill[white] (8,1) rectangle (9,2);
			\fill[white] (10,0) rectangle (11,1);

			\fill[yellow!50!orange] (1,2) rectangle (2,3);
			\fill[white] (9,2) rectangle (10,3);
			\fill[pattern=north east lines, pattern color=blue!60] (9,2) rectangle (10,3);

			\foreach \x in {0,...,10} { \foreach \y in {0,...,3} { \draw (\x,\y) rectangle (\x+1,\y+1); } }
		\end{tikzpicture}
		\caption{\linespread{1.0}\selectfont Initial layout at \(t=0\) with scattered ancillas (white).
			The yellow qubit will be moved to the hatched location.
		}
	\end{subfigure}
	\hfill
	\begin{subfigure}[t]{0.48\textwidth}
		\centering
		\begin{tikzpicture}[scale=0.7]
			\foreach \x in {0,...,10} { \foreach \y in {0,...,3} { \fill[gray!20] (\x,\y) rectangle (\x+1,\y+1); } }
			\fill[white] (8,3) rectangle (11,4); %
			\fill[white] (0,1) rectangle (1,2);
			\fill[white] (2,0) rectangle (3,1);
			\fill[white] (2,1) rectangle (3,2);
			\fill[white] (3,3) rectangle (4,4);
			\fill[white] (3,0) rectangle (4,1);
			\fill[white] (6,2) rectangle (7,3);
			\fill[white] (7,2) rectangle (8,3);
			\fill[white] (8,1) rectangle (9,2);
			\fill[white] (10,0) rectangle (11,1);
			\fill[yellow!50!orange] (1,2) rectangle (2,3);
			\fill[white] (9,2) rectangle (10,3);
			\fill[pattern=north east lines, pattern color=blue!60] (9,2) rectangle (10,3);

			\fill[red, opacity=0.3] (4,0) rectangle (5,1); %
			\draw[-{Stealth[length=3mm]}, very thick, red] (4,0.5) -- (3.4,0.5);
			\fill[red, opacity=0.3] (5,0) rectangle (10,1); %
			\draw[-{Stealth[length=3mm]}, very thick, red] (10,0.5) -- (10.6,0.5);

			\foreach \x in {0,...,10} { \foreach \y in {0,...,3} { \draw (\x,\y) rectangle (\x+1,\y+1); } }
		\end{tikzpicture}
		\caption{\linespread{1.0}\selectfont A horizontal shuffle moves six data qubits (shaded red) to prepare for corridor creation.
			Ancilla Volume: 2 blocks, 1 logical timestep (we assume a faster movement speed for walking codes in the optimistic model, based on \cite{Gidney2024-fy}).
		}
	\end{subfigure}

	\vspace{1em}

	\begin{subfigure}[t]{0.48\textwidth}
		\centering
		\begin{tikzpicture}[scale=0.7]
			\foreach \x in {0,...,10} { \foreach \y in {0,...,3} { \fill[gray!20] (\x,\y) rectangle (\x+1,\y+1); } }

			\fill[white] (8,3) rectangle (11,4); %
			\fill[white] (0,1) rectangle (1,2);
			\fill[white] (2,1) rectangle (3,2);
			\fill[white] (2,0) rectangle (3,1);
			\fill[white] (3,3) rectangle (4,4);
			\fill[white] (4,0) rectangle (5,1);      %
			\fill[white] (5,0) rectangle (6,1);      %
			\fill[white] (6,2) rectangle (7,3);
			\fill[white] (7,2) rectangle (8,3);
			\fill[white] (8,1) rectangle (9,2);

			\fill[yellow!50!orange] (1,2) rectangle (2,3);
			\fill[white] (9,2) rectangle (10,3);
			\fill[pattern=north east lines, pattern color=blue!60] (9,2) rectangle (10,3);

			\foreach \x in {0,...,10} { \foreach \y in {0,...,3} { \draw (\x,\y) rectangle (\x+1,\y+1); } }
		\end{tikzpicture}
		\caption{\linespread{1.0}\selectfont Layout at \(t=1\), after the horizontal shuffle.}
	\end{subfigure}
	\hfill
	\begin{subfigure}[t]{0.48\textwidth}
		\centering
		\begin{tikzpicture}[scale=0.7]
			\foreach \x in {0,...,10} { \foreach \y in {0,...,3} { \fill[gray!20] (\x,\y) rectangle (\x+1,\y+1); } }
			\fill[white] (8,3) rectangle (11,4); %
			\fill[white] (0,1) rectangle (1,2);
			\fill[white] (2,1) rectangle (3,2);
			\fill[white] (2,0) rectangle (3,1);
			\fill[white] (3,3) rectangle (4,4);
			\fill[white] (4,0) rectangle (5,1);
			\fill[white] (5,0) rectangle (6,1);
			\fill[white] (6,2) rectangle (7,3);
			\fill[white] (7,2) rectangle (8,3);
			\fill[white] (8,1) rectangle (9,2);
			\fill[yellow!50!orange] (1,2) rectangle (2,3);
			\fill[white] (9,2) rectangle (10,3);
			\fill[pattern=north east lines, pattern color=blue!60] (9,2) rectangle (10,3);

			\foreach \x in {2,3,8} { \fill[red, opacity=0.3] (\x,2) rectangle (\x+1,3); }
			\foreach \x in {4,5} { \fill[red, opacity=0.3] (\x,1) rectangle (\x+1,3); }

			\draw[-{Stealth[length=3mm]}, very thick, red] (2.5,2) -- (2.5,1.4);    %
			\draw[-{Stealth[length=3mm]}, very thick, red] (3.5,3) -- (3.5,3.6);    %
			\draw[-{Stealth[length=3mm]}, very thick, red] (8.5,2) -- (8.5,1.4);    %
			\draw[-{Stealth[length=3mm]}, very thick, red] (4.5,1) -- (4.5,0.4);    %
			\draw[-{Stealth[length=3mm]}, very thick, red] (5.5,1) -- (5.5,0.4);    %

			\foreach \x in {0,...,10} { \foreach \y in {0,...,3} { \draw (\x,\y) rectangle (\x+1,\y+1); } }
		\end{tikzpicture}
		\caption{\linespread{1.0}\selectfont A vertical shuffle moves seven data qubits (shaded red) to create a corridor.
			Ancilla Volume: 5 blocks, 1 logical timestep.
		}
	\end{subfigure}

	\vspace{1em}

	\begin{subfigure}[t]{0.48\textwidth}
		\centering
		\begin{tikzpicture}[scale=0.7]
			\foreach \x in {0,...,10} { \foreach \y in {0,...,3} { \fill[gray!20] (\x,\y) rectangle (\x+1,\y+1); } }

			\fill[white] (8,3) rectangle (11,4); %
			\fill[white] (0,1) rectangle (1,2);      %
			\fill[white] (2,0) rectangle (3,1);      %
			\fill[white] (2,2) rectangle (3,3);      %
			\fill[white] (3,2) rectangle (4,3);      %
			\fill[white] (4,2) rectangle (5,3);      %
			\fill[white] (5,2) rectangle (6,3);      %
			\fill[white] (6,2) rectangle (7,3);      %
			\fill[white] (7,2) rectangle (8,3);      %
			\fill[white] (8,2) rectangle (9,3);      %

			\fill[yellow!50!orange] (1,2) rectangle (2,3);
			\fill[white] (9,2) rectangle (10,3); %
			\fill[pattern=north east lines, pattern color=blue!60] (9,2) rectangle (10,3);

			\foreach \x in {0,...,10} { \foreach \y in {0,...,3} { \draw (\x,\y) rectangle (\x+1,\y+1); } }
		\end{tikzpicture}
		\caption{\linespread{1.0}\selectfont Layout at \(t=2\), with a clear corridor to the target.}
	\end{subfigure}
	\hfill
	\begin{subfigure}[t]{0.48\textwidth}
		\centering
		\begin{tikzpicture}[scale=0.7]
			\foreach \x in {0,...,10} { \foreach \y in {0,...,3} { \fill[gray!20] (\x,\y) rectangle (\x+1,\y+1); } }
			\fill[white] (8,3) rectangle (11,4);
			\fill[white] (0,1) rectangle (1,2);
			\fill[white] (2,0) rectangle (3,1);
			\fill[white] (2,2) rectangle (3,3);
			\fill[white] (3,2) rectangle (4,3);
			\fill[white] (4,2) rectangle (5,3);
			\fill[white] (5,2) rectangle (6,3);
			\fill[white] (6,2) rectangle (7,3);
			\fill[white] (7,2) rectangle (8,3);
			\fill[white] (8,2) rectangle (9,3);

			\fill[yellow!50!orange] (1,2) rectangle (10,3);

			\fill[pattern=north east lines, pattern color=blue!60] (9,2) rectangle (10,3);

			\foreach \x in {0,...,10} { \foreach \y in {0,...,3} { \draw (\x,\y) rectangle (\x+1,\y+1); } }
		\end{tikzpicture}
		\caption{\linespread{1.0}\selectfont The logical qubit patch grows to the target location via lattice surgery.
			Ancilla Volume: 7 blocks, 1 logical timestep.
		}
	\end{subfigure}

	\vspace{1em}

	\begin{subfigure}[t]{0.48\textwidth}
		\centering
		\begin{tikzpicture}[scale=0.7]
			\foreach \x in {0,...,10} { \foreach \y in {0,...,3} { \fill[gray!20] (\x,\y) rectangle (\x+1,\y+1); } }

			\fill[white] (8,3) rectangle (11,4); %
			\fill[white] (0,1) rectangle (1,2);      %
			\fill[white] (2,0) rectangle (3,1);      %
			\fill[white] (1,2) rectangle (2,3);      %
			\fill[white] (2,2) rectangle (3,3);
			\fill[white] (3,2) rectangle (4,3);
			\fill[white] (4,2) rectangle (5,3);
			\fill[white] (5,2) rectangle (6,3);
			\fill[white] (6,2) rectangle (7,3);
			\fill[white] (7,2) rectangle (8,3);
			\fill[white] (8,2) rectangle (9,3);

			\fill[yellow!50!orange] (9,2) rectangle (10,3);

			\foreach \x in {0,...,10} { \foreach \y in {0,...,3} { \draw (\x,\y) rectangle (\x+1,\y+1); } }
		\end{tikzpicture}
		\caption{\linespread{1.0}\selectfont Final layout at \(t=3\).
			The patch has shrunk from its original location, completing the move and leaving a scattered ancilla layout.
		}
	\end{subfigure}

	\caption{\linespread{1.0}\selectfont An example of a `Move` operation with an estimate of the ancilla volume. (a) The qubits are initially in a disordered layout (as if we are in the middle of a computation that keeps shuffling them around locally). (b-e) A corridor is created through a series of local shuffles of neighboring data qubits. (f) The yellow logical qubit is moved via a 'grow' and 'shrink' operation. (g) The operation completes, leaving the surrounding data qubits in a new, shuffled configuration.
		The total ancilla volume of \(14 \approx 2 p(q_1, q_2) = 16\) is lower because it does not include the cost of restoring the surrounding qubits to their initial positions and we assume a faster movement speed via improved walking surface codes.
	}
	\label{fig:move_optimistic}
\end{figure}

Our heuristic estimate of the ancilla volume required for a Move operation depends on how we account for these displaced qubits.
This accounting differs between the optimistic and conservative cost models.
In the conservative model, we explicitly capture the cost of restoring these qubits to their original locations and we assume that it takes two timesteps to move a group of logical qubits over by one position.
In the optimistic model, we only account for the cost of moving the displaced qubits aside, assuming they can remain in their new, shuffled positions.
Furthermore, the optimistic costs assume we can move qubits twice as fast, taking only one logical timestep per unit of distance on the grid of logical qubits.
These two scenarios are illustrated in \Cref{fig:move_conservative} and \Cref{fig:move_optimistic}, respectively.

\subsection{CNOT / CZ gate}

\begin{figure}
	\centering
	\begin{subfigure}[t]{.32\textwidth}
		\centering
		\includegraphics[width=.5\linewidth]{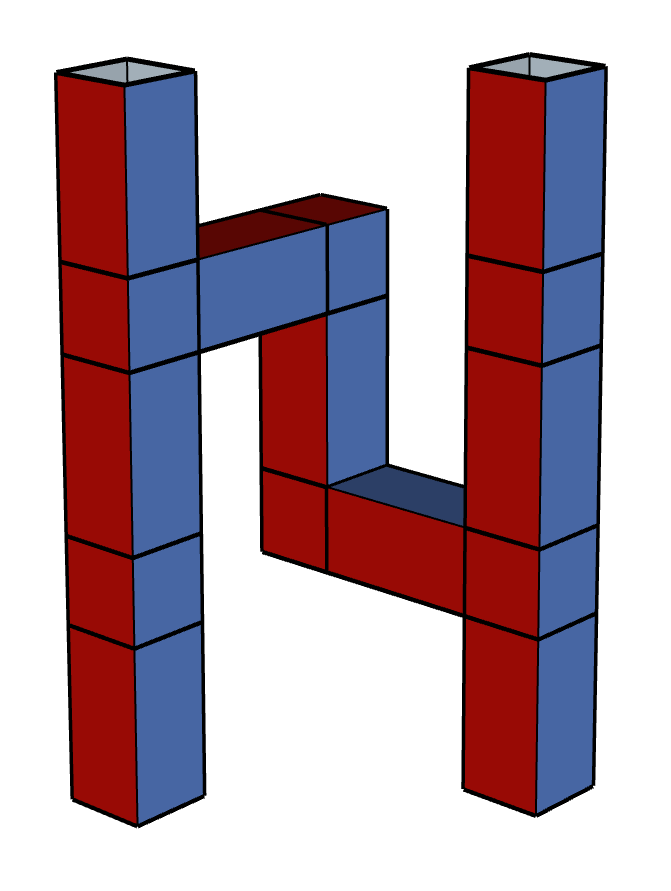}
		\caption{Standard CNOT gate between two qubits each adjacent to an ancilla.}
		\label{fig:sub1}
	\end{subfigure}
	\hfill
	\begin{subfigure}[t]{.32\textwidth}
		\centering
		\includegraphics[width=.55\linewidth]{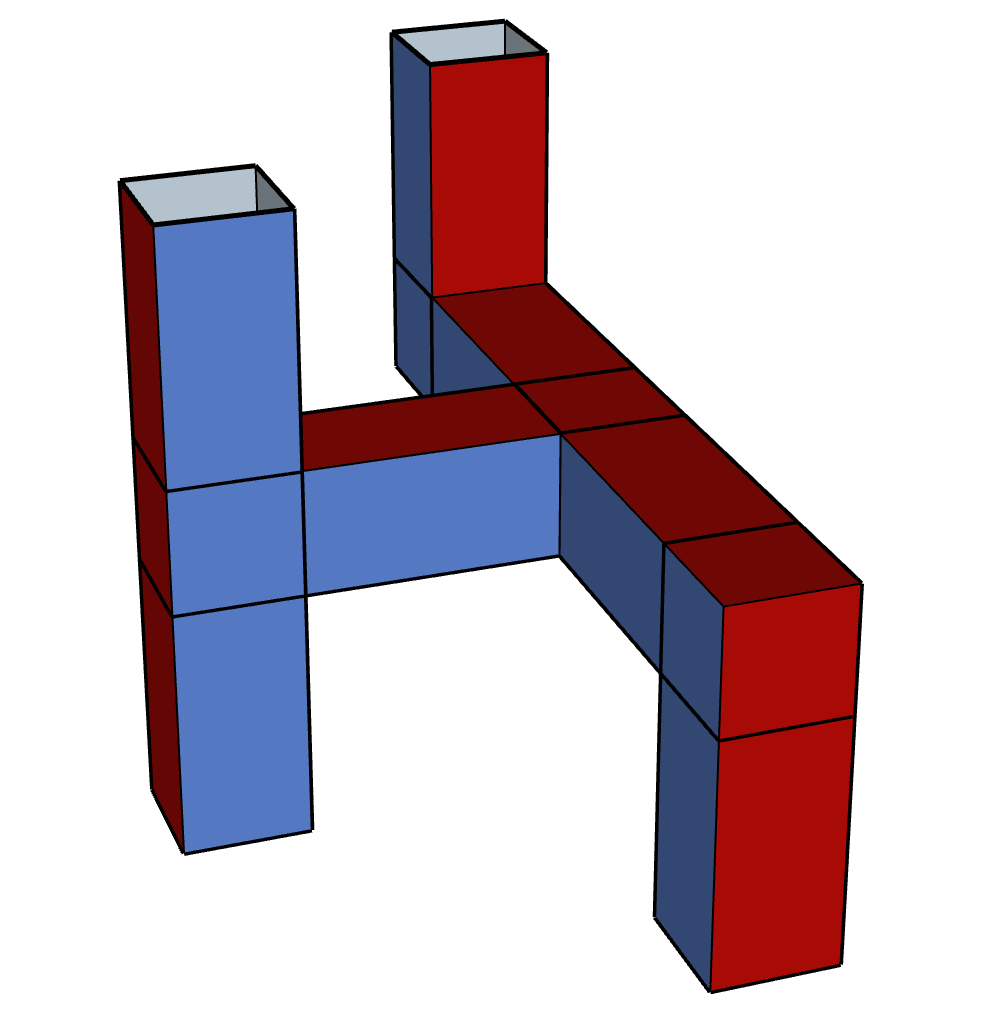}
		\caption{CNOT gate between one stationary qubit and another qubit moving past it.}
		\label{fig:sub2}
	\end{subfigure}
    \hfill
    \begin{subfigure}[t]{.32\textwidth}
		\centering
		\includegraphics[width=.5\linewidth]{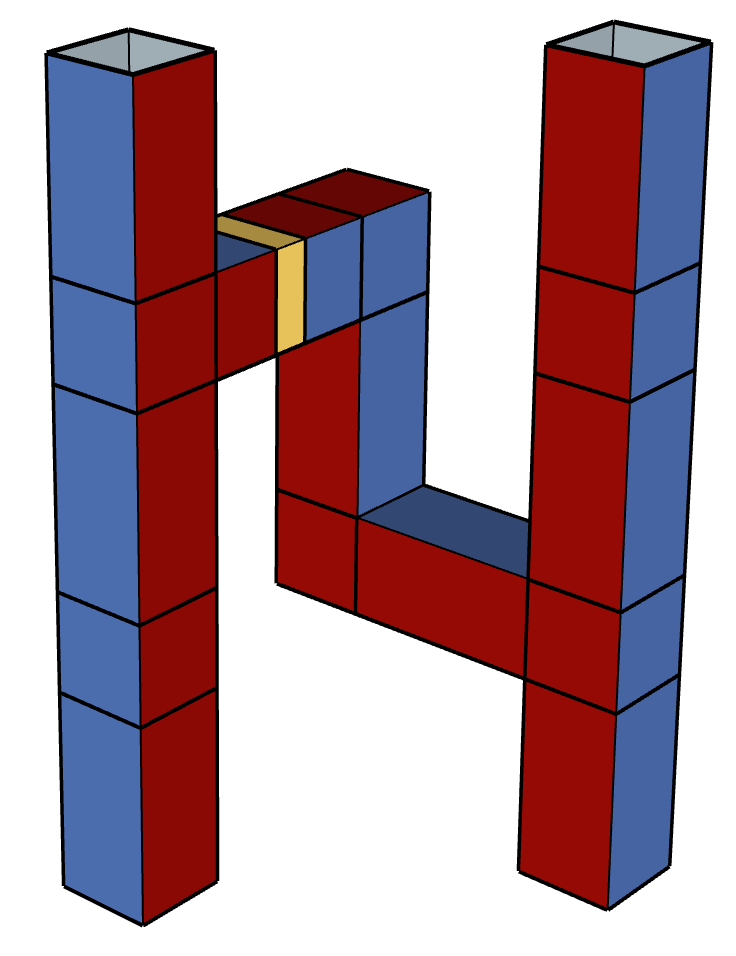}
		\caption{A CZ gate between two qubits each adjacent to an ancilla.}
		\label{fig:sub3}
	\end{subfigure}
	\caption{Pipe diagrams for several standard two-qubit gates.}
	\label{fig:cnot_pipe_diagrams}
\end{figure}

The ancilla volume of a CNOT or CZ gate depends on the distance between the qubits and whether the intervening space is occupied.
In the ideal case where two stationary qubits are separated by an empty region of ancillas, a CNOT can be implemented with a standard lattice surgery operation, as illustrated in \Cref{fig:sub1}.
The ancilla spacetime volume of this base interaction is exactly \(p(q_1, q_2)\), the Manhattan distance between the qubits.
Other, more optimized constructions also exist; for example, a CNOT can be performed with potentially lower ancilla volume when one qubit is ``moving'' past another, as shown in \Cref{fig:sub2}.
Our model does not explicitly account for these dynamic optimizations, but they illustrate the types of savings a sophisticated compiler might find.

In the more realistic scenario, the \(p(q_1, q_2) - 1\) logical qubits in the path between \(q_1\) and \(q_2\) are occupied and must be moved to create a temporary corridor for the CNOT.
According to our model's principles, the ancilla volume of moving \(k\) qubits is \(k\) in the optimistic case (moving them out of the way takes a duration of \(1\) logical timestep) and \(4k\) in the conservative case (moving them out and back each takes a duration of \(2\) logical timesteps).
For simplicity, we approximate the number of qubits to move as \(p(q_1, q_2)\) rather than the more precise \(p(q_1, q_2) - 1\).
This simplification means we slightly overestimate the routing volume in this general case, but we accept this for the sake of a simple formula (and because certain special cases are more expensive, as we explain below).
Combining the base interaction volume with this routing volume gives the total ancilla volume: \(p(q_1, q_2) + p(q_1, q_2) = 2 p(q_1, q_2)\) for the optimistic estimate, and \(p(q_1, q_2) + 4 p(q_1, q_2) = 5 p(q_1, q_2)\) for the conservative estimate.

This simplification is further justified by considering the special case where the two qubits lie in the same row or column.
Here, the standard lattice surgery cannot be performed directly.
One of the qubits must first be moved to an adjacent row or column to create the necessary caddy-corner geometry.
This initial move requires its own ancilla volume, and the subsequent CNOT interaction occurs over a new, longer path of length \(p(q_1, q_2) + 1\).
By using \(p(q_1, q_2)\) as the base interaction volume in our formula, we slightly underestimate the true ancilla volume in this specific scenario.
This simplification balances the overestimation of volume in the general case with the underestimation in the same-line case.

CZ gates are implemented similarly, except that they involve a ``domain wall,'' visualized as a yellow slice in the pipe diagram in \Cref{fig:sub3}.
These domain walls indicate either i) transversal Hadamard gates applied across a logical qubit or ii) a modification of the lattice surgery protocol~\cite{Hirano2025-rl}.
In either case, the ancilla spacetime volume required by the CZ gate is similar to the CNOT gate.

\subsection{T\(_X\) gate (\(T_X, T_X^\dagger\))}
\label{app:T_X_gate_cost}

The implementation of a \(T\) gate in the \(X\) basis (i.e., \(H T H\)) is performed using nearly the same steps as the normal \(T\).
Specifically, a \(T = T \ket{+}\) state is cultivated as normal.
Applying a transversal Hadamard yields \(T_X = H T \ket{+}\) (it also switches the X and Z boundaries of the qubit, but this can be accounted for in how the \(T\) state is cultivated).
An \(XX\) parity measurement is then used instead of a \(ZZ\) parity measurement to perform the teleportation step~\cite{Gidney2025-bn}.
As with the standard \(T\) gate, one obtains \(T_X\) or \(T_X^\dagger\) randomly with equal probability and then applies a correction if necessary by choosing whether to measure the ancilla qubit (the one that had the magic state) in the \(X\) or \(Y\) basis.

\subsection{R\(_Z(\theta)\) gate}
\label{app:rotation_synthesis_details}

Arbitrary rotations are a key component of many applications and there is a large body of work on various techniques for implementing such rotations~\cite{kliuchnikov_synthesis_2013,Ross2014-zf,Bocharov2015-fd,Bocharov2015-gj,kliuchnikov_practical_2016,Kliuchnikov2023-vm}.
In this work, we assume that rotations are synthesized about a single axis, applying the operation
\begin{equation}
	R_Z(\theta) =
	\begin{pmatrix}
		e^{-i \theta / 2} & 0
		\\
		0                 & e^{i \theta / 2}
	\end{pmatrix}
	,
\end{equation}
or the related \(R_X(\theta) = H R_Z{\theta} H\).
Any arbitrary single-qubit gate can be decomposed into three such rotations using a standard Euler angle decomposition, but there is also work that suggests that one can obtain more efficient approximations directly~\cite{Hao2025-ub}.

In this section, we briefly review results from \cite{Kliuchnikov2023-vm}, which contains a mixture of state-of-the-art techniques and a comprehensive presentation of older techniques.
Table 1 of \cite{Kliuchnikov2023-vm} provides estimates of the T-count for several different rotation synthesis strategies.
We reprint their estimates for the Clifford + T gate set in \Cref{tab:rotation_synthesis_costs_appendix}.

\begin{table*}
	\centering %
	\caption{The number of T gates required for rotation synthesis in a Clifford + T gate set, adapted from Table 1 of \cite{Kliuchnikov2023-vm}.
		Mean and max T-counts were obtained by fitting from numerical data for \(\epsilon < 10^{-4}\).
		Except for the diagonal synthesis method, all methods provide a guaranteed error in the diamond norm rather than the operator norm, although the fallback approach could be adapted to have nearly deterministic error.
	}
	\label{tab:rotation_synthesis_costs_appendix}
	\resizebox{\textwidth}{!}{%
		\setlength{\tabcolsep}{8pt} %
		\begin{tabular}{@{}l l l l l@{}} %
			\hline\hline %

			Protocol                                            & Mean T-count                       & Max T-count                         & Heuristic T-count                           & Norm
			\\
			\hline

			Diagonal \cite{Ross2014-zf}                         & $3.02\log_2(\epsilon^{-1}) + 1.77$ & $3.02\log_2(\epsilon^{-1}) + 9.19$  & $3.0\log_2(\epsilon^{-1}) + \mathcal{O}(1)$ & Operator
			\\
			Fallback \cite{Bocharov2015-gj, Kliuchnikov2023-vm} & $1.03\log_2(\epsilon^{-1}) + 5.75$ & $1.05\log_2(\epsilon^{-1}) + 11.83$ & $1.0\log_2(\epsilon^{-1}) + \mathcal{O}(1)$ & Diamond\(^*\)
			\\
			Mixed diagonal \cite{Kliuchnikov2023-vm}            & $1.52\log_2(\epsilon^{-1}) - 0.01$ & $1.54\log_2(\epsilon^{-1}) + 6.85$  & $1.5\log_2(\epsilon^{-1}) + \mathcal{O}(1)$ & Diamond
			\\
			Mixed fallback \cite{Kliuchnikov2023-vm}            & $0.53\log_2(\epsilon^{-1}) + 4.86$ & $0.57\log_2(\epsilon^{-1}) + 8.83$  & $0.5\log_2(\epsilon^{-1}) + \mathcal{O}(1)$ & Diamond
			\\
			\hline
		\end{tabular}
	}
\end{table*}

\begin{figure}
	\centering
	\begin{quantikz}
		\lstick{$ \ket{0} $}    & \targ{}   & \qw      & \targ{}   & \meter{} & \ctrl[vertical wire=c]{1}  \setwiretype{c} & \cw & \ctrl[vertical wire=c]{1} & \setwiretype{n}
		\\
		\lstick{$ \ket{\psi} $} & \ctrl{-1} & \gate{V} & \ctrl{-1} & \qw      & \gate{Y}                                   & \qw & \gate{\mathcal{B}}        & \qw
		\\
	\end{quantikz}
	\caption{A reproduction of Figure 7 from \cite{Kliuchnikov2023-vm}, which shows the form of the circuits used for the fallback and mixed fallback protocol.
		The single-qubit operations \(V\) and \(\mathcal{B}\) are both instances of the diagonal unitary approximation discussed in this section, and \(Y\) is a Pauli \(Y\) gate.
		Furthermore, \(\mathcal{B}\) is implemented at most \(1\%\) of the time in the protocols they considered.
	}
	\label{fig:fallback_protocol}
\end{figure}
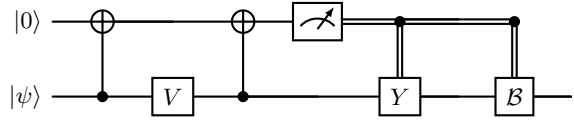

All of the resource estimates we present using the FLASQ model assume the used of the ``mixed fallback'' rotation synthesis strategy.
This strategy uses a probabilistic mixture of circuits of the form shown in \Cref{fig:fallback_protocol} to approximate the desired rotation to within some \(\epsilon\) in the diamond norm.
This use of a probabilistic compilation results in a significant savings compared with deterministic approaches.
The other innovation compared to a direct unitary synthesis is the ``fallback'' component of the construction.
The desired unitary is synthesized accurately except that a signifcant rotation away from the \(Z\) axis of the block sphere is allowed (this is the \(V\) in the circuit diagram).
Then one measures the ancilla qubit and either obtains \(0\), in which case this error is projected out, or obtains \(1\), in which case another compensating rotation (\(\mathcal{B}\)) is performed.
A substantial savings in the expected T-count can be obtained while keeping the probability of failure below \(1 \%\)~\cite{Kliuchnikov2023-vm}.

A direct deterministic unitary synthesis of an \(R_Z\) rotation (the Diagonal protocol in \Cref{tab:rotation_synthesis_costs_appendix}) and the \(V\) and \(\mathcal{B}\) gates required by the mixed fallback protocol illustrated in \Cref{fig:fallback_protocol} all involve circuits with the same structure, which we now explain.
For the Clifford + T gate set, \cite{Kliuchnikov2023-vm} focuses on approximations that use Hadamard and CNOT gates, as well as gates of the form
\begin{equation}
	T_P = \frac{1}{\sqrt{2 + \sqrt{2}}} \left( I + \frac{I - iP}{\sqrt{2}} \right),
\end{equation}
where \(P\) is a single-qubit Pauli operator, \(P \in \pm\left\{X, Y, Z \right\}\).
Note that \(T_Z\) is the normal T gate and \(T_{-Z}\) is equivalent to \(T^{\dagger}\).

\begin{figure}
	\centering
			\includegraphics[width=0.15\textwidth]{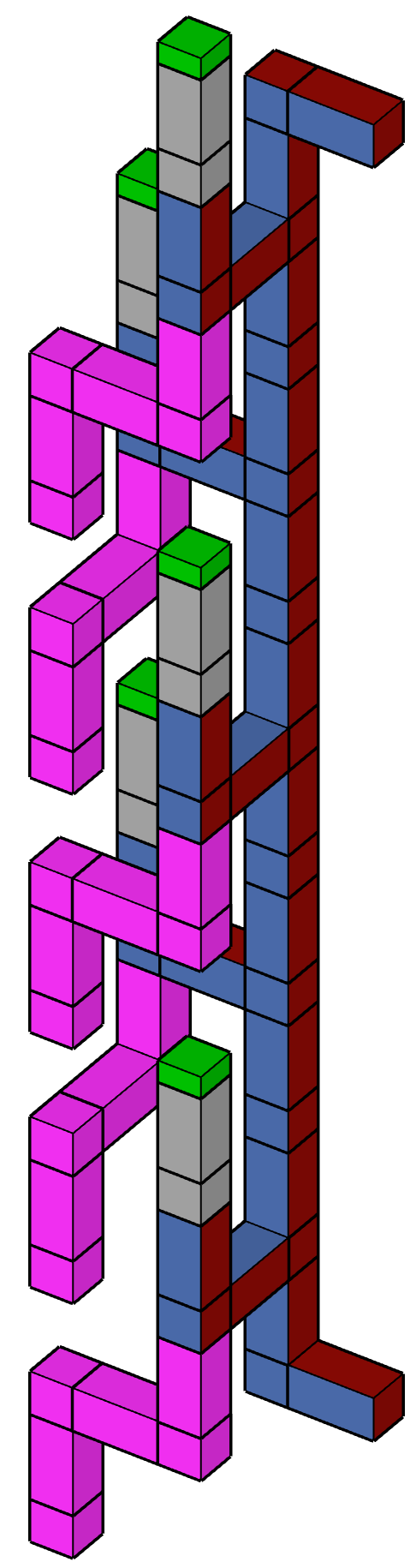}
			\caption{
				Lattice surgery pipe diagram showing a $2\times 2$ rotation synthesis unit which implements the single qubit rotation $V$ as an alternating sequence of $T^{\pm 1}_X$ and $T^{\pm 1}_Z$ gates.
				Pink boxes correspond to spacetime used for magic state cultivation in the X or Z basis~\cite{Craig2024-kv}.
				Grey boxes correspond to waiting time needed for the classical control system to process the results of previous measurements.
				Green boxes represent a choice of \(X\) or \(Y\) basis measurements used to apply the phase correction to the injected T state if necessary~\cite{Gidney2024-fy}.
				The rotation synthesis unit consumes a \(T\) state every \(2d\) surface code cycles.
				The horizontal poles on the bottom and top correspond to CNOT used to connect the rotation synthesis unit to the system qubit on which the rotation is being applied.
			}
			\label{fig:compact_rotation_unit_appendix}
\end{figure}

Using the normal form of \cite{Matsumoto2008-tc}, we can express any product of single-qubit Clifford gates and \(t\) of these generalized T gates as a product \(W_t T W_{t-1} T \cdots W_1 T W_0\), where \(W_t \in \left\{ I , H , SH \right\}\), \(W_k \in \left\{ H, SH \right\}\) for \(1 \leq k < t\), and \(W_0\) is an arbitrary single-qubit Clifford.
This is convenient because we can write \(T S = T^\dagger Z\) and \(H T H = T_X\) to express the circuit as a product of zero or one \(S\) gates, \(t\) generalized \(T\) gates of the form \(T, T^\dagger, T_X, T_\dagger\), some number of Pauli gates, and an additional arbitrary single-qubit Clifford gate.
As we discussed above in \Cref{app:T_X_gate_cost}, we can implement \(T_X\) and \(T_X^\dagger\) gates as easily as their \(Z\) basis counterparts.
In \Cref{fig:compact_rotation_unit_appendix}, we reproduce a panel from \Cref{fig:mixed_fallback_pipes} in the main text, which shows the application of an alternating series of \(T / T^\dagger\) and \(T_X / T_X^\dagger\) gates to a single qubit.

Now we estimate the ancilla volume required for implementing \(V\) and \(\mathcal{B}\) (or any other rotation synthesized this way) before returning to the task of estimating the volume for the mixed fallback protocol overall.
The overall ancilla spacetime volume is bounded by
\begin{equation}
	\textsc{Vol} \left(V( \epsilon )\right) \leq \textsc{Vol}\left( S \right) + t \textsc{Vol}\left( T \right) + \textsc{Vol}\left( \mathcal{C}_1 \right),
\end{equation}
where \(t\) is the number of T gates required for a precision \(\epsilon\).
The arbitrary single-qubit Clifford can be written as a product \(AB\), where \(A \in \left\{ I, H, S, HS, SH, HSH \right\}\) and \(B \in {I, X, Y, Z}\).
For the conservative estimates, we use the bound \(\textsc{Vol}\left( \mathcal{C}_1 \right) \leq 2 \textsc{Vol}\left( H \right) + \textsc{Vol}\left( S \right)\) and we also reserve space for implementing the \(S\) gate, resulting in an overall cost of
\begin{equation}
	\textsc{Vol}(V) = 2 \cdot \textsc{Vol} \left( S \right) + 2 \cdot \textsc{Vol} (H) + t \cdot \textsc{Vol} (T).
\end{equation}

The estimates in the optimistic case are similar, but we use the expected number of gates rather than the upper bound (since these numbers can be determined at compile time and, in principle, an optimizing compiler could take advantage of the fact that some sequences would be shorter).
We therefore assume that each single-qubit Clifford occurs with even probability and we take \(\ev{\textsc{Vol}\left( \mathcal{C}_1 \right)} = \frac{2}{3} \cdot \textsc{Vol}\left( S \right) + \frac{5}{6} \cdot \textsc{Vol}\left( H \right)\).
Similarly, with the initial \(S\) gate, we assume it is only required \(50\%\) of the time for the optimistic model.
The estimates for the optimistic model are therefore
\begin{equation}
	\textsc{Vol}(V) = \frac{7}{6} \cdot \textsc{Vol} \left( S \right) + \frac{5}{6} \cdot \textsc{Vol} (H) + t \cdot \textsc{Vol} (T).
\end{equation}

Now we turn towards estimates of the ancilla volume for the mixed fallback rotation synthesis approach overall.
The main difficulty in accounting for the cost is that we temporarily use an ancilla qubit.
In general in the FLASQ model, we account for ancillas either by considering them as ``algorithmic ancilla'' that exist at the circuit level or we handle them by considering them as ``fluid ancilla'' and we estimate the ancilla spacetime volume by considering a representative compilation of the operation to basic surface code instructions.
Here we will follow a blended approach and propose a heuristic for the (fluid) ancilla spacetime volume of the whole \(R_Z\left( \theta \right)\) operation by considering its circuit implementation and how one might realize it as a pipe diagram.

In order to account for the spacetime volume occupied by this ancilla qubit, we need to understand the number of logical timesteps that it is in use for.
Provided with sufficient \(T\) states, it is possible to consume one every logical timestep.
For the optimistic estimate, we assume that we consume the states at this rate, and so we add a factor of \(t\) units of ancilla spacetime volume to account for holding this ancilla qubit active, plus another \(4\) to account for the expected number of Clifford gates (assuming they are each performed in a single time step as well), plus another \(1\) to account for the cost of freeing an ancilla qubit's worth of space.
For the conservative estimate, we assume that we consume states at half of this speed (see \Cref{fig:compact_rotation_unit_appendix} for an example), add an additional \(6\) to account for the maximum number of Clifford gates, plus another \(4\) to account for the cost of allocating and deallocating an ancilla qubit's worth of space.
We leave it to a future version of the FLASQ model to improve these estimates by providing a variety of hand-compiled constructions suitable for different cases.

In the mixed fallback construction, the \(\mathcal{B}\) gate is generally taken to be a much longer sequence of rotations but it occurs with a probability of less than \(1\%\), so it contributes only a small amount to the \(T\) count.
We make the assumption that a compiler could dynamically adapt to the measurement result on the ancilla qubit in order to allow our model to take advantage of the fact that this \(\mathcal{B}\) operation is performed rarely and so we only account for the expected number of \(T\) gates necessary rather than the maximum number (otherwise we would have to resort to the much more costly diagonal or mixed diagonal approaches).
Furthermore, since this operation is performed rarely, we choose to use a formula for the ancilla volume of a rotation that acts as though all of the \(T\) gates are performed when the ancilla qubit is active (since this simplifies the formula and the difference is small).

Overall then, for the conservative volume estimates of the \(R_Z(\theta)\) gate implemented to within a precision \(\epsilon\) in the diamond norm, we have
\begin{equation}
	\textsc{Vol}(R_Z(\theta, \epsilon)) \approx t \left( 2 + \textsc{Vol}\left( T \right) \right) + 2 \cdot \textsc{Vol}\left( S \right) + 2 \cdot \textsc{Vol}(H) + 2 \cdot \textsc{Vol}\left( CNOT \right) + 10,
\end{equation}
where the number of \(T\) gates \(t\), is given by
\begin{equation}
	t = 4.86 + 0.53 \log_2 \left( \epsilon^{-1} \right).
\end{equation}
In the optimistic case, we have:
\begin{equation}
	\textsc{Vol}(R_Z(\theta, \epsilon)) \approx t \left( 1 + \textsc{Vol}\left( T \right) \right) + \frac{7}{6} \cdot \textsc{Vol}\left( S \right) + \frac{5}{6} \cdot \textsc{Vol}(H) + 2 \cdot \textsc{Vol}\left( CNOT \right) + 5.
\end{equation}

\subsection{SWAP gate}

We imagine implementing the SWAP gate by performing two nearly simultaneous Move operations.
While we cannot actually perform them both completely at the same time, we expect that routing them around each other in space and time will have a cost that is negligible (compared with the precision of the FLASQ model).

Therefore, we expect that the ancilla volume consumed by a SWAP operation will be \(2 p(q_1, q_2)\) plus the volume required to account for clearing a path between the qubits (\(4 p(q_1, q_2)\) in the default conservative framework and \(1\) in the optimistic framework).

\subsection{AND gate}

\begin{figure}
	\centering
	\begin{subfigure}[b]{0.48\textwidth}
		\centering
		\begin{quantikz}[row sep={0.8cm,between origins}]
			\lstick{$x$}       & \ctrl{2} & \qw      & \gate{T^\dagger} & \targ{}   & \qw      & \qw      & \rstick{$x$}
			\\
			\lstick{$y$}       & \qw      & \ctrl{1} & \gate{T^\dagger} & \targ{}   & \qw      & \qw      & \rstick{$y$}
			\\
			\lstick{$\ket{T}$} & \targ{}  & \targ{}  & \gate{T}         & \ctrl{-2} & \gate{H} & \gate{S} & \rstick{$xy$}
		\end{quantikz}
		\caption{A circuit diagram for the temporary logical-AND gate from \cite{Gidney2018-xg}.
			The circuit on the left computes the logical AND of qubits \(x\) and \(y\) into an ancilla qubit initialized to the magic state \(\ket{T}\).
			The result is that the ancilla qubit becomes \(xy\).
			Because the three T gates are implemented in parallel, the measurement depth of this circuit is one despite requiring four T states.
		}
		\label{fig:and_gate_circuit_quantikz}
	\end{subfigure}
	\hfill
	\begin{subfigure}[b]{0.48\textwidth}
		\centering
		\includegraphics[width=.9\textwidth]{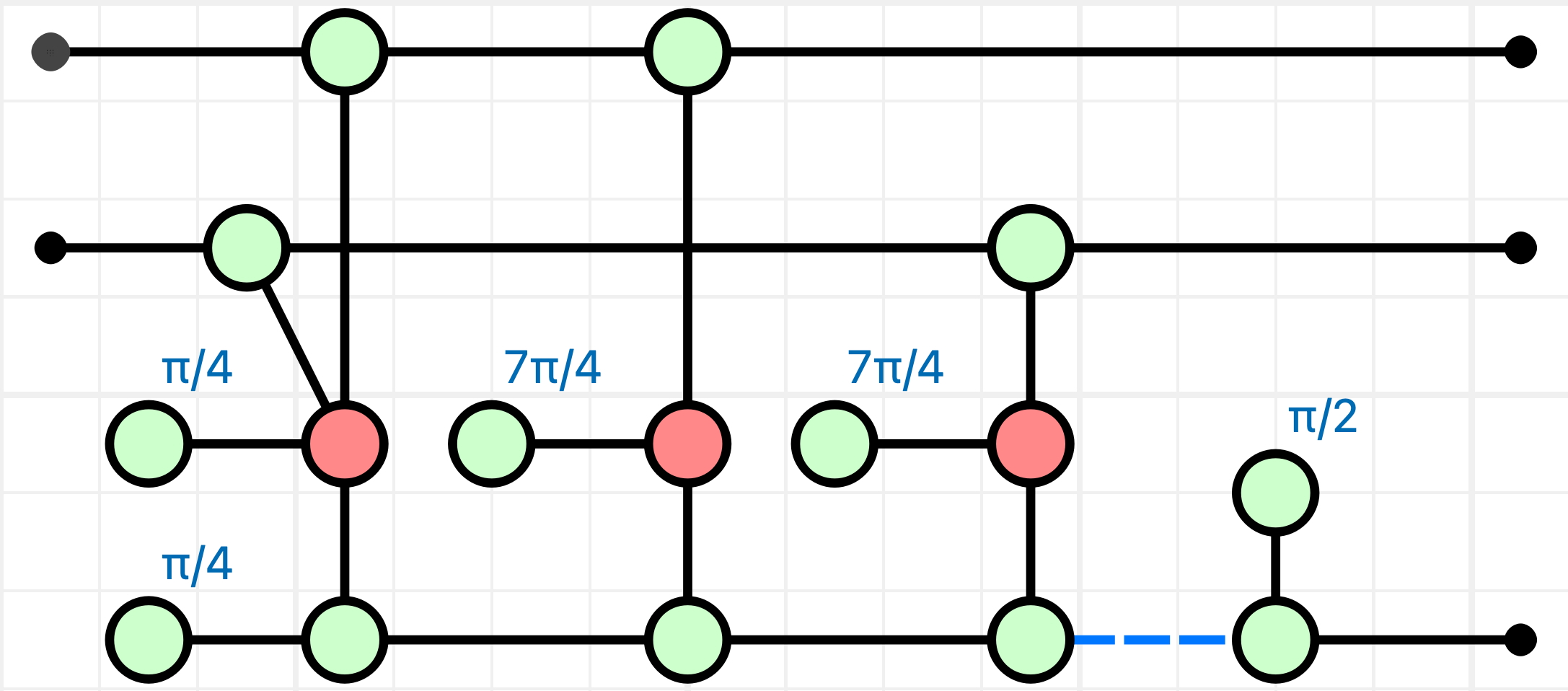}
		\caption{ZX graph for the AND gate shown in \cref{fig:and_gate_circuit_quantikz}.
			Here is a \href{https://algassert.com/quirk\#circuit=\%7B\%22cols\%22\%3A\%5B\%5B\%22Y\%5Et\%22\%2C\%22X\%5E-t\%22\%2C\%22Z\%5E\%C2\%BC\%22\%2C\%22Z\%5E-\%C2\%BC\%22\%2C\%22Z\%5E-\%C2\%BC\%22\%2C\%22Z\%5E\%C2\%BC\%22\%5D\%2C\%5B\%22QFT2\%22\%5D\%2C\%5B\%22\%E2\%80\%A6\%22\%2C\%22\%E2\%80\%A6\%22\%2C\%22\%E2\%80\%A6\%22\%5D\%2C\%5B\%22Chance2\%22\%5D\%2C\%5B\%22Amps2\%22\%5D\%2C\%5B\%5D\%2C\%5B\%22\%E2\%80\%A6\%22\%2C\%22\%E2\%80\%A6\%22\%2C\%22\%E2\%80\%A6\%22\%5D\%2C\%5B1\%2C\%22zpar\%22\%2C\%22zpar\%22\%2C\%22zpar\%22\%2C1\%2C1\%2C\%22X\%22\%5D\%2C\%5B\%22zpar\%22\%2C1\%2C\%22zpar\%22\%2C1\%2C\%22zpar\%22\%2C1\%2C1\%2C\%22X\%22\%5D\%2C\%5B\%22zpar\%22\%2C\%22zpar\%22\%2C\%22zpar\%22\%2C1\%2C1\%2C\%22zpar\%22\%2C1\%2C1\%2C\%22X\%22\%5D\%2C\%5B1\%2C1\%2C\%22H\%22\%5D\%2C\%5B1\%2C1\%2C\%22Z\%5E\%C2\%BD\%22\%5D\%2C\%5B\%22\%E2\%80\%A6\%22\%2C\%22\%E2\%80\%A6\%22\%2C\%22\%E2\%80\%A6\%22\%5D\%2C\%5B1\%2C1\%2C1\%2C1\%2C1\%2C1\%2C\%22Measure\%22\%2C\%22Measure\%22\%2C\%22Measure\%22\%5D\%2C\%5B1\%2C1\%2C1\%2C1\%2C1\%2C\%22H\%22\%2C1\%2C1\%2C\%22\%E2\%97\%A6\%22\%5D\%2C\%5B1\%2C1\%2C1\%2C1\%2C1\%2C\%22X\%5E\%C2\%BD\%22\%2C1\%2C1\%2C\%22\%E2\%80\%A2\%22\%5D\%2C\%5B1\%2C1\%2C1\%2C1\%2C1\%2C\%22Measure\%22\%5D\%2C\%5B1\%2C1\%2C1\%2C1\%2C\%22H\%22\%2C1\%2C1\%2C\%22\%E2\%97\%A6\%22\%5D\%2C\%5B1\%2C1\%2C1\%2C1\%2C\%22X\%5E-\%C2\%BD\%22\%2C1\%2C1\%2C\%22\%E2\%80\%A2\%22\%5D\%2C\%5B1\%2C1\%2C1\%2C1\%2C\%22Measure\%22\%5D\%2C\%5B1\%2C1\%2C1\%2C\%22H\%22\%2C1\%2C1\%2C\%22\%E2\%97\%A6\%22\%5D\%2C\%5B1\%2C1\%2C1\%2C\%22X\%5E-\%C2\%BD\%22\%2C1\%2C1\%2C\%22\%E2\%80\%A2\%22\%5D\%2C\%5B1\%2C1\%2C1\%2C\%22Measure\%22\%5D\%2C\%5B1\%2C1\%2C\%22X\%22\%2C\%22zpar\%22\%2C\%22zpar\%22\%2C\%22zpar\%22\%5D\%2C\%5B1\%2C\%22Z\%22\%2C\%22Z\%22\%2C\%22\%E2\%80\%A2\%22\%5D\%2C\%5B\%22Z\%22\%2C1\%2C\%22Z\%22\%2C1\%2C\%22\%E2\%80\%A2\%22\%5D\%2C\%5B\%22Z\%22\%2C\%22Z\%22\%2C\%22Z\%22\%2C1\%2C1\%2C\%22\%E2\%80\%A2\%22\%5D\%2C\%5B\%22Chance\%22\%2C\%22Chance\%22\%2C\%22\%E2\%80\%A2\%22\%5D\%2C\%5B\%22\%E2\%80\%A6\%22\%2C\%22\%E2\%80\%A6\%22\%2C\%22\%E2\%80\%A6\%22\%5D\%2C\%5B\%22\%E2\%80\%A2\%22\%2C\%22\%E2\%80\%A2\%22\%2C\%22X\%22\%5D\%2C\%5B\%22QFT\%E2\%80\%A02\%22\%5D\%2C\%5B\%22Y\%5E-t\%22\%2C\%22X\%5Et\%22\%5D\%2C\%5B\%22Chance2\%22\%5D\%5D\%2C\%22init\%22\%3A\%5B0\%2C0\%2C\%22\%2B\%22\%2C\%22\%2B\%22\%2C\%22\%2B\%22\%2C\%22\%2B\%22\%5D\%7D}{quirk link} that shows how the AND gate can be implemented by consuming 4 $\ket{T}$ states and performing 3 multi-qubit parity measurements.
		}
		\label{fig:and_gate_zxgraph}
	\end{subfigure}
	\begin{subfigure}[b]{.7\textwidth}
		\centering
		\includegraphics[width=0.5\textwidth]{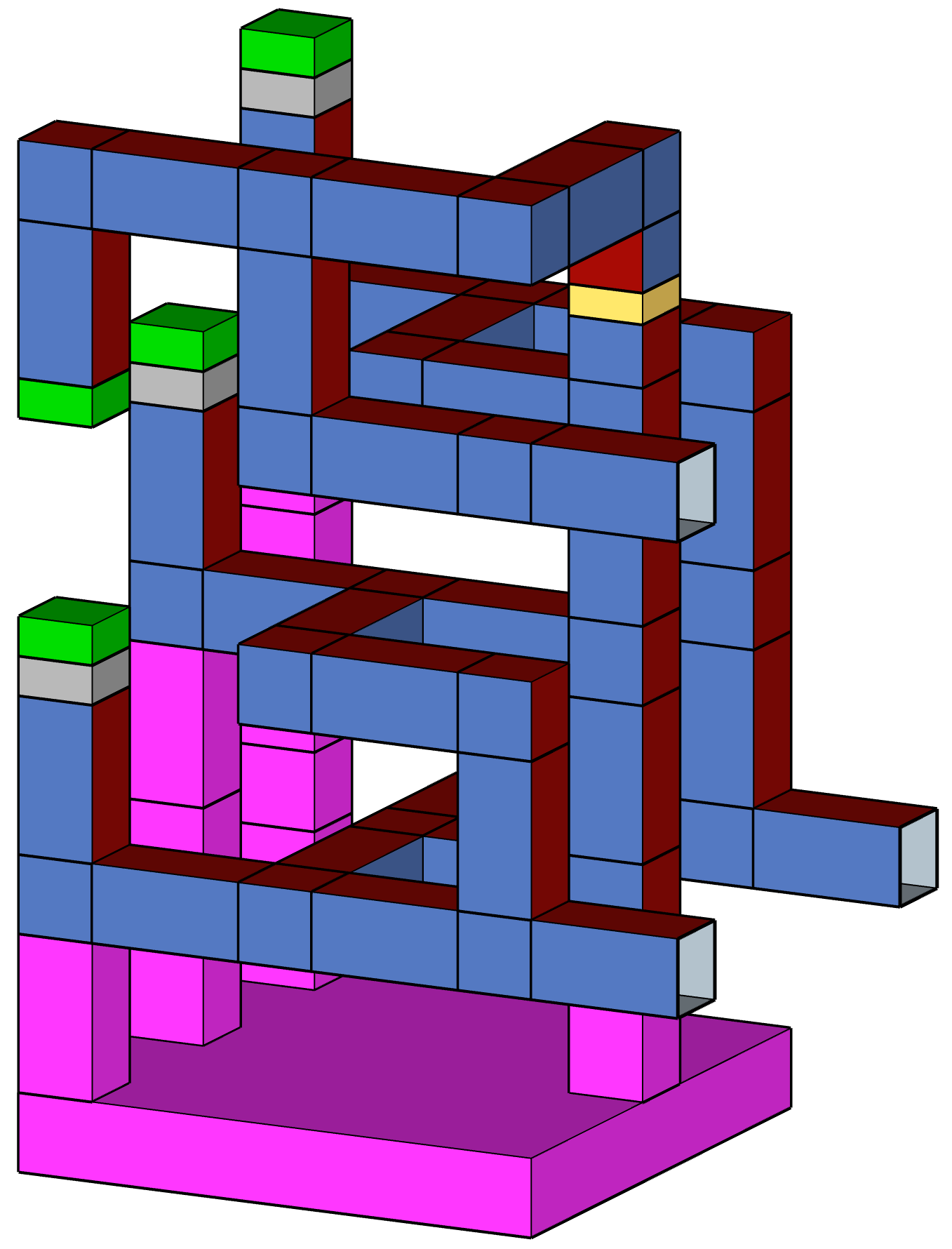}
		\hfill
		\includegraphics[width=0.4\textwidth]{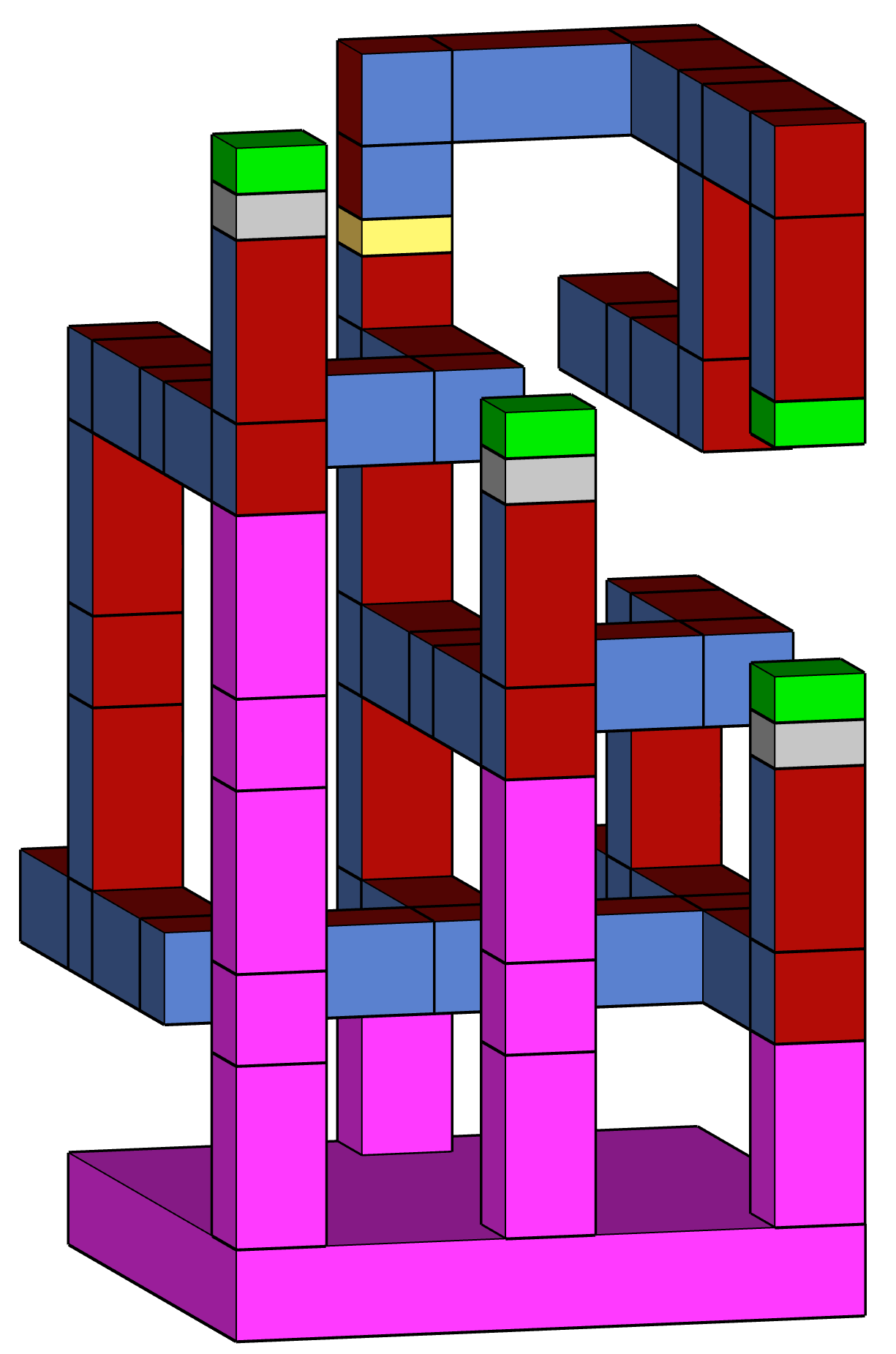}
		\caption{
			Two perspectives on the same spacetime layout of an AND gate in a $3\times3\times 5$ region using Z ports.
			Bottom two ports should be connected to the controls and top port correspond to the target.
			Pink boxes correspond to space used for magic state cultivation.
			For every $\ket{T}$ state to be consumed, there is $3d^3$ spacetime available for cultivation in this layout.
			Grey boxes correspond to the reaction time required for the decoder to finish processing measurement results and determine if S corrections are needed.
			Green boxes correspond to Y basis initialization and measurement.
		}
		\label{fig:and_gate_pipe_diagram}
	\end{subfigure}
	\caption{An AND gate expressed in three ways, as a quantum circuit diagram, as a ZX calculus graph, and as a lattice surgery pipe diagram.}
	\label{fig:and_gate_circuit}
\end{figure}

We present a circuit diagram (a), ZX calculus graph (b), and lattice surgery pipe diagram for the AND gate in \Cref{fig:and_gate_circuit}.
Because this is a commonly-used primitive, we provide a direct estimate of its ancilla volume by analyzing the pipe diagram, but it is also instructive to compare this estimate with the volume that would be obtained from applying the FLASQ model to the quantum circuit that implements the AND gate using simpler primitive operations.
In this subsection, we will focus on the volumes under a conservative model, giving the optimistic volumes in parentheses when they differ.

The pipe diagram construction uses a ``Z port'' construction~\cite{Craig2024-kv}, meaning that the input and output qubits are connected to the ports externally.
For our purposes, this means that the entire diagram shown consists of ancilla volume.
We now proceed bottom to top, explaining our accounting.
First of all, we need to account for providing the nine units of ancilla space, which requires a volume of \(36\) in the conservative cost model (\(9\) in the optimistic model).
We account for the purple regions that represent magic state cultivation with a volume of \(4 \textsc{Vol}\left( Cultivate \right)\).
The first level of the diagram that isn't entirely purple has \(7\) blocks occupied by normal lattice surger operations, as does the second level.
Note that we treat the column that has both a \(Y\) basis measurement and an initialization as fully utilized (since any empty space would be too small to make use of).
The third level has \(8\) fully occupied blocks, plus a grey box that represents waiting for at least \(t_{react}\) followed by a possible \(Y\) basis measurement, which we assign a cost of \(t_{react} + 1\) (\(t_{react} + .5\)). Similarly, the total volume of the top level is \(t_{react} + 5\) (\(t_{react} + 4.5\)).

When the qubits are spatially separated, we also expect a distance-dependent component to the cost.
The Z-port construction gives us a good deal of freedom - essentially we just need to open a path that connects the qubits, place the construction at some point along this path, and perform one operation connecting each qubit to the central location.
In the conservative model, the ancilla volume required for opening the path is \(4 p(q_1, q_2, q_3)\) (\(p(q_1, q_2, q_3)\) in the optimistic model).
In both models the operation itself that connects the input and output qubits to the central structure requires an ancilla volume of \(p(q_1, q_2, q_3)\).
The distance dependent components of the cost are therefore just an additive \(5 p(q_1, q_2, q_3)\) in the conservative model (\(2 p(q_1, q_2, q_3)\) in the optimistic model).

This gives a total ancilla volume of
\begin{equation}
	\textsc{Vol}\left( AND \right) = 4 \textsc{Vol}\left( \text{Cultivate} \right) + 2 t_{react} + 5 p(q_1, q_2, q_3) + 64 \label{eq:and_vol_conservative}
\end{equation}
in the conservative model and
\begin{equation}
	\textsc{Vol}\left( AND \right) = 4 \textsc{Vol}\left( \text{Cultivate} \right) + 2 t_{react} + 2 p(q_1, q_2, q_3) + 36 \label{eq:and_vol_optimistic}
\end{equation}
in the optimistic model.

As an exercise, we apply the FLASQ model directly to the circuit for the AND gate shown in \Cref{fig:and_gate_circuit_quantikz}.
We find that the overall estimate of the ancilla volume is \(4 \textsc{Cultivate} + 3 t_{react} + 70.5\) in the conservative model, whereas in the optimistic model we get a volume of \(4 \textsc{Cultivate} + 3 t_{react} + 27.75\). Applying the formulas \Cref{eq:and_vol_conservative} and \Cref{eq:and_vol_optimistic} to this arrangement of input qubits would instead yield a volume of \(4 \textsc{Vol}\left( \text{Cultivate} \right) + 2 t_{react} + 79\) in the conservative model and \(4 \textsc{Vol}\left( \text{Cultivate} \right) + 2 t_{react} + 42\) in the optimistic model.

These costs reveal a significant variation between the direct estimation and the estimation one obtains from looking at the circuit gate-by-gate.
The underestimation in the gate-by-gate case is particularly concerning, although we note that this may be an artifact of a generic construction optimized for minimal routing overhead (using the Z port construction).

\subsection{AND\(^\dagger\) gate}

\begin{figure}
	\centering
	\begin{quantikz}[row sep={0.8cm,between origins}]
		\lstick{$x$}  & \qw                  &                                      & \ctrl{1}  & \rstick{$x$}
		\\
		\lstick{$y$}  & \qw                  &                                      & \gate{Z}  & \rstick{$y$}
		\\
		\lstick{$xy$} & \gate{H} \wire[r]{q} & \meter{} \setwiretype{n} \wire[r]{c} & \ctrl{-1}
	\end{quantikz}
	\caption{A circuit diagram for the inverse of the temporary logical-AND gate, which performs measurement-based uncomputation~\cite{Gidney2018-xg}.
		The circuit takes inputs \(x\), \(y\), and the result \(xy\), and uncomputes the third qubit back to a clean state using a measurement-based strategy that avoids the use of T gates for uncomputation.
	}
	\label{fig:and_dagger_gate_circuit}
\end{figure}
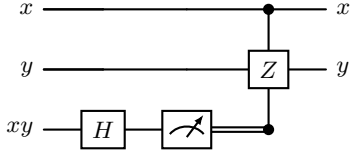

We can use measurement-based uncomputation to implement the conjugate of the AND gate, which reduces the ancilla volume and simplifies the implementation.
We show the circuit diagram for this strategy in \Cref{fig:and_dagger_gate_circuit}.
Note that the CZ correction will be required \(50\%\) of the time but that this won't be known until execution time.
Because the Hadamard followed by the measurement is just an \(X\) basis measurement, the CZ correction is the only gate that we must account for.
We account for its full volume under the assumption that a compiler would have to react unreasonably quickly to make use of the extra volume in the case where it can be neglected.
Therefore, we take the ancilla volume for the AND\(^\dagger\) gate to be equal to the volume for a CZ gate between the first two qubits.
Its measurement depth is one because deciding on the implementation of the CZ gate requires waiting for the \(X\) basis measurement to be resolved.
We note that the delayed choice CZ of \cite{Gidney2019-qi} has a similar but slightly larger volume, so the FLASQ model should still be reasonably accurate when this construction is preferred.

\subsection{Toffoli gate}

\begin{figure}
	\centering
	\begin{subfigure}[b]{0.48\textwidth}
		\centering
		\begin{quantikz}[wire types={n,q,q,q}, row sep={0.8cm,between origins}]
			\\
			 & \ctrl{2} &
			\\
			 & \ctrl{1} &
			\\
			 & \targ{}  &
		\end{quantikz}
		$=$
		\begin{quantikz}[wire types={n,q,q,q}, row sep={0.8cm,between origins}]
			             & \wire[r]{q} & \ctrl{3} & \wire[l]{q} &
			\\
			\lstick{$x$} & \ctrl{-1}   & \qw      & \ctrl{-1}   & \rstick{$x$}
			\\
			\lstick{$y$} & \ctrl{-2}   & \qw      & \ctrl{-2}   & \rstick{$y$}
			\\
			\lstick{$z$} & \qw         & \targ{}  & \qw         & \rstick{$z \oplus xy$}
		\end{quantikz}
		\caption{A circuit diagram for the Toffoli gate using an AND, CNOT, AND$^\dagger$ gate.
			The decomposition consumes 4 $\ket{T}$ states, 1 ancilla qubit, measurements and classically controlled S and CZ phase corrections \cite{Jones2013-vm}.
			Here is a \href{https://algassert.com/quirk\#circuit=\%7B\%22cols\%22\%3A\%5B\%5B\%22Y\%5Et\%22\%2C\%22Y\%5E-t\%22\%2C\%22Y\%5Et\%22\%2C\%22Z\%5E\%C2\%BC\%22\%2C\%22Z\%5E-\%C2\%BC\%22\%2C\%22Z\%5E-\%C2\%BC\%22\%2C\%22Z\%5E\%C2\%BC\%22\%5D\%2C\%5B\%22QFT3\%22\%5D\%2C\%5B\%22\%E2\%80\%A6\%22\%2C\%22\%E2\%80\%A6\%22\%2C\%22\%E2\%80\%A6\%22\%5D\%2C\%5B\%22Chance3\%22\%5D\%2C\%5B\%22Amps3\%22\%5D\%2C\%5B\%5D\%2C\%5B\%22\%E2\%80\%A6\%22\%2C\%22\%E2\%80\%A6\%22\%2C\%22\%E2\%80\%A6\%22\%5D\%2C\%5B1\%2C\%22zpar\%22\%2C1\%2C\%22zpar\%22\%2C\%22zpar\%22\%2C1\%2C1\%2C\%22X\%22\%5D\%2C\%5B\%22zpar\%22\%2C1\%2C1\%2C\%22zpar\%22\%2C1\%2C\%22zpar\%22\%2C1\%2C1\%2C\%22X\%22\%5D\%2C\%5B\%22zpar\%22\%2C\%22zpar\%22\%2C1\%2C\%22zpar\%22\%2C1\%2C1\%2C\%22zpar\%22\%2C1\%2C1\%2C\%22X\%22\%5D\%2C\%5B1\%2C1\%2C1\%2C\%22H\%22\%5D\%2C\%5B\%22\%E2\%80\%A6\%22\%2C\%22\%E2\%80\%A6\%22\%2C\%22\%E2\%80\%A6\%22\%2C\%22\%E2\%80\%A6\%22\%5D\%2C\%5B1\%2C1\%2C\%22X\%22\%2C\%22zpar\%22\%5D\%2C\%5B1\%2C1\%2C1\%2C\%22Z\%5E\%C2\%BD\%22\%5D\%2C\%5B1\%2C1\%2C1\%2C\%22H\%22\%5D\%2C\%5B1\%2C1\%2C1\%2C\%22Measure\%22\%2C1\%2C1\%2C1\%2C\%22Measure\%22\%2C\%22Measure\%22\%2C\%22Measure\%22\%5D\%2C\%5B1\%2C1\%2C1\%2C1\%2C1\%2C1\%2C\%22H\%22\%2C1\%2C1\%2C\%22\%E2\%97\%A6\%22\%5D\%2C\%5B1\%2C1\%2C1\%2C1\%2C1\%2C1\%2C\%22X\%5E\%C2\%BD\%22\%2C1\%2C1\%2C\%22\%E2\%80\%A2\%22\%5D\%2C\%5B1\%2C1\%2C1\%2C1\%2C1\%2C1\%2C\%22Measure\%22\%5D\%2C\%5B1\%2C1\%2C1\%2C1\%2C1\%2C\%22H\%22\%2C1\%2C1\%2C\%22\%E2\%97\%A6\%22\%5D\%2C\%5B1\%2C1\%2C1\%2C1\%2C1\%2C\%22X\%5E-\%C2\%BD\%22\%2C1\%2C1\%2C\%22\%E2\%80\%A2\%22\%5D\%2C\%5B1\%2C1\%2C1\%2C1\%2C1\%2C\%22Measure\%22\%5D\%2C\%5B1\%2C1\%2C1\%2C1\%2C\%22H\%22\%2C1\%2C1\%2C\%22\%E2\%97\%A6\%22\%5D\%2C\%5B1\%2C1\%2C1\%2C1\%2C\%22X\%5E-\%C2\%BD\%22\%2C1\%2C1\%2C\%22\%E2\%80\%A2\%22\%5D\%2C\%5B1\%2C1\%2C1\%2C1\%2C\%22Measure\%22\%5D\%2C\%5B\%22\%E2\%80\%A2\%22\%2C\%22Z\%22\%2C1\%2C\%22zpar\%22\%2C\%22zpar\%22\%2C\%22zpar\%22\%2C\%22zpar\%22\%5D\%2C\%5B1\%2C\%22Z\%22\%2C\%22X\%22\%2C1\%2C\%22\%E2\%80\%A2\%22\%5D\%2C\%5B\%22Z\%22\%2C1\%2C\%22X\%22\%2C1\%2C1\%2C\%22\%E2\%80\%A2\%22\%5D\%2C\%5B\%22Z\%22\%2C\%22Z\%22\%2C\%22X\%22\%2C1\%2C1\%2C1\%2C\%22\%E2\%80\%A2\%22\%5D\%2C\%5B\%22\%E2\%80\%A6\%22\%2C\%22\%E2\%80\%A6\%22\%2C\%22\%E2\%80\%A6\%22\%2C\%22\%E2\%80\%A6\%22\%5D\%2C\%5B\%22\%E2\%80\%A2\%22\%2C\%22\%E2\%80\%A2\%22\%2C\%22X\%22\%5D\%2C\%5B\%22QFT\%E2\%80\%A03\%22\%5D\%2C\%5B\%22Y\%5E-t\%22\%2C\%22Y\%5Et\%22\%2C\%22Y\%5E-t\%22\%5D\%2C\%5B\%22Chance3\%22\%5D\%5D\%2C\%22init\%22\%3A\%5B0\%2C0\%2C0\%2C\%22\%2B\%22\%2C\%22\%2B\%22\%2C\%22\%2B\%22\%2C\%22\%2B\%22\%5D\%7D}{quirk link} that shows the decomposition.
		}
		\label{fig:toffoli_circuit_quantikz}
	\end{subfigure}
	\hfill
	\begin{subfigure}[b]{0.48\textwidth}
		\centering
		\includegraphics[width=.9\textwidth]{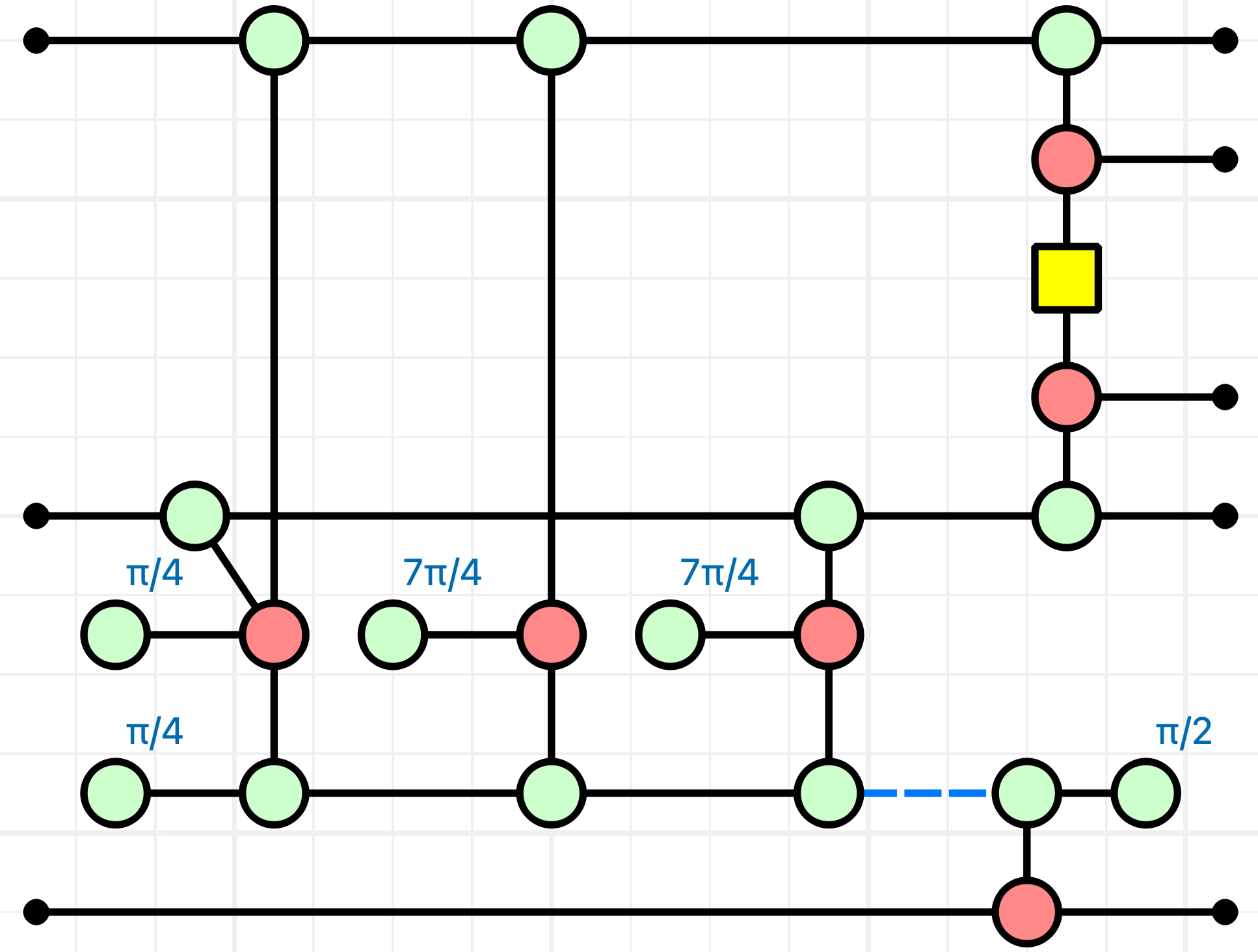}
		\caption{
			ZX graph for Toffoli gate using 4 T gates.
			Resembles the ZX graph from \cref{fig:and_gate_zxgraph}, but includes a delayed choice CZ gadget from \cite{Gidney2019-qi} to account for phase corrections.
		}
		\label{fig:toffoli_zx_graph}
	\end{subfigure}
	\begin{subfigure}[b]{.7\textwidth}
		\centering
		\includegraphics[width=0.45\textwidth]{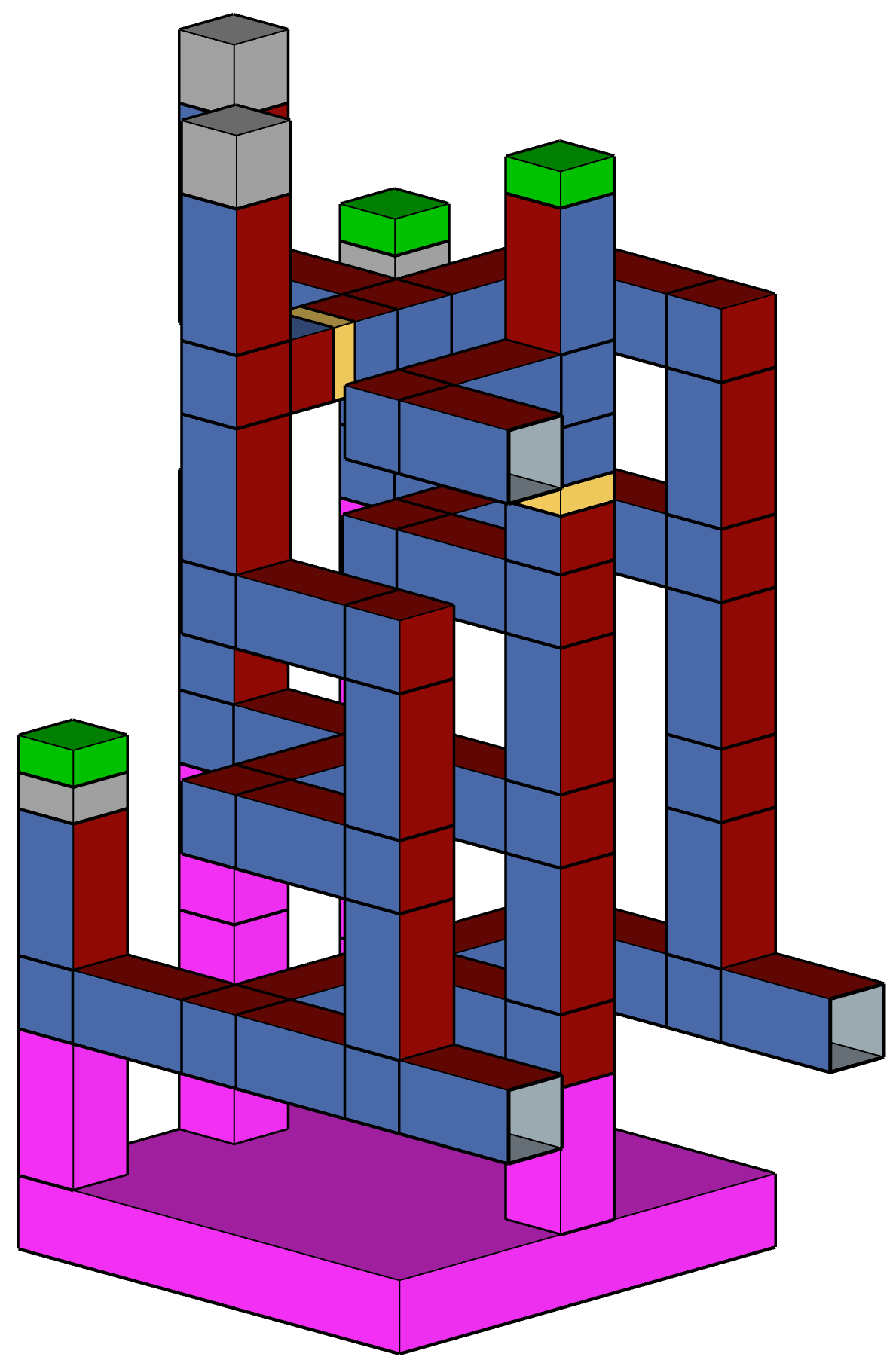}
		\hfill
		\includegraphics[width=0.45\textwidth]{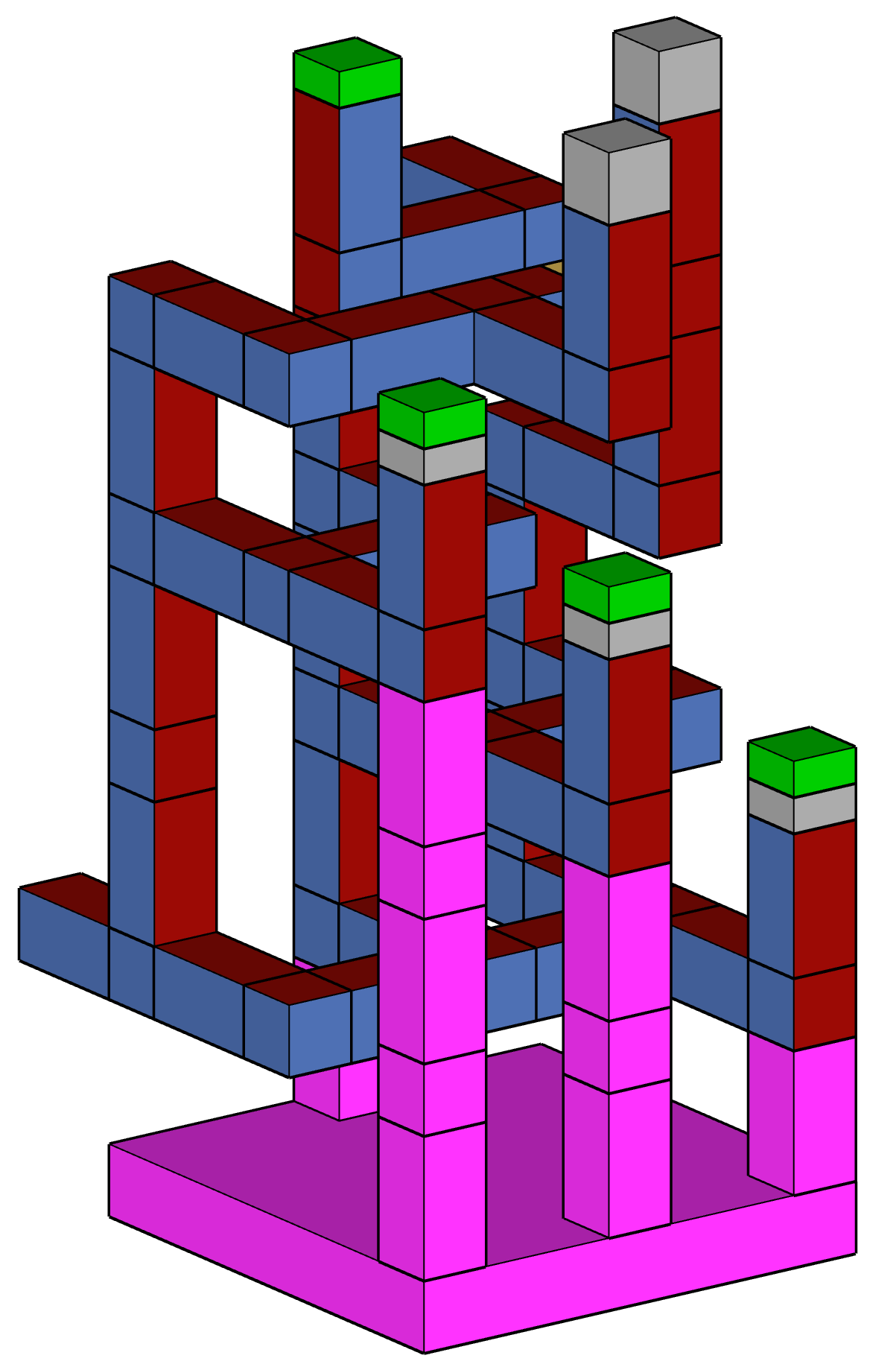}
		\caption{
			Space time layout of the Toffoli gate in a $3\times3\times 6$ region, which extends the layout of of an AND gate from \cref{fig:and_gate_pipe_diagram} with a delayed choice CZ correction \cite{Gidney2019-qi}.
			Pink boxes correspond to space used for magic state cultivation.
			For every $\ket{T}$ state to be consumed, there is $3d^3$ spacetime available for cultivation.
			Grey boxes correspond to the reaction time required for the decoder to finish processing measurement results and determine if S and CZ corrections are needed.
			Green boxes correspond to Y basis initialization and measurement.
		}
		\label{fig:toffoli_pipe_diagram}
	\end{subfigure}
	\caption{A Toffoli gate expressed in three ways, as a quantum circuit diagram, as a ZX calculus graph, and as a lattice surgery pipe diagram.}
	\label{fig:toffoli_gate_diagrams}
\end{figure}

To assign an ancilla volume to the Toffoli gate we will follow a similar strategy to the one we followed for the AND gate, first analyzing the volume of the concrete pipe diagram implementation in \Cref{fig:toffoli_pipe_diagram}.
This construction broadly follows the strategy outlined in the quantum circuit diagram of \Cref{fig:toffoli_circuit_quantikz}, computing the AND of two input qubits, XORing this to the output, and using a delayed-choice CZ construction to uncompute the temporary AND.
As a result, the measurement depth is two, although other constructions may have a lower measurement depth (potentially at the expense of increasing the T-count).
Like the AND gate we considered in \Cref{fig:and_gate_circuit}, the Toffoli gate uses a Z port construction, so the entire volume present in the figure (which shows the same structure from two sides) is fluid ancilla volume.

For clarity, we work through the pipe diagram in two stages, first addressing the components other than the blocks consisting of red and blue boxes indicating standard lattice surgery operations.
The T state cultivation takes a total of \(4 \textsc{Vol}\left( Cultivate \right)\).
There are five grey boxes (which require a height at least equal to the reaction time), contributing a volume of \(5 t_{react}\).
The four green boxes denoting \(Y\) basis measurements contribute \(4\) blocks in the conservative model (\(2\) in the optimimistic model).
Additionally, we need to account for the ancilla volume required to provide this ancilla space, which is \(36\) in the conservative model (\(9\) in the optimistic model).
The remaining volume can be accounted for by counting the occupied space.
Going bottom to top, we use \(7 + 6 + 7 + 7\), plus there is a single block worth of space where an ancilla is only unused for a single timestep and we therefore include it as a cost since it is unlikely to be usable for anything else (visible in the second row of the figure on the righthand side).

As with the AND gate, the distance dependent portions of the cost are \(5 p(q_1, q_2, q_3)\) in the conservative model and \(2 p(q_1, q_2, q_3)\) in the optimistic model.
This is because the only operations that are required are connecting this structure to the input and output qubits (each operation can be done in a single timestep with a single ``pipe''), plus the cost of opening and potentially closing the path.
The overall requirements for the ancilla volume of the Toffoli gate are therefore
\begin{equation}
	\textsc{Vol}\left( Toffoli \right) = 4 \textsc{Vol}\left( Cultivate \right) + 5 t_{react} + 5 p(q_1, q_2, q_3) + 68
\end{equation}
in the conservative model and
\begin{equation}
	\textsc{Vol}\left( Toffoli \right) = 4 \textsc{Vol}\left( Cultivate \right) + 5 t_{react} + + 2 p(q_1, q_2, q_3) + 39
\end{equation}
in the optimistic model.

As a consistency check for the FLASQ model, we compare the volume estimates obtained directly from the pipe diagram with those we obtain by applying the FLASQ model to the circuit construciton in \Cref{fig:toffoli_circuit_quantikz}.
As shown in \Cref{fig:toffoli_circuit_quantikz}, the circuit consists of an `AND` gate that computes the conjunction of the controls into a temporary ancilla, a `CNOT` that applies this result to the target, and finally an `AND\(^\dagger\)` gate to uncompute the ancilla.
By summing the individual gate costs for this specific layout, we obtain a total ancilla volume of \(4 \cdot \textsc{Vol}\left( \text{Cultivate} \right) + 2 t_{react} + 94\) for the conservative model and \(4 \cdot \textsc{Vol}\left( \text{Cultivate} \right) + 2 t_{react} + 46.25\) for the optimistic model.
These values are in reasonably close agreement with the volumes of \(4 \cdot \textsc{Vol}\left( \text{Cultivate} \right) + 5 t_{react} + 78\) (conservative) and \(4 \cdot \textsc{Vol}\left( \text{Cultivate} \right) + 5 t_{react} + 40\) (optimistic) predicted by the formulas derived from the pipe diagram, although we note that we treat the temporary ancilla qubit as a fluid ancilla qubit in the context of the Toffoli implementation (but have not done so in simply summing the gate costs).

\subsection{Y basis initialization and measurement}

The fundamental ancilla volume for a Y-basis measurement is \(0.5\) blocks, based on the methods in \cite{Gidney2024-fy}.
The optimistic volume uses this \(0.5\) value directly.
The conservative volume is \(1\), which is obtained by rounding the base volume of \(0.5\) up to the nearest whole number.
 
\section{Estimating the number of fluid ancilla}
\label{app:max_qubit_heuristic}

The FLASQ model requires an estimate of the number of qubits available for use as fluid ancilla, $A$.
Given a total number of available logical qubits $N_{tot}$, we calculate \(A\) as $A = N_{tot} - Q$, where $Q$ is a heuristic estimate of the maximum number of data qubits and algorithmic ancilla simultaneously active during the execution of the circuit $\mathcal{C}$.
This estimate is generated for one particular way of implementing the circuit that attempts to serialize the usage of algorithmic ancilla qubits.
For simple circuits that act on a fixed set of qubits with unitary gates, it is trivial to determine \(Q\) and setting \(A\) this way involves no approximation.
For more general algorithms however, we make several approximations.

Many algorithms involve the temporary use of ancilla qubits.
When this ancilla usage is defined at the circuit level, we call these ancilla ``algorithmic ancilla'' to distinguish them from the ``fluid ancilla'' that are implicitly managed by FLASQ.
When an algorithm requires a fresh ancilla qubit in the \(\ket{0}\) state, we say that the ancilla is ``allocated,'' and when the ancilla may be measured or traced out we say that it is ``deallocated.''
For algorithms that use algorithmic ancillla, setting \(Q\) to be the maximum number of data qubits and algorithmic ancilla simultaneously in use is an approximation.
If the algorithm's usage of ancilla qubits is uneven (i.e., sometimes it needs many and sometimes it needs few) then this approximation may underestimate the amount of fluid ancilla spacetime available.
As a result, it may overestimate the number of logical timesteps required to implement the circuit as well as the overall spacetime volume required.

Even under this approximation, the actual value of \(Q\) could be challenging to determine and will, in general, depend on how the computation is laid out in space and time.
In order to resolve this, we make another approximation by employing a heuristic implemented in the Qualtran software library~\cite{Harrigan2024-rj}.
For simplicity, we explain how this heuristic functions when using FLASQ as it is currently implemented.
Qualtran is designed with a hierarchical structure (so that subroutines can be encapsulated and efficiently processed by the library, even for very large algorithms), but we do not take advantage of this capability when using it to analyze early fault-tolerant algorithms at present.

Like our calculation of measurement depth, the qubit-counting heuristic analyzes the circuit as a directed acyclic graph (the ``compute graph''), where the nodes are gates or other operations and the edges represent the qubits those nodes act on.
To estimate the maximum qubit usage, Qualtran begins by topologically sorting the graph.
The outcome of this sorting is one of many valid orderings for performing the operations in the circuit.
The sorting attempts to greedily minimize the number of qubits in use, trying to allocate qubits as late as possible and deallocate them as early as possible, but it otherwise leaves operations in the order they were defined (usually by importing from a 2D circuit).

Qualtran iterates through the sequence of operations obtained from sorting the compute graph and calculates the following for each operation:
\begin{enumerate}
	\item The number of qubits the operation acts on (which is trivial to determine for all of the primitive gates and operations we consider here).
	\item The number of other qubits in use at that point in the sequence.
	      This is straightforward to determine by keeping track of a list of outgoing edges as the sequences is traversed.
\end{enumerate}
The heuristic sets \(Q\) to be the maximum value obtained this way while iterating through the sequence of operations.

This approach to determining \(Q\) effectively minimizes the estimate of the required space by assuming that operations that temporarily use ancilla qubits are executed as serially as possible.
This approximation risks overestimating the amount of fluid ancilla space for algorithms that use many temporary ancilla qubits in a way that is highly parallelizable.
When sufficient space is available and the FLASQ model implicitly parallelizes these operations, it will not properly account for the fact that each parallel operation has its own ancilla qubit requirement.
This will lead to an underestimation of \(Q\), an overestimation of \(A\) (the amount of fluid ancilla space), and therefore an underestimation of the cost in the FLASQ model.
We expect this approximation to be reliable in the space-limited regimes that will characterize the earliest parts of early fault-tolerance, or when the usage of algorithmic ancilla is serial in nature anyway.

Taken together, these two approximations enable us to apply the FLASQ model automatically and scalably to circuits that use algorithmic ancilla.
We foresee future iterations of the FLASQ model including tools that help researchers manage their drawbacks.
For example, one could manually set a minimum number of algorithmic ancilla to account for some parallelism, or even do this in a semi-automated way for particular applications.
Or, to deal with the issue of uneven algorithmic ancilla usage, one could apply the FLASQ model to various stages of an algorithm separately and then combine the results to obtain more accurate results overall.

\section{Applying the FLASQ model to finite field multiplication}
\label{app:validation_details}

In this section, we compare the estimates of the FLASQ model with the spacetime volume realized by hand-optimized constructions for several arithmetic circuits.
In \cite{Khattar2025-yr}, the authors estimated the resources for a particular realization of the decoded quantum interferometry algorithm of \cite{Jordan2024-hj}.
One of the key technical contributions of \cite{Khattar2025-yr} was providing optimized lattice surgery implementations for arithmetic over finite fields of size \(2^n\) (commonly referred to as Galois fields of size \(2^n\), or more succinctly as \(GF(2^n)\)).
We present our comparison in \Cref{tab:gf_circuit_comparison}.

\begin{table*}
	\centering
	\setlength{\tabcolsep}{8pt}
	\renewcommand{\arraystretch}{1.5}
	\begin{tabular}{@{}l l r@{}}
		\toprule
		\textbf{Instance}                                  & \textbf{Methodology}                 & \makecell[r]{\textbf{Spacetime}
		\\
			\textbf{volume}}
		\\
		\midrule
		Karatsuba multiplier
		                                                   & FLASQ (conservative)                 & 30,536
		\\
		(51 logical qubits)                                & FLASQ (optimistic)                   & 14,456
		\\
		                                                   & Hand-compiled                        & \(13{,}005\)
		\\
		                                                   & Ratio (Conservative / Hand-compiled) & 2.35
		\\
		                                                   & Ratio (Optimistic / Hand-compiled)   & 1.11
		\\
		\specialrule{0.1pt}{2pt}{2pt}
		Karatsuba multiplier                               & FLASQ (conservative)                 & 21,184                          
		\\
		(102 logical qubits)                               & FLASQ (optimistic)                   & 9,781
		\\
		                                                   & Hand-compiled                        & \(7{,}956\)
		\\
		                                                   & Ratio (Conservative / Hand-compiled) & 2.66
		\\
		                                                   & Ratio (Optimistic / Hand-compiled)   & 1.23
		\\
		\specialrule{0.1pt}{2pt}{2pt} Quadratic multiplier & FLASQ (conservative)                 & 23,765                         
		\\
		(105 logical qubits)                               & FLASQ (optimistic)                   & 12,040
		\\
		                                                   & Hand-compiled                        & 8{,}400
		\\
		                                                   & Ratio (Conservative / Hand-compiled) & 2.83
		\\
		                                                   & Ratio (Optimistic / Hand-compiled)   & 1.43
		\\
		\bottomrule
	\end{tabular}
	\caption{ A comparison of resource estimates from the FLASQ model against several hand-compiled lattice surgery implementations of multiplication over \(GF(2^{10})\).
		All spacetime volumes are in units of blocks.
		The hand-compiled values for the multipliers are derived from the pipe diagrams presented in \Cref{fig:quadratic_multiplication,fig:karatsuba_small_footprint,fig:karatsuba_large_footprint}.
		The conservative FLASQ model substantially overestimates the required spacetime volume.
		The optimistic model also overestimates the volumes, although it is considerably more accurate.
		For the hand-compiled multipliers, the ratio of the spacetime volumes for the small footprint Karatsuba multiplier, the large footprint Karatsuba multiplier, and the quadratic multiplier is approximately \(1.63 : 1.00 : 1.06\).
		Whereas for the FLASQ model, the ratios are approximately \(1.44 : 1.00 : 1.12\) (conservative) and \(1.48 : 1.00 : 1.23\) (optimistic).
	}
	\label{tab:gf_circuit_comparison}
\end{table*}

\begin{figure}
	\centering
	\includegraphics[width=0.8\textwidth]{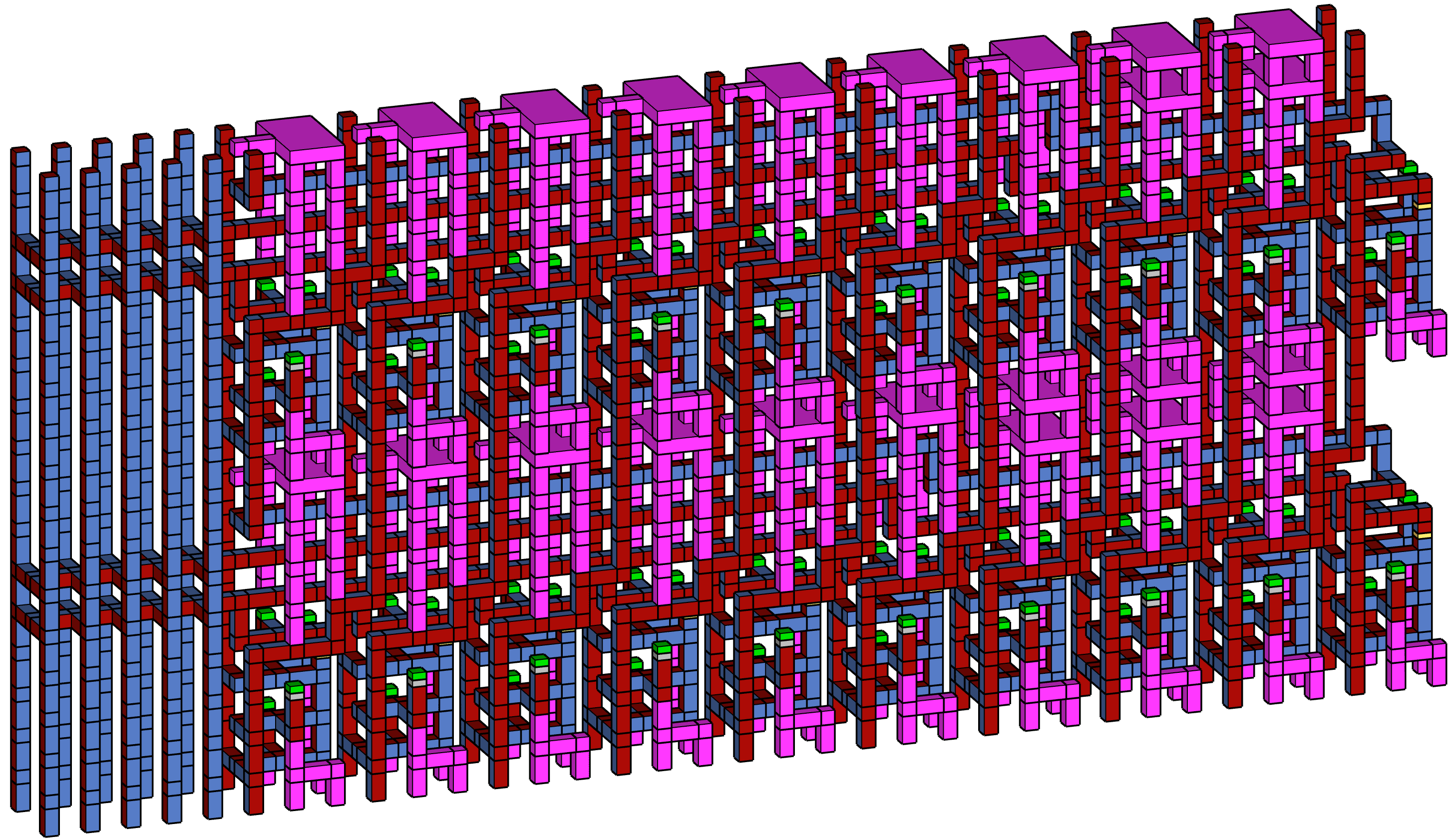}
	\caption{A hand-optimized pipe diagram showing a compilation of a Mastrovito multiplier over \(GF(2^{10})\).
		This figure shows \(2\) of the \(10\) layers of Toffoli gates used in the complete compilation~\cite{Harrigan2024-rj}.
		The entire multiplier fits within a spatial footprint of \(3 \times 35\), and can be performed in \(80\) logical timesteps for a total spacetime volume of \(8{,}400\).
	}
	\label{fig:quadratic_multiplication}
\end{figure}

\begin{figure}
	\centering
	\includegraphics[width=0.8\textwidth]{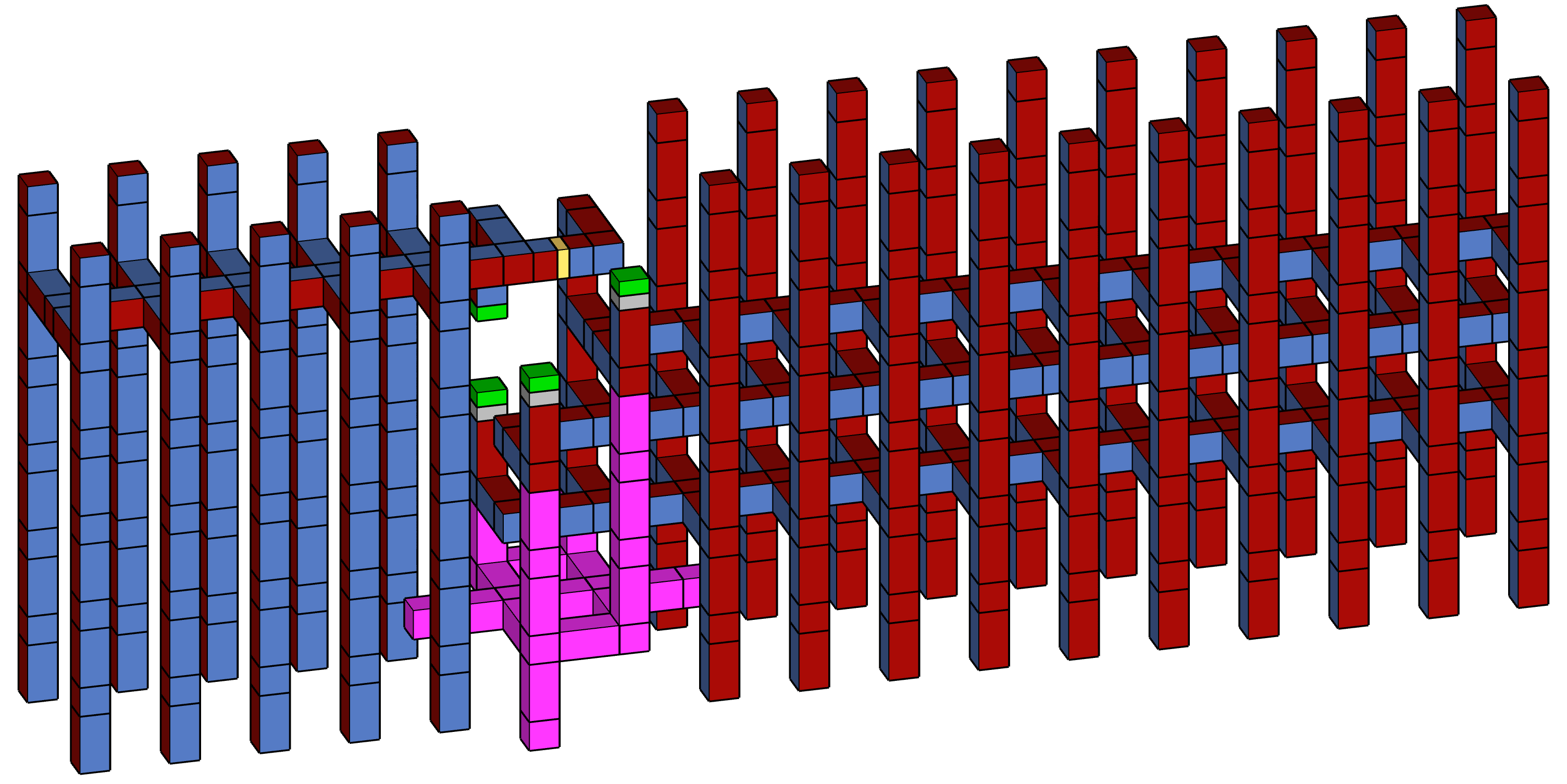}
	\caption{
		A pipe diagram that shows a hand-optimized compilation of one component of a Karatsuba multiplier over \(GF(2^{10})\) designed to occupy a small spatial footprint.
		This figure shows \(1\) of the \(51\) parity-control Toffoli gates used in the complete compilation~\cite{Harrigan2024-rj}.
		The entire multiplier fits within a spatial footprint of \(3 \times 17\), and can be performed in \(255\) logical timesteps for a total spacetime volume of \(13{,}005\).}
		\label{fig:karatsuba_small_footprint}
\end{figure}

\begin{figure}
	\centering
	\includegraphics[width=0.8\textwidth]{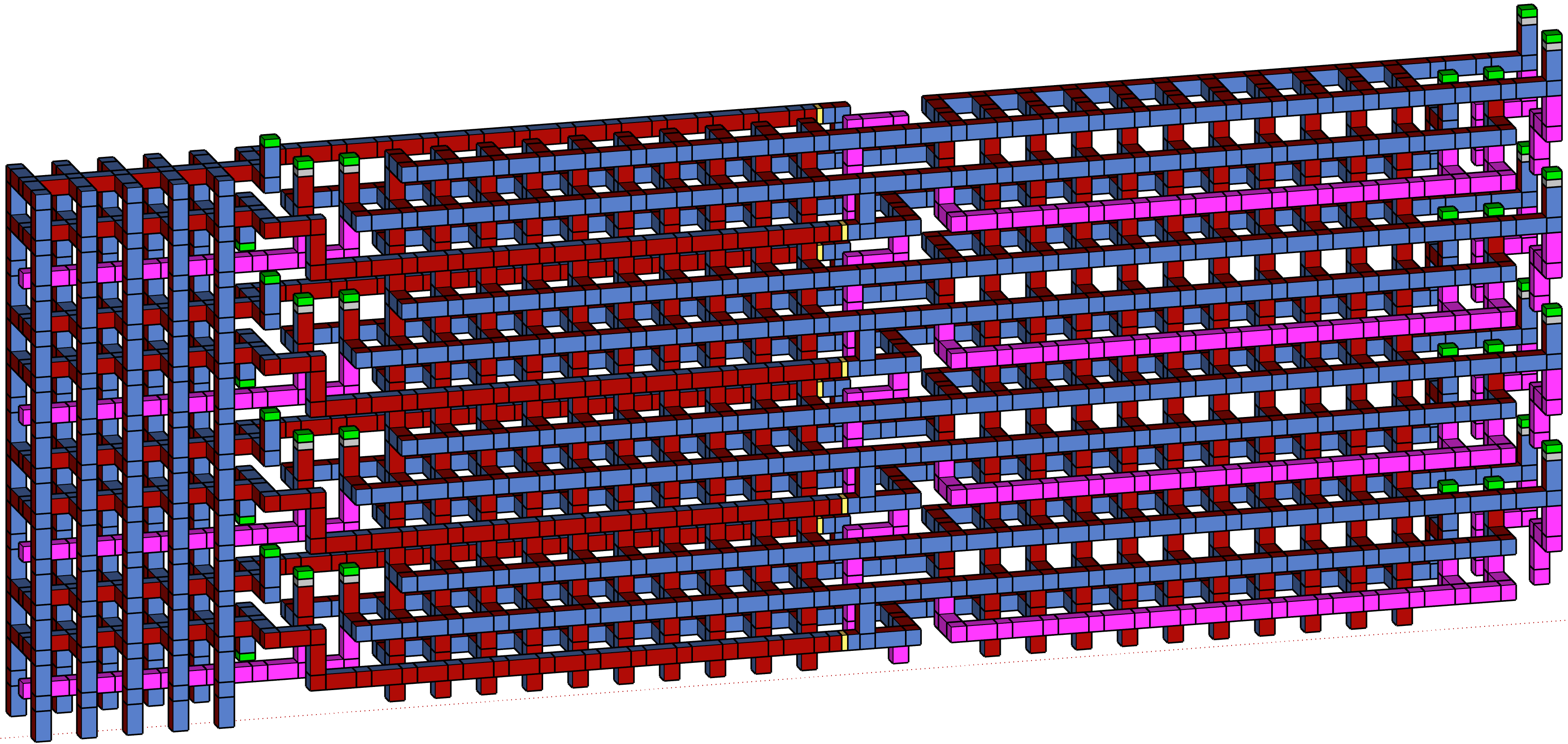}
	\caption{
		A pipe diagram that shows a hand-optimized compilation of a Karatsuba multiplier over \(GF(2^{10})\) originally presented in \cite{Khattar2025-yr}.
		This figure shows \(8\) of the \(51\) parity-control Toffoli gates used in the complete compilation~\cite{Harrigan2024-rj}.
		The entire multiplier fits within a spatial footprint of \(3 \times 34\), and can be performed in \(78\) logical timesteps for a total spacetime volume of \(7{,}956\).}
		\label{fig:karatsuba_large_footprint}
\end{figure}

We focus our comparisons on three different circuits for multiplication over \(GF(2^{10})\).
The first, shown in \Cref{fig:quadratic_multiplication}, shows a construction based on the Mastrovito multiplier that uses \(n^2 = 100\) Toffoli gates together with a relatively small number of CNOTs~\cite{Cheung2008-tk}.
The other two constructions are based on Karatsuba multiplication and use fewer Toffoli gates (\(\bigo{n^{\log_2 3}}\)) at the expense of requiring a more complicated set of Clifford operations~\cite{van-Hoof2019-bq}.
In \Cref{fig:karatsuba_small_footprint} and \Cref{fig:karatsuba_large_footprint}, we illustrate these two constructions, one of which is optimized to use a smaller number of ancilla qubits and one of which is optimized to use a larger number.
All three figures assume that three blocks of ancilla spacetime volume are sufficient for T state cultivation.

When performing the FLASQ estimates, we use the conservative and optimistic models as explained in the main text and in \Cref{app:flasq_volume_details}, except for a few variations.
To match the methodology used in the hand-compiled constructions, we assume that 3 blocks of ancilla spacetime are sufficient for cultivation in both models (i.e., we set \(\textsc{Vol}\left( \textsc{Cultivate} \right) = 3\)).
We assume a code distance of \(20\), which only impacts our estimates by setting \(t_{react} = .5\) (in units of logical timesteps).
We lay out the logical circuit analyzed by FLASQ in a way that emulates the basic structure of the hand-compiled layouts but excludes the ancilla space.
For the Mastrovito multiplier and the compact version of the Karatsuba multiplier, we arrange the output qubits in a \(2 \times 5 \) rectangle with corners at \((0,0)\) and \((1,4)\) and we arrange the input qubits in two side-by-side lines, together occupying the rectangle with corners at \((0, 5)\) and \((10, 14)\).
For the larger Karatsuba multiplier, we use the same layout of the output qubits, but arrange the input qubits in two lines end-to-end, spanning from a qubit at \((0, 5)\) to one at \(0, 24\).

Overall, we find that the conservative model consistently overestimates the spacetime volume required by more than a factor of \(2\), although it predicts the relative costs of the three constructions much more accurately.
The optimistic model provides spacetime volume estimates that are much closer, overestimating the volumes by \(11\%\) to \(43\%\).
Both models are more accurate as tools for comparing the relative costs than the absolute costs, although the variation in the error suggests that one should be cautious with quantitative comparisons.

The divergence between the conservative FLASQ spacetime volume estimates and the hand-compiled spacetime volumes highlights the impact of our assumptions about managing ancilla space.
The conservative model assigns a high overhead for allocating ancilla space for each operation.
This is based, roughly, on the assumption that we are compiling our algorithm by consistently restoring to some canonical configuration between applying gates.
More generally, it reflects as assumption that the costs of routing are high and that there are significant packing inefficiencies.
This contrasts with the optimistic model, which assumes these overheads are minimal.
The fact that the optimistic model aligns much more closely with the hand-compiled results indicates that these dense arithmetic circuits achieve a level of efficiency not captured by the conservative assumptions, reusing the same working space rather than repetedly paying a cost to provide it.

\section{Applying the FLASQ model to obtain resource estimates}
\label{app:resource_estimates}

The FLASQ model provides estimates for the logical timesteps ($L$), the total spacetime volume ($S$), and the number of magic states consumed ($M$).
In this appendix, we detail how these outputs are used to derive practical resource estimates for executing a quantum algorithm, particularly in the context of early fault-tolerance where we may not have sufficient resources to guarantee that logical errors happen with a negligible probability.
We begin by outlining the assumptions used in our analysis before explaining how we apply these assumptions to calculate several useful quantities.

We focus on two primary metrics for resource estimation.
The first is the expected time to successful execution, $T_{success}$.
This metric is relevant when errors can be detected and removed with postselection (e.g., when the output is classically verifiable), representing a ``best case'' scenario for handling residual errors.
The second metric is the total time required to estimate an expectation value to within some target precision using Probabilistic error cancellation (PEC), $T_{PEC}$.

Probabilistic error cancellation is a convenient error mitigation method for our purposes because it is straightforward to analytically determine its performance given a simple, known, noise model.
Furthermore, given that the noise model is known, it provides an unbiased estimator of the true expectation value.
These properties come at the cost of a relatively high sampling overhead.
Roughly speaking, PEC requires a number of samples that scales with the fourth power of the inverse fidelity, whereas postselection (when possible) scales linearly with the inverse fidelity~\cite{Cai2022-um}.
We expect that FLASQ could be used to analyze the performance of other error mitigation methods that offer different tradeoffs between their cost, performance, and ease of analysis.

\subsection{Default assumptions}
\label{app:standard_assumptions_details}

Many of our concrete resource estimates share some common assumptions.
Unless otherwise noted, our wall-clock time calculations assume a surface code cycle time of \(\SI{1}{\us}\) and a reaction time of \(\qty{10}{\us}\).

To analyze the impact of errors in logical qubits, we adopt a simple noise model based on the probability of an error per qubit per surface code cycle (\(p_{cyc}\)) defined in \Cref{sec:surface_code}.
We abstract away the physical error rate into a single parameter \(p_{phys}\), and make the assumption that this parameter is related to the surface code error suppression factor \(\Lambda\) through the equation
\begin{equation}
	\Lambda = \frac{.01}{p_{phys}}.
\end{equation}
Here \(.01\) is an estimate of the surface code threshold~\cite{Raussendorf2007-sa,Wang2011-ud,Fowler2012-li}.
We further assume that the logical qubits experience an error per cycle with probability
\begin{equation}
	p_{cyc} = c_{cyc} \Lambda^{-(d + 1) / 2},
\end{equation}
where \(c_{cyc} \approx .03\) is a constant~\cite{Fowler2012-li}.
More specifically, we assume that bit and phase flip errors each happen with a probability \(\tilde{p}_{cyc}/2\),
\begin{align}
	\rho \rightarrow \left( 1 - \frac{\tilde{p}_{cyc}}{2} \right) \rho + \frac{\tilde{p}_{cyc}}{2} Z \rho Z^{\dagger}, \label{eq:phase_flip_channel}
	\\
	\rho \rightarrow \left( 1 - \frac{\tilde{p}_{cyc}}{2} \right) \rho + \frac{\tilde{p}_{cyc}}{2} X \rho X^{\dagger}, \label{eq:bit_flip_channel}
\end{align}
where \(\tilde{p}_{cyc} = p_{cyc} + \mathcal{O}\left( p_{cyc}^2 \right) = 2 - 2\sqrt{ 1 - p_{cyc}}\).
Frequently we will neglect the second order term and conflate \(p_{cyc}\) and \(\tilde{p}_{cyc}\).

As we discuss in \Cref{app:cultivation_details}, we analyze the performance of magic state cultivation using a particular uniform depolarizing noise model, following closely the methodology of \cite{Craig2024-kv}.
Using simulated data, we determine an error probability \(p_{mag}\) for magic state cultivation.
We assume the use of twirling, such that instead of obtaining the ideal state \(T \ket{+}\), cultivation produces the state
\begin{equation}
	\rho_{mag} = \left( 1 - p_{mag} \right) T \ketbra{+} T^\dagger + p_{mag} Z T \ketbra{+} T^\dagger Z.
\end{equation}
In our resource estimates, we conflate the noise strength of the uniform depolarizing noise model used to analyze magic state cultivation with the \(p_{phys}\) error parameter used to calculate \(\Lambda\).

When analyzing the cost of rotation synthesis, we assume the use of the ``mixed fallback'' method of \cite{Kliuchnikov2023-vm}, as reviewed in \Cref{app:rotation_synthesis_details}.
This synthesis strategy has a low cost, but this cost comes at the expense of error guarantees in the diamond norm rather than the operator norm.
We generally account for this by setting an overall error tolerance and dividing by the total number of rotations to obtain an error tolerance for each rotation.
Additionally, we do not use the Hamming weight phasing strategy to reduce the cost of parallel rotations.
See \Cref{sec:hwp_example} for a discussion of this approach and its performance in the FLASQ cost model.

In many of our resource estimates, we provide estimates (of, e.g., runtime) obtained by optimizing over two parameters: the code distance used for the logical qubits and the postselection probability used for magic state cultivation.
Reducing the code distance or the postselection probability increases the probability of a logical error (in normal surface code qubit operations and in magic state cultivation respectively), but decreases the required spacetime.
This decrease in resources can lead to lower resource requirements despite the increased error probabilities, and we generally perform a grid search over parameter choices to minimize the resource requirements unless otherwise specified.

\subsection{Wall-Clock Time and Success Probability}

We can translate the number of logical timesteps (\(L\)) directly into an estimate of the wall-clock time required for a single execution:
\begin{equation}
	W = t_{cyc} d L,
\end{equation}
where \(t_{cyc}\) is the surface code cycle time and we have assumed that \(d\) surface code cycles are required for a logical timestep (to ensure fault-tolerance).

Using the error model defined in \Cref{sec:surface_code} (specifically $p_{cyc}$ and $p_{mag}$), we can calculate the probability that an algorithm executes successfully.
Letting \(S_{cliff} = S - v_{cult} M\) denote the spacetime volume used for all operations except magic state cultivation (in units of blocks), we have
\begin{equation}
	P_{success} = \left( 1 - p_{cyc} \right)^{d S_{cliff}}\left( 1 - p_{mag} \right)^M,
\end{equation}
where the quantity \(d S_{cliff}\) is the number of surface code cycles performed by all logical qubits.
To build intuition, we can consider the approximation
\begin{equation}
	P_{success} \approx \exp \left( -p_{cyc} d S_{cliff} -p_{mag} M\right) = \exp\left(-c_{cyc} \Lambda^{-\left( d + 1 \right) / 2} d S_{cliff} - p_{mag} M\right), \label{eq:P_success_approximation}
\end{equation}
which is accurate when the overall number of expected errors is \(\mathcal{O}(1)\) and the individual error probabilities (\(p_{cyc}\) and \(p_{mag}\)) are small.
Then the expected time to successfully execute the algorithm once is given by
\begin{equation}
	T_{success} \approx \underbrace{t_{cyc} d L}_{\text{time per execution}} \exp \left( \underbrace{c_{cyc} \Lambda ^{- (d + 1) / 2} d S_{cliff}}_{\text{clifford overhead}} + \underbrace{p_{mag} M}_{\text{T overhead}} \right).
	\label{eq:T_success_approximation}
\end{equation}

\subsection{Resource Estimation with Probabilistic Error Cancellation (PEC)}
\label{app:pec_details}

In situations where we can detect errors, then the $T_{success}$ calculation is sufficient to tell us the overall time required.
However, this is frequently not the case.
In many applications, we have a quantum circuit \(U\) that prepares a state \(\ket{\psi} = U \ket{0}\) and we want to estimate the expectation value of some observable \(O\) to within a precision \(\epsilon\).
With a noiseless quantum computer, we can do this with high probability by repeated state preparation and measurement using a number of samples that scales as \(M \propto \frac{\norm{O}^2}{\epsilon^2}\).
Repeated executions of the circuit on a noisy quantum computer will yield a biased estimate of the expectation value but we may not be able to detect errors in any particular execution of the circuit.
There are a variety of error mitigation techniques that help us construct an unbiased (or less biased) estimator for \(\ev{O}\) using a noisy quantum computer~\cite{Cai2022-um}.
In general, these approaches reduce the bias at the expense of increasing the variance.

One well-studied type of error mitigation, probabilistic error cancellation (PEC)~\cite{Temme2017-hj}, is known to be rather costly in practice but it has two appealing properties for our purposes: Given a known noise model, it yields an unbiased estimator of the expectation value and it is straightforward to bound the increased variance (and, therefore, the cost).

Probabilistic error cancellation works by writing the ideal channel as a linear combination of achievable noisy channels.
If this linear combination were convex (i.e., all of the coefficients were positive), then we could efficiently sample from the ideal channel by sampling from the achievable noisy channels.
This is not typically the case, but we can still sample from the ideal channel with some overhead by treating the sum over noisy channels as a quasi-probability distribution.

For example, consider the phase flip channel \(\mathcal{E}\)
\begin{equation}
	\mathcal{E}\left( \rho \right) = \left( 1 - p \right) \rho + Z \rho Z^{\dagger}.
\end{equation}
It is straightforward to verify that we can obtain the identity channel as a linear combination of \(\mathcal{E}\) and \(\mathcal{Z} \circ \mathcal{E},\) where \(\mathcal{Z}\left( \rho \right) = Z \rho Z^{\dagger}\), i.e.,
\begin{equation}
	\rho = \frac{1 - p}{1 - 2p} \mathcal{E}\left( \rho \right) - \frac{p}{1 - 2p} \mathcal{Z} \circ \mathcal{E}\left( \rho \right).
\end{equation}
We can recover expectation values with respect to \(O\) by observing that
\begin{equation}
	\tr \left[ O \rho \right] = \mathbf{\left( 1 - p\right)} \frac{1}{1 - 2p}\tr \left[ O \mathcal{E}\left( \rho \right) \right] + \mathbf{p} \frac{-1}{1 - 2p} \tr \left[ \mathcal{Z} \circ \mathcal{E} \left( \rho \right) \right].
\end{equation}
Specifically, we can use measurements to sample from \(\tr\left[ O \mathcal{E}\left( \rho \right) \right]\) with probability \(1 - p\) and \(\tr\left[ \mathcal{Z} \circ \mathcal{E} \left( \rho \right) \right]\) with probability \(p\), weighting the measurement outcomes by \(\frac{1}{1 - 2p}\) and \(\frac{-1}{1 - 2p}\) respectively.
This allows us to construct an unbiased estimator for \(\tr\left[ O \rho \right]\) with an increased number of samples.
The overhead is captured by a quantity \(\Gamma\), referred to as the sampling overhead.
In this example, the overhead scales as \(\Gamma^2 = \left( 1 - 2p \right)^{-2}\).
The error mitigation overhead from PEC is multiplicative.

We can encapsulate the total overhead into \(\Gamma\), so that variance of a measurement of an observable \(O\) is bounded by
\begin{equation}
	\left( \Delta O \right)_{noisy}^2 \leq \norm{O}^2 \Gamma^2.
\end{equation}
In our case, we have two components to our noise model and the overall overhead \(\Gamma^2\) is the product of \(\Gamma^2_{cliff}\) and \(\Gamma^2_{mag}\), which are given by the expressions
\begin{align}
	\Gamma^2_{cliff} & = \left( 1 - p_{cyc} \right)^{-4 d S_{cliff}},
	\\
	\Gamma^2_{mag}   & = \left( 1 - 2p_{mag} \right)^{-2 M}.
\end{align}
Approximating these quantities as exponentials, we can derive an expression for the time required to estimate the noiseless expectation value of an observable \(O\) to within a standard error \(\epsilon\),
\begin{equation}
	T_{PEC} \approx \underbrace{t_{cyc} d L}_{\text{time per sample}} \exp \left( \underbrace{4 c_{cyc} \Lambda ^{- (d + 1) / 2} d S_{cliff}}_{\text{clifford overhead}} + \underbrace{4 p_{mag} M}_{\text{T overhead}} \right) \underbrace{\norm{O}\epsilon^{-2}.
	}_{\text{\# noiseless samples}}
\end{equation}
This equation is similar to the time-to-success of \Cref{eq:T_success_approximation}, but the factor in the exponent is four times larger.
As a result, the sampling overhead grows much faster with the error rate than the time-to-success.

\section{Methodologies and additional data for the Ising model case studies}
\label{app:ising_details}

This appendix provides detailed methodologies, parameters, and the Trotter formulas used for the analysis presented in Section~\ref{sec:ising_model}.

\subsection{Trotter-Suzuki Formulas}
\label{app:trotter_formulas}

To simulate the time evolution of the Ising Hamiltonian, we employ Trotter-Suzuki product formulas.
The Hamiltonian is divided into \(H = A + B\), where \(A\) contains the \(X_i\) terms (external field) and \(B\) contains the \(Z_i Z_j\) terms (interactions).
The second-order formula \(U_2\) and the recursive construction for the fourth-order formula \(U_4\) are given by:

\begin{align}
	U_2\left( \Delta t\right) =  & e^{-i A \Delta t / 2} e^{- i B \Delta t} e^{-i A \Delta t / 2}, \nonumber
	\\
	U_4\left( \Delta t \right) = & U_2\left(\gamma \Delta t\right) U_2\left(\gamma \Delta t\right) U_2\left(\left( 1 - 4 \gamma \right) \Delta t\right) U_2\left(\gamma \Delta t\right) U_2\left(\gamma \Delta t\right),
	\\
	\gamma =                     & \left( 4 - 4^{1/3} \right)^{-1}.
	\nonumber
\end{align}

The circuits for implementing the time evolution under \(A\) and \(B\) are structured as follows.
The evolution under \(A\) consists of applying \(R_X\) rotations to all \(N\) qubits in parallel.
To implement the evolution under \(B\), the \(Z_i Z_j\) interaction terms are partitioned into four groups based on the geometry of the square lattice: interactions between a qubit and its neighbor above, below, to the right, and to the left.
For even sized lattices this yields four non-overlapping groups of terms.
For odd-sized lattices, we need to include the terms that wrap around the lattice in their own separate group to obtain a collection of non-overlapping groups (although this only affects the calculation of the measurement depth, not the amount of fluid ancilla spacetime volume required).
The evolution for each group of non-overlapping interactions is performed in parallel.
Each individual \(e^{-i\theta Z_i Z_j}\) term is implemented as a single-qubit \(R_Z(2\theta)\) rotation on one of the qubits, conjugated by CNOT gates between qubits \(i\) and \(j\).
The first and last terms of adjacent \(U_2\) blocks can be merged, resulting in exactly \(T + 1\) evolutions under \(A\) and \(T\) evolutions under \(B\) for the 2nd order evolution, or \(5T + 1\) evolutions under \(A\) and \(5T\) evolutions under \(B\) for the 4th order evolution.

\subsection{FLASQ resource estimates for the Ising model in the classically challenging regime}
\label{app:flasq_ising_classical_benchmark}

This section contains additional details about the calculations presented in \Cref{sec:ising_classical_benchmark_comparison}.

In the main text, we presented FLASQ estimates for simulating the 2D TFIM on an \(11 \times 11\) lattice.
The specific task was to estimate the expectation value of the observable \(Z_{tot}^2 = \frac{1}{N^2} \sum_{j,k} Z_j Z_k\) to within an absolute error of \(.01\) (with high probability) after performing \(20\) second-order Trotter steps.
We assume the use of error mitigation to remove (in expectation) the residual errors from implementing surface code operations and from magic state cultivation, but there is some error from compiling the rotations to a finite gate set that remains.
We set the total rotation error to a maximum of \(.001\) (in the diamond norm) and then calculated the sample complexity targeting a standard deviation of \(\sigma = .0045\), so that the overall error in the expectation is bounded by \(.01\) with \(\approx 95\%\) confidence.
In the resource estimates shown in \Cref{fig:ising_heatmap}, we assumed a surface code cycle time of \(\qty{1}{\us}\) and used the standard ``conservative'' FLASQ assumptions, which include an assumption that the reaction time is \(\qty{10}{\us}\).
As we described in the main text, the final runtimes were calculated by optimizing over the code distance and magic state cultivation parameters, with residual errors handled by probabilistic error cancellation as described in \Cref{app:pec_details}.

While we focused on the particular example studied in Refs.~\citenum{Haghshenas2025-kd,Mandra2025-sa}, the same estimates can be applied to the simulation task studied in \cite{Begusic2025-pk}, which also studied the 2D TFIM using a comparable number of Trotter steps with a much shorter total evolution time.
Using a Trotter step size of \(.04\), they found that simulations based on sparse Pauli dynamics could accurately estimate the expectation value of a single Pauli \(Z\) observable up to a total time of roughly \(T=.8\). 
For this variation of the simulation task, a single large workstation running sparse Pauli dynamics for more than a day was able to achieve an overall (estimated) absolute error of less than \(.01\), including a Trotter error of \(.003\) or less.
In order to significantly outperform classical methods in this setting, some combination of targeting a higher precision or a longer total evolution time would be required.

\begin{figure}
	\centering
	\includegraphics[width=.75\textwidth]{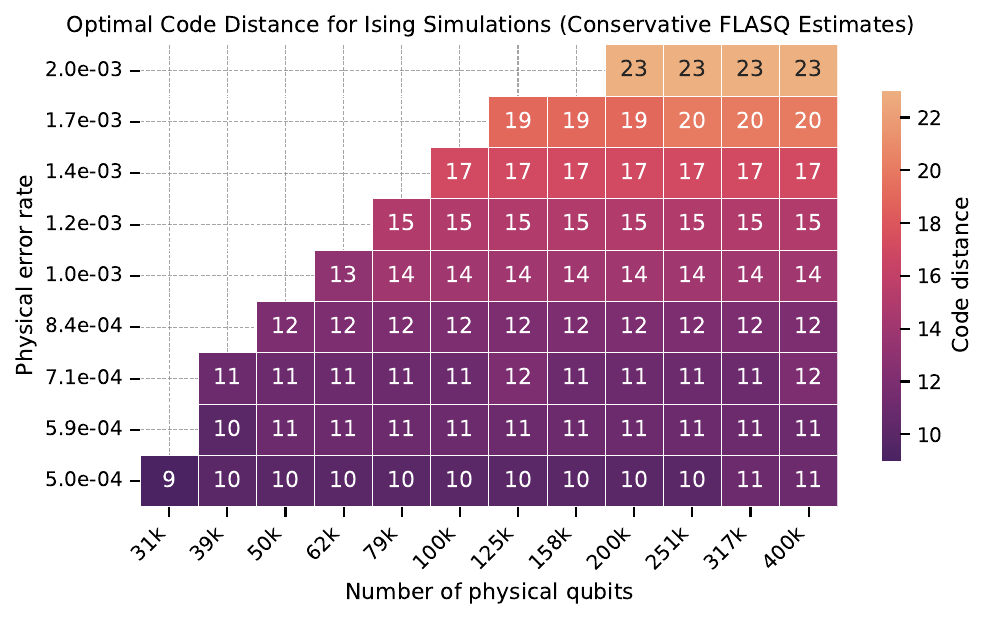}
	\caption{Optimal code distance for the conservative FLASQ simulation of the 2D TFIM.
		The plot shows the code distance that minimizes the total runtime for a given number of physical qubits and physical error rate.
	}
	\label{fig:optimal_code_distance}
\end{figure}

\begin{figure}
	\centering
	\includegraphics[width=.75\textwidth]{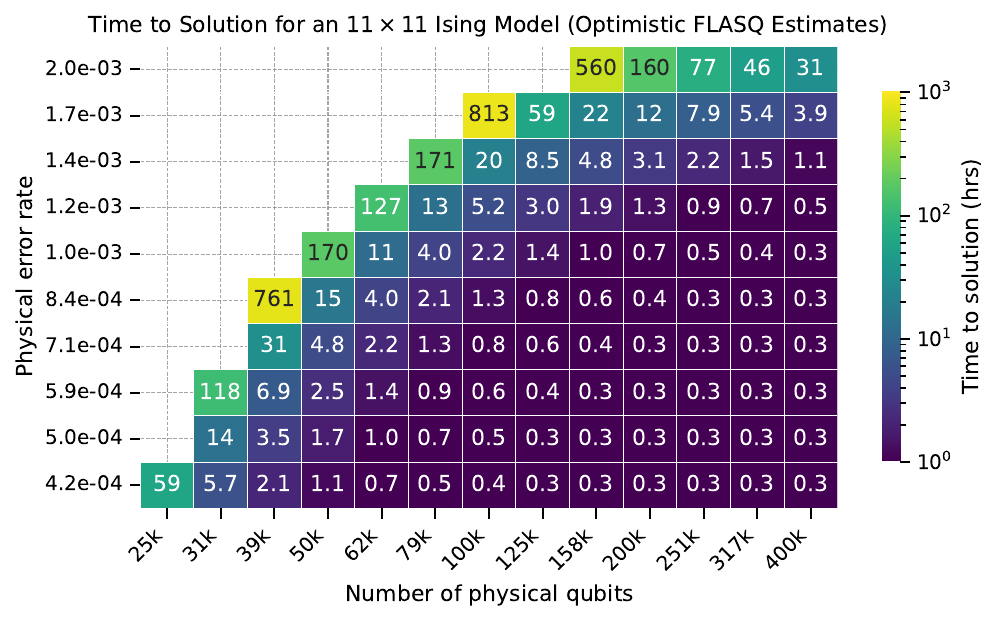}
	\caption{FLASQ estimates for the total runtime (in hours) required to estimate a diagonal observable with unit norm for an \(11\times11\) 2D TFIM, using the optimistic gate cost model from \Cref{tab:gate_costs_large}.
		The runtime is shown as a function of the number of available physical qubits and the physical gate error rate \(p_{phys}\).
	}
	\label{fig:ising_heatmap_optimistic}
\end{figure}

To better understand qualitative behavior of the FLASQ model for this application, we present some additional data.
In \Cref{fig:optimal_code_distance}, we plot the optimal code distance (the code distance that minimizes the total runtime) as a function of the physical error rate and the number of physical qubits.
This data is generated under the same assumptions used to generate \Cref{fig:ising_heatmap} in the main text.
In \Cref{fig:ising_heatmap_optimistic}, we generate an alternative version of the time-to-solution plot from the main text that uses the optimistic FLASQ cost estimates.
Comparing the two plots, we see that FLASQ predicts that the minimum number of physical qubits required for the simulation task is relatively insensitive to the exact estimates of the ancilla volume.

\subsection{Hand-compiled resource estimates}
\label{app:hand_compiled_ising_details}

In this appendix, we provide a detailed, step-by-step derivation of the hand-compiled resource estimates for the 2D Ising model simulation presented in \Cref{sec:hand_compiled_ising}, as well as a second example that parallels the simulation task considered in \Cref{sec:ising_beverland_comparison}.
This serves both to validate the estimates from our FLASQ model and to provide a concrete, state-of-the-art benchmark for this task.
We first outline the general compilation strategy before detailing the specific calculations for the two examples.

\subsubsection{General Compilation Strategy}

\begin{figure}
\includegraphics[width=0.6\linewidth]{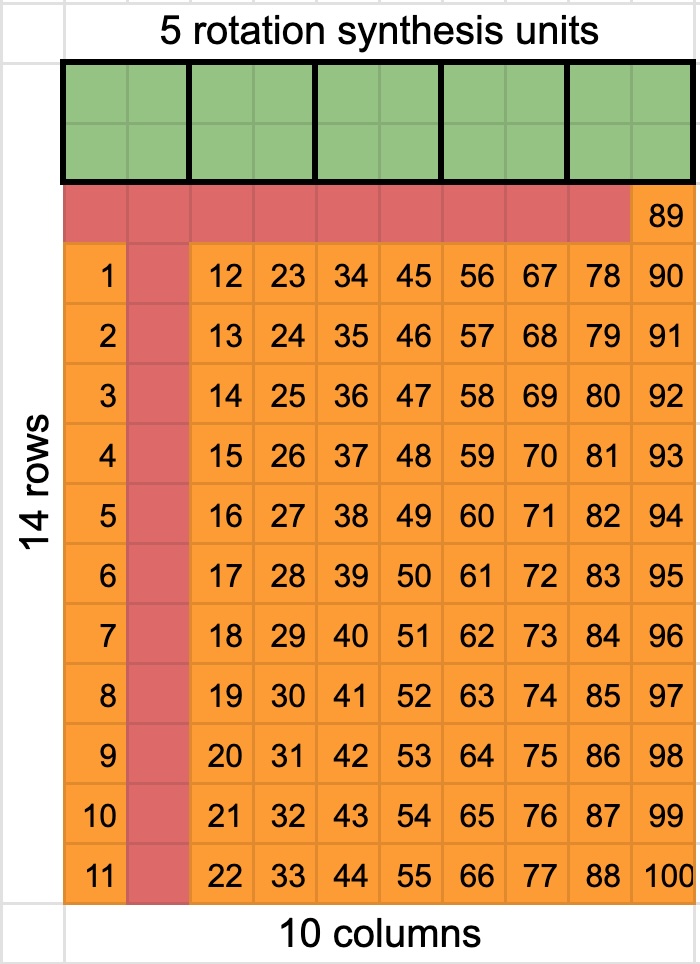} 
\caption{ Mockup of the physical layout for the \(10 \times 10\) 2D Ising model simulation, realized using a \(14 \times 10\) rectangular grid of logical qubits.
		Orange patches correspond to system qubits and are numbered according to their ordering (top to bottom, then left to right) in the \(10 \times 10\) Ising model lattice.
		Green boxes denote single qubit rotation synthesis units from \cref{fig:mixed_fallback_pipes}.
		Red patches are ancilla qubits used as routing space for connecting the orange system qubits with green boxes.
		The red ancilla region will sweep from left to right and right to left as the computation proceeds.
		This sweeping can be performed between timesteps when the rotation synthesis gadgets have to be connected to the system qubits in almost all cases.
		Most of the two qubit interactions happen between pairs of orange patches that lie within adjacent columns.
		With $d=14$, this requires a total of \num{63000} physical qubits.
	}
	\label{fig:10_x_10_ising_model_layout}
\end{figure}

Our compilation strategy is built around a rectangular grid of logical qubits with dedicated zones for performing arbitrary rotations.
The system qubits for the Ising model simulation are laid out in a zig-zag pattern to keep nearest-neighbor interactions in the simulation relatively local on the 2D grid of logical qubits.
The remaining space is populated with routing ancillas and several parallel rotation synthesis units, as illustrated in \Cref{fig:11_x_11_ising_model_layout} and \Cref{fig:10_x_10_ising_model_layout}.
Both \(R_X\) and \(R_{ZZ}\) rotations are implemented using the same underlying rotation synthesis gadget, which is based on the mixed fallback protocol of \cite{Kliuchnikov2023-vm} (see \Cref{fig:mixed_fallback_circuit} and \Cref{app:rotation_synthesis_details}).
The routing space is sufficient to connect the qubits involved in any particular rotation to one of the rotation synthesis gadgets.
As shown in the pipe diagrams in \Cref{fig:mixed_fallback_pipes}, these gadgets occupy a \(2 \times 2\) block of logical qubits and can synthesize a rotation in \(2m+2\) logical timesteps, where \(m\) is the number of T gates required.
With several of these units operating in parallel, we can pipeline the synthesis of the many rotations required by the algorithm.

\Cref{tbl:ising_model_physical_estimates} provides a comprehensive summary of the physical resource estimation parameters for the two examples we consider.
The following subsections detail how these parameters are derived.

\begin{table*}[]
	\centering
	\resizebox{\textwidth}{!}{%
		\renewcommand{\arraystretch}{1.5} %
		\begin{tabular}
			{@{}llcc@{}}
			\toprule
			\textbf{Symbol}                       & \textbf{Explanation}                                                                        & \textbf{10x10, 4th order} & \textbf{11x11, 2nd order}
			\\
			\midrule
			$N$                                   & Number of logical qubits in the system                                                      & 100                       & 121
			\\
			$A$                                   & Number of ancilla qubit patches for cultivation and routing                                 & 40                        & 39
			\\
			$N_{tot}$                             & Total logical qubit patches ($=N + A$)                                                      & 140                       & 160
			\\
			\midrule
			$n_{rot}$                             & Number of single qubit rotations in the logical circuit                                     & \num{30100}               & \num{7381}
			\\
			$m$                                   & Avg.
			T gates per rotation (mixed fallback) & 19                                                                                          & 17
			\\
			$n_\text{cycles\_per\_rot}$           & Rate of rotation synthesis (logical timesteps/rotation)                                     & 9                         & 10
			\\
			$L_{hand}$                            & Total logical timesteps ($= n_{rot} \times n_\text{cycles\_per\_rot}$)                      & \num{270900}              & \num{73810}
			\\
			$S_{hand}$                            & Total spacetime volume in blocks ($= N_{tot} \times L_{hand}$)                              & \num{3.79e7}              & \num{1.18e7}
			\\
			\midrule $d$                          & Code distance                                                                               & 14                        & 13
			\\
			$t_{cyc}$                             & Surface code cycle time                                                                     & \qty{400}{\ns}            & \qty{1}{\us}
			\\
			$t_{react}$                           & Reaction time (in units of cycles)                                                          & 10                        & 10
			\\
			$p_{phys}$                            & Physical error rate (assumed)                                                               & \num{1e-3}                & \num{1e-3}
			\\
			$p_{cyc}$                             & Logical error rate per qubit per cycle ($ \approx 0.03 \cdot (p_{th}/p_{phys})^{-(d+1)/2}$) & \num{9.5e-10}             & \num{3e-9}
			\\
			$P_\text{lattice\_surgery}$           & Fidelity from lattice surgery ($=(1 - p_{cyc})^{S_{hand} \times d}$)                        & \num{0.6038}              & \num{0.6309}
			\\
			\midrule $p_{mag}$                    & Target logical error rate of cultivated $\ket{T}$ states                                    & \num{2e-7}                & \num{3e-7}
			\\
			$v_\text{cult, physical}$             & Spacetime volume to cultivate a $\ket{T}$ state (physical qubits \(\times\) cycles)         & \num{1.8e4}               & \num{1.5e4}
			\\
			$v_\text{cult}$                       & Spacetime volume for cultivation in blocks ($=v_\text{cult, physical} / (2(d+1)^2 d)$)      & $\approx 3$               & $\approx 3$
			\\
			$P_\text{cultivation}$                & Fidelity from cultivation ($=(1 - p_{mag})^{n_{rot}\times m}$)                              & \num{0.8919}              & \num{0.9630}
			\\
			\midrule $P_\text{success}$           & Total fidelity per shot ($=P_\text{lattice\_surgery} \times P_\text{cultivation}$)          & \num{0.5385}              & \num{0.6075}
			\\
			$\Gamma^2$                            & Sampling overhead from PEC                                                                  & \num{11.89}               & \num{7.34}
			\\
			$W$                                   & Wall-clock time per shot ($= L_{hand} \times d \times t_{cyc}$)                             & \qty{1.52}{\s}            & \qty{0.96}{\s}
			\\
			$T_{PEC}$                             & Time per effective error-mitigated sample ($= W \times \Gamma^2$)                           & \qty{18.1}{\s}            & \qty{7.0}{\s}
			\\
			\midrule \# Physical Qubits           & Total physical qubits needed ($= N_{tot} \times 2(d + 1)^2$)                                & \num{63000}               & \num{62720}
			\\
			\bottomrule
		\end{tabular}
	} \caption{Parameters for the hand-compiled physical resource estimation of the 2D Ising model simulations.
		We consider a $10\times 10$ Ising model with $4$th order Trotterization and an $11 \times 11$ Ising model with $2$nd order Trotterization, both with 20 Trotter steps.
		The layouts are shown in \Cref{fig:10_x_10_ising_model_layout} and \Cref{fig:11_x_11_ising_model_layout}.
	}
	\label{tbl:ising_model_physical_estimates}
\end{table*}

\subsubsection{11x11 Ising Model with 2nd-order Trotterization}

We first consider the simulation of an \(11 \times 11\) Ising model using 20 second-order Trotter steps, as discussed in \Cref{sec:ising_classical_benchmark_comparison}.
This requires a total of \(n_{rot} = 7{,}381\) arbitrary rotations.
Using the mixed fallback method, an average of \(m=17\) T gates per rotation is sufficient to achieve an overall rotation synthesis error below \num{0.001}. The rotation synthesis gadgets shown in \Cref{fig:mixed_fallback_pipes} require \(2m + 2 = 36\) logical timesteps per rotation.
With four units working in parallel and a routing buffer of 4 logical timesteps, we can complete a new rotation every \(n_\text{cycles\_per\_rot}\) logical timesteps, where: \[ n_\text{cycles\_per\_rot} = \frac{36 + 4}{4} = 10 \text{ logical timesteps/rotation} \] The total number of logical timesteps, \(L_{hand}\), is the product of the number of rotations and the rate of synthesis: \[ L_{hand} = n_{rot} \times n_\text{cycles\_per\_rot} = 7{,}381 \times 10 = 73{,}810 \] With a total of \(N_{tot}=160\) logical qubits, the spacetime volume is \(S_{hand} = N_{tot} \times L_{hand} = 1.18 \times 10^7\) blocks.
We choose a code distance of \(d=13\), which yields a logical error rate per cycle of \(p_{cyc} \approx \qty{3e-9}{}\).
The probability of success, considering only lattice surgery errors, is: \[ P_\text{lattice\_surgery} = (1 - p_{cyc})^{S_{hand} \times d} \approx 0.6309 \] We target a magic state error of \(p_{mag} = \qty{3e-7}{}\).
The total probability of success from cultivation errors is: \[ P_\text{cultivation} = (1 - p_{mag})^{n_{rot} \times m} \approx 0.9630 \] The total success probability is \(P_\text{success} = P_\text{lattice\_surgery} \times P_\text{cultivation} \approx 0.6075\).
The wall-clock time per shot is \(W = L_{hand} \times d \times \qty{1}{\us} \approx \qty{0.96}{s}\).

\subsubsection{10x10 Ising Model with 4th-order Trotterization}

Next, we consider the simulation of a \(10 \times 10\) Ising model using 20 fourth-order Trotter steps, as discussed in \Cref{sec:ising_beverland_comparison}.
This requires a total of \(n_{rot} = 30{,}100\) arbitrary rotations.
Using the mixed fallback method, an average of \(m=19\) T gates per rotation is sufficient.
The rotation synthesis gadgets require \(2m + 2 = 40\) logical timesteps per rotation.
With five units working in parallel and a routing buffer of 5 logical timesteps, we can complete a new rotation every \(n_\text{cycles\_per\_rot}\) logical timesteps, where: \[ n_\text{cycles\_per\_rot} = \frac{40 + 5}{5} = 9 \text{ logical timesteps/rotation} \] The total number of logical timesteps, \(L_{hand}\), is the product of the number of rotations and the rate of synthesis: \[ L_{hand} = n_{rot} \times n_\text{cycles\_per\_rot} = 30{,}100 \times 9 = 270{,}900 \] With a total of \(N_{tot}=140\) logical qubits, the spacetime volume is \(S_{hand} = N_{tot} \times L_{hand} = 3.79 \times 10^7\) blocks.
We choose a code distance of \(d=14\), which yields a logical error rate per cycle of \(p_{cyc} \approx \qty{9.5e-10}{}\).
The probability of success from lattice surgery errors is: \[ P_\text{lattice\_surgery} = (1 - p_{cyc})^{S_{hand} \times d} \approx 0.6038 \] We target a magic state error of \(p_{mag} = \qty{2e-7}{}\).
The total probability of success from cultivation errors is: \[ P_\text{cultivation} = (1 - p_{mag})^{n_{rot} \times m} \approx 0.8919 \] The total success probability is \(P_\text{success} = P_\text{lattice\_surgery} \times P_\text{cultivation} \approx 0.5385\).
The wall-clock time per shot is \(W = L_{hand} \times d \times \qty{400}{\ns} \approx \qty{1.52}{s}\).

\subsection{Comparing fault-tolerant and NISQ operating modes of a quantum processor (\Cref{sec:nisq_ft_crossover})}
\label{app:flasq_ising_nisq_details}

This section contains some additional details on the analysis of the crossover between NISQ and fault-tolerance for the Ising model calculation presented in \Cref{sec:nisq_ft_crossover}.
As we discussed in the main text, we compare the performance of a fault-tolerant machine performing a single calculation at a time with a NISQ machine that uses all available extra space to reduce the time required by running parallel copies of the calculation.
In both cases, the task we first considered is to perform \(20\) Trotter steps of an \(11 \times 11\) Ising model and then measure a diagonal observable with unit norm to within a precision of \(.01\) (with high probability).
Crucially, for this comparison, we focus on the case with open boundary conditions to allow for a fairer comparison with a NISQ processor that lacks long-range connectivity (although this makes the classical simulation task significantly easier).

\begin{figure}
	\centering
	\includegraphics[width=.75\textwidth]{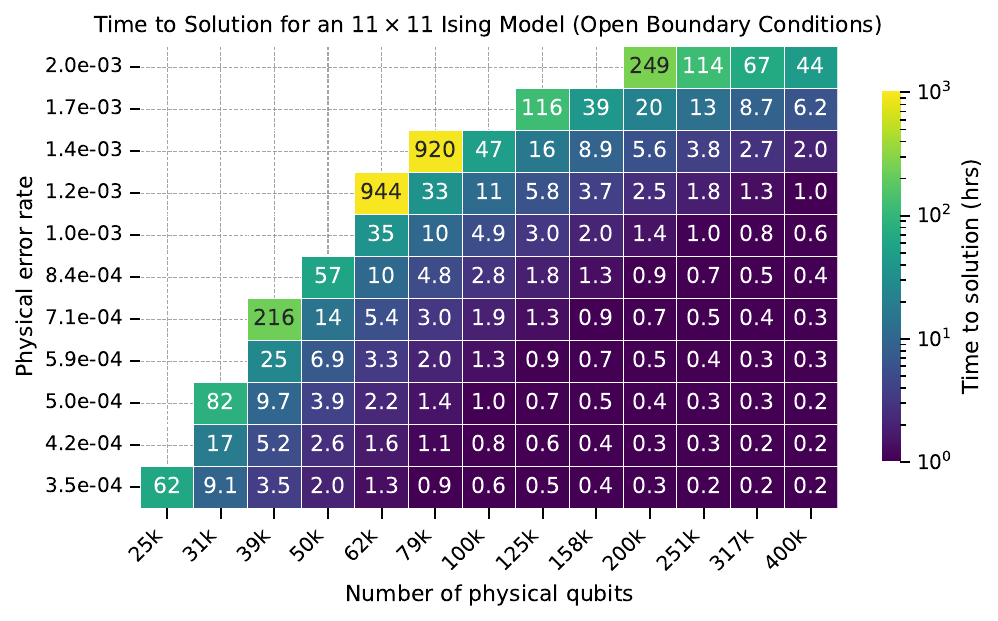}
	\caption{
		Estimated time-to-solution (in hours) for the Ising simulation task described in the text on a fault-tolerant machine, assuming open boundary conditions for the \(11 \times 11\) Ising Hamiltonian.
		We assume that residual error are corrected using probabilistic error cancellation.
		The code distance and magic state cultivation postselection probabilities were optimized using the FLASQ model to minimize the overall runtime.
		The x-axis represents the total number of physical qubits available, and the y-axis represents the physical error rate.
		We omit points where the runtime is estimated to be more than \(1{,}000\) hours.
        We find that considering open boundary conditions instead of periodic boundary conditions results in a relatively small reduction in the time to solution.
	}
	\label{fig:ising_heatmap_obc}
\end{figure}

The runtime for the fault-tolerant machine to perform this task was already presented in \Cref{fig:ising_heatmap}, assuming periodic boundary conditions.
The shift to open boundary conditions only marginally reduces the runtime on a fault-tolerant machine because, while the long-range gates require some additional ancilla volume, the majority of the ancilla volume is used to synthesize the rotations accurately.
To establish a baseline for the comparison, we plot the absolute runtime for the fault-tolerant machine with open boundary conditions in \Cref{fig:ising_heatmap_obc}.
Except for the boundary conditions, we adopt the simulation parameters detailed in \Cref{app:flasq_ising_classical_benchmark}: we target a total rotation synthesis error budget of \(0.001\) and set the target standard deviation for the expectation value estimation to \(\sigma = 0.0045\).
We use the default conservative assumptions for the ancilla volumes of individual gates and we assume a surface code cycle time of \(\qty{1}{\us}\) and a reaction time of \(\qty{15}{\us}\).

\begin{figure}
	\centering
	\includegraphics[width=.75\textwidth]{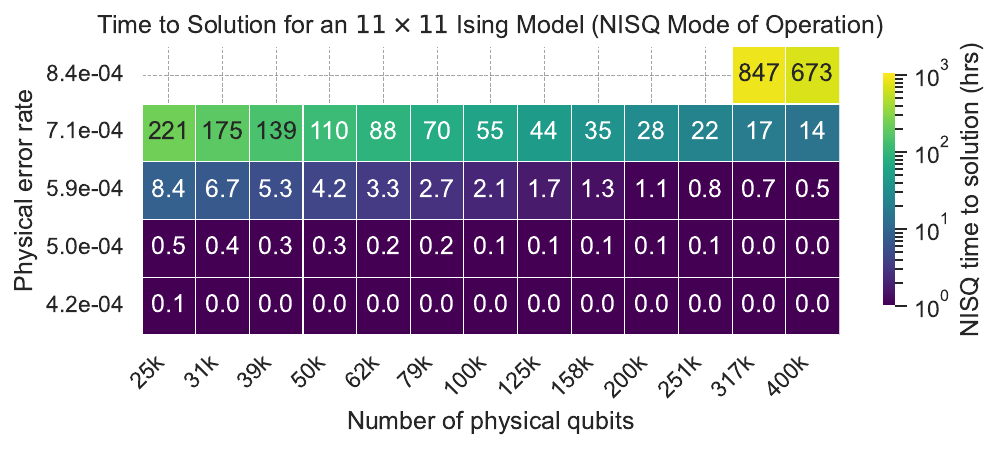}
	\caption{
		Estimated time-to-solution (in hours) for the \(11 \times 11\) Ising model simulation on a device running many parallel copies of the calculation in a NISQ fashion, assuming open boundary conditions.
		The cost model assumes the use of probabilistic error cancellation (PEC) to mitigate errors from a simplified depolarizing noise model.
		The x-axis represents the total number of physical qubits available, and the y-axis represents the physical error rate.
		We omit points where the runtime is estimated to be more than \(1{,}000\) hours.
		This heatmap provides the NISQ baseline runtime used to generate the ratio plot in \Cref{fig:nisq_ft_crossover}.
	}
	\label{fig:nisq_ising_heatmap}
\end{figure}

We plot our estimate of the absolute runtimes for the NISQ machine in \Cref{fig:nisq_ising_heatmap}.
As discussed in the main text, this estimate assumes a simplified error model where each two-qubit gate layer is followed by single-qubit depolarizing noise with a probability equal to the physical error rate, \(p_{phys}\).
This estimate relies on calculating the sampling overhead required for probabilistic error cancellation (PEC)~\cite{Temme2017-hj}.
For the NISQ calculation, we use a simplified noise model that assumes that the noise in the circuit is dominated by the two-qubit gate errors.

We model the noise such that every qubit experiences a single-qubit depolarizing channel after each layer of two-qubit gates.
The depolarizing channel \(\mathcal{D}\) is defined as:
\begin{equation}
	\mathcal{D}(\rho) = (1 - \epsilon)\rho + \epsilon \frac{I}{2},
\end{equation}
where the noise strength \(\epsilon\) is taken to be equal to the abstract physical error rate \(p_{phys}\) (the same one used as the y axis in many of our plots).
The overhead factor, \(\gamma\), required to invert this noise channel using PEC is:
\begin{equation}
	\gamma = \frac{1 + \epsilon/2}{1 - \epsilon}.
\end{equation}
The total sampling overhead, \(\Gamma^2\), scales exponentially with the total number of gate opportunities (the product of the number of qubits \(N=121\) and the number of two-qubit gate layers, \(D_{NISQ}=80\)):
\begin{equation}
	\Gamma^2 = \gamma^{2 \times N \times D_{NISQ}}.
\end{equation}
The total number of samples required is \(N_{samples} = \Gamma^2 / \sigma^2\).
For the NISQ model, we take \(\sigma = .005\), so that the overall error is less than \(.01\) with approximately the same probability as the fault-tolerant machine.

To estimate the total runtime \(T_{NISQ}\), we assume the time per sample is the number of two-qubit gate layers (\(D_{NISQ} = 80\)) multiplied by a rough estimate of the two-qubit gate time, \(\qty{50}{\ns}\).
We also assume that the available physical qubits \(N_{phys}\) can be perfectly partitioned to run \(N_{parallel} = \lfloor N_{phys} / 121 \rfloor\) simulations in parallel, neglecting tiling constraints.
The total runtime is then:
\begin{equation}
	T_{NISQ} = (D_{NISQ} \cdot \qty{50}{\ns}) \times \frac{N_{samples}}{N_{parallel}}.
\end{equation}

\begin{figure}
	\centering
	\includegraphics[width=.75\textwidth]{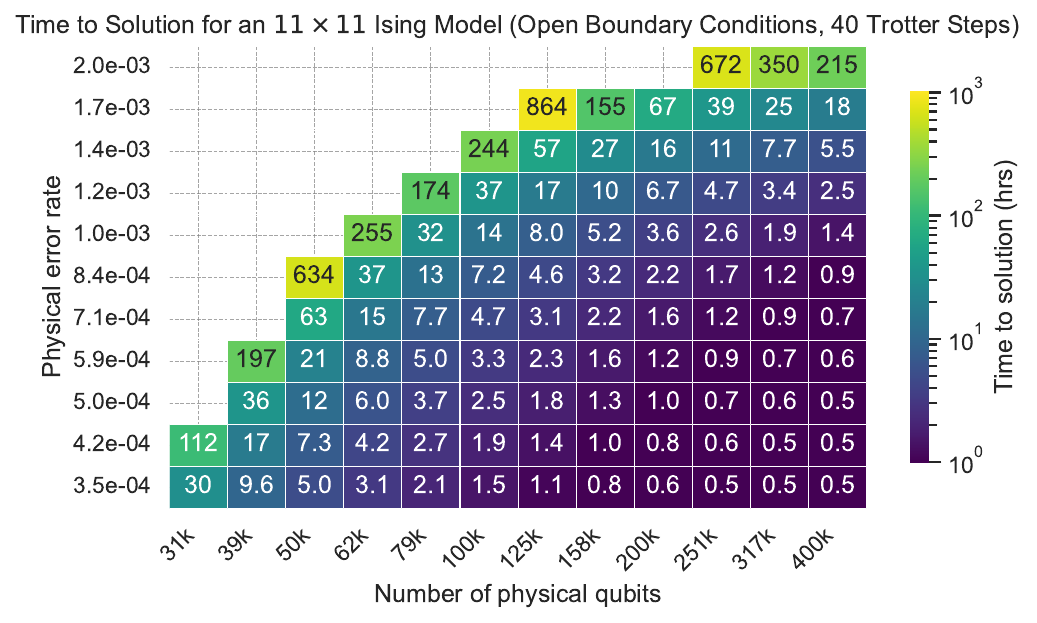}
	\caption{
		Estimated time-to-solution (in hours) for a fault-tolerant machine to estimate a diagonal observable to high precision after performing \(40\) second-order Trotter steps for an \(11 \times 11\) 2D TFIM Hamiltonian with open boundary conditions.
		We assume that residual error are corrected using probabilistic error cancellation.
		The code distance and magic state cultivation postselection probabilities were optimized using the FLASQ model to minimize the overall runtime.
		The x-axis represents the total number of physical qubits available, and the y-axis represents the physical error rate.
		We omit points where the runtime is estimated to be more than \(1{,}000\) hours.
	}
	\label{fig:ising_heatmap_obc_deep}
\end{figure}

We presented a second comparison using the same methodology in the main text, except that we considered a circuit that performs \(40\) second-order Trotter steps.
For completeness, we plot the FLASQ estimates for the time required by the device operating in a fault-tolerant mode below in \Cref{fig:ising_heatmap_obc_deep}.
The same device operating in a NISQ mode would require more than \(1{,}000\) hours to complete the task at any of the error rates or qubit numbers we considered, so we do not plot that data here.

\subsection{Comparing FLASQ estimates with a previously reported fault-tolerant resource estimate (\Cref{sec:ising_beverland_comparison})}
\label{app:flasq_ising_beverland_details}

This section details the methodology and parameters used for the analysis in \Cref{sec:ising_beverland_comparison}, comparing FLASQ against Beverland et al.
(2022)~\citenum{Beverland2022-dh}.

The benchmark analyzed is the "quantum dynamics" application: a \(10 \times 10\) spin Ising model (N=100; $J=1.0, g=0.5, dt=0.1$) evolved using a 4th-order Trotter-Suzuki formula (see \Cref{app:trotter_formulas}) for 20 steps.
This results in 30,100 total rotations.
We aligned our hardware assumptions with their "nanosecond" qubit profiles ($p_{phys}=10^{-3}$ and $10^{-4}$).
The surface code cycle time ($t_{cyc}$) was set to \(\qty{400}{\ns}\), derived from their specified gate timing assumptions (4 layers of 2Q gates @ \(\qty{50}{\ns}\), 2 layers of 1Q gates @ \(\qty{50}{\ns}\), and 1 layer of Measurement/Reset @ \(\qty{100}{\ns}\)).
We assumed an optimistic reaction time of 10 cycles (\(\qty{4}{\us}\)).

The comparison targets are the specific estimates provided in Table IV of \cite{Beverland2022-dh}.
At $p_{phys} = 10^{-3}$, these are Time-Optimal (8.2M qubits, 1.1s) and Space-Optimal (0.94M qubits, 13s).
At $p_{phys} = 10^{-4}$, they are Time-Optimal (0.68M qubits, 0.56s) and Space-Optimal (0.11M qubits, 6.7s).

To generate the FLASQ estimates, we performed sweeps over available physical qubits (50k to 10M), optimizing code distance and cultivation parameters, utilizing the mixed fallback protocol for rotation synthesis (\Cref{app:rotation_synthesis_details}).

For the High-Fidelity comparison (Comparison 1), we constrained FLASQ to match the pure-QEC methodology.
The total error budget ($\epsilon_{total} = 0.001$) was partitioned equally across synthesis, logical, and cultivation errors ($\epsilon = 0.001 / 3$ each).
The metric used is Wall Clock Time (s).
This required a total T-count of approximately 571,900.
Notably, at $p_{phys}=10^{-3}$, no feasible solution was found as cultivation could not meet the required fidelity.

For the FLASQ+PEC comparison (Comparison 2), we employed an error mitigation strategy.
The error budget was allocated entirely to synthesis bias ($\epsilon_{syn} = 0.001$), with PEC used to manage variance.
The metric used is Effective Runtime per Sample (s).
The resulting T-count is approximately 546,700.

We note a discrepancy regarding the number of T gates in the circuits considered in our work and in \cite{Beverland2022-dh}.
Our analysis, based on the 30,100 rotations required for this benchmark, yields T-counts (e.g., approx.
571,900 for the high-fidelity budget) roughly four times lower than reported in \cite{Beverland2022-dh}.
This is despite the fact that both works consider similar ``mixed fallback'' rotation synthesis strategies and calculate the same number of rotations (30,100) in the circuit.
We are unsure of the source of this discrepancy, or if the number of T gates mentioned in the tables of \cite{Beverland2022-dh} accurately represent the number used to generate their final runtime numbers.
However, even if the estimates of \cite{Beverland2022-dh} were revised downward by this factor of four, the qualitative conclusions of our comparison remain robust.

\section{Comparing FLASQ with two constructive approaches to compilation}
\subsection{Overview}
\label{app:ising_gosc_csc_comparison}

The FLASQ model attempts to provide a reasonable estimate of the resources required to perform a fault-tolerant quantum computation, but it does not provide a constructive upper bound.
Instead, it attempts to predict the runtime that could be achieved by a future highly-optimized compiler.
There are alternative compilation paradigms that provide constructive recipes for compiling algorithms.
These provide concrete upper bounds on the execution cost, often at the expense of neglecting some of the optimizations anticipated by our model.
In this section, we compare our FLASQ estimates to the outputs of two such approaches.
We focus on the resources required for a single Trotter step of the Ising model simulation described previously in \Cref{sec:ising_classical_benchmark_comparison}, with some small modifications that are described in \Cref{app:flasq_ising_constructive_comparison_details}.
We study sizes from \(2\times 2\) to \(8 \times 8\).

\textbf{Game of Surface Codes.}
The first alternative we consider here is the ``Game of Surface Codes'' (GoSC) framework introduced in \cite{Litinski2019-nu}.
This framework provides a high-level abstraction for designing and analyzing surface code computations based purely on Pauli-Based Computation (PBC).
In the GoSC cost model, the quantum runtime is determined exclusively by the number of T gates in the circuit.
This is because all of the Clifford operations are absorbed into the final measurements.
By eliminating the Clifford operations from the primary computation, this approach removes many of the complexities of the compilation problem at the expense of serializing the implementation of non-Clifford gates.
The non-Clifford T gates are replaced by multi-qubit generalized T gates (phasing the eigenstates of arbitrary Pauli operators) implemented by lattice surgery measurements that generally span the whole system and therefore must be performed one at a time.

\textbf{Constructive Surface Code Compiler.}
The second approach provides a constructive upper bound using a compilation model more closely related to the lattice surgery assumptions underlying FLASQ.
We utilize a novel heuristic Constructive Surface Code (CSC) compiler~\cite{Xu2026-am}.
This framework represents a significant step towards automated, validated fault-tolerant compilation.
It extends the MaxState solver of \cite{molavi_generating_2025} to support the automated solving of mapping and routing problems that arise in compiling to the surface code.
The MaxState algorithm is a generic search algorithm for qubit mapping and routing problems that uses simulated annealing as a subroutine to find reasonable (but not necessarily optimal) solutions.

The CSC compiler is a large-scale automated compiler that constructs a proof of physical realizability in the form of a valid pipe diagram.
It achieves this by extracting a standalone intermediate representation (IR) for expressing mapping and routing solutions.
Programs in this IR are validated according to the constraints defining the mapping and routing problem, which guarantees that the output corresponds to a physically realizable execution plan.
If a program is valid, then the compiler translates it to the LaSre format~\cite{tan_sat_2024}, which supports exporting to a glTF pipe diagram.

While this approach provides a constructive upper bound, we emphasize that the MaxState algorithm is heuristic and therefore not guaranteed to be optimal given the general hardness of mapping and routing problems.
Furthermore, the CSC compiler does not yet incorporate advanced optimizations such as the use of walking surface codes or the application of ZX-calculus identities.
Lastly, due to the limitations of downstream tooling, CSC is currently only able to validate --- but not export pipe diagrams for --- solutions that use constructs beyond the standard one square logical patch per logical qubit with orthogonal Z/X boundaries.
For example, the Lintinski-style patch rotation (Figure 11(a) in \cite{Litinski2019-nu}) relies on growing a patch into a rectangle and deforming its boundaries.

\begin{figure}
	\centering
	\includegraphics[width=\textwidth]{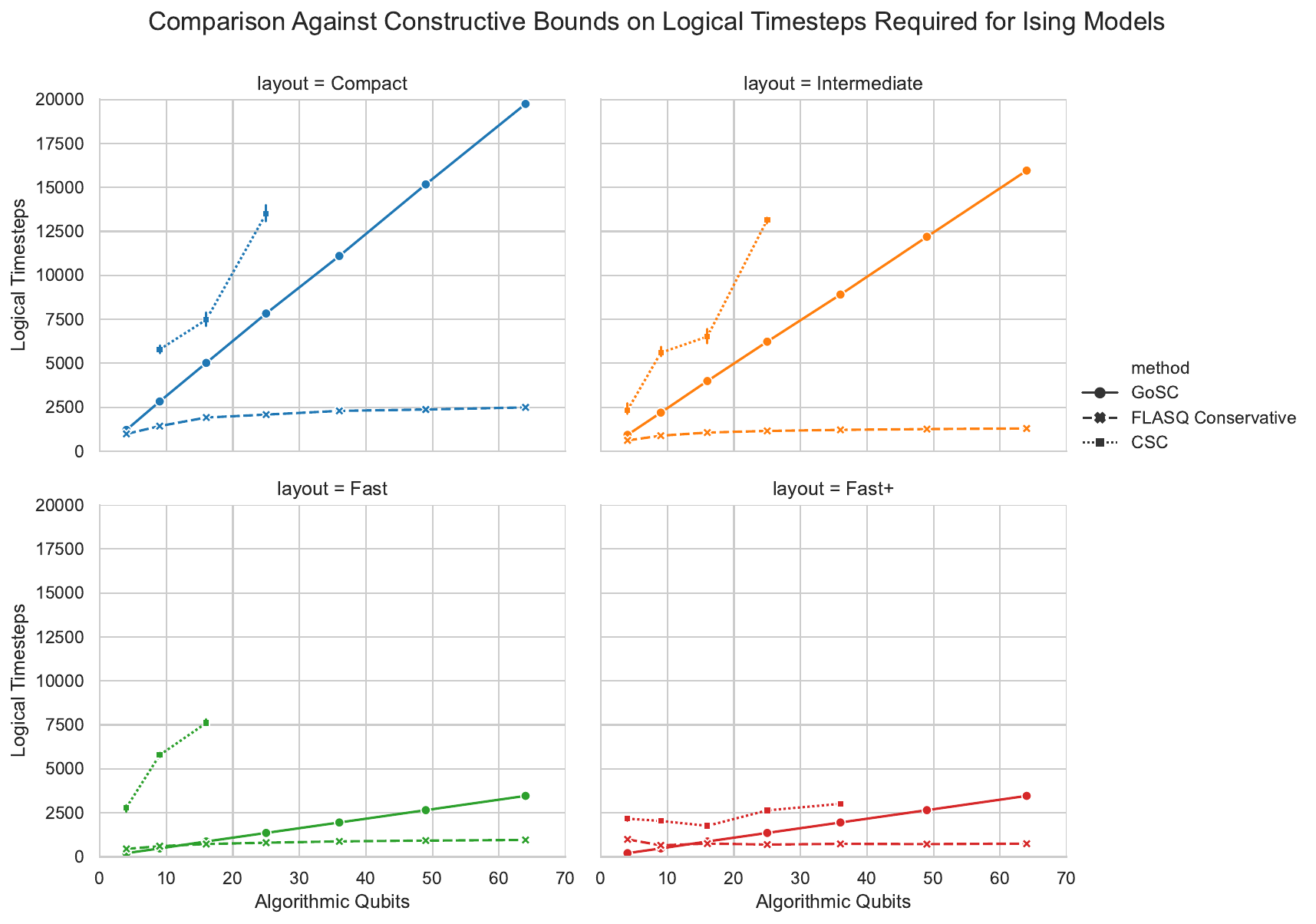}
	\caption{
		A comparison of the total time (y axis, in terms of logical timesteps) required to perform a single Trotter step of the 2D Ising model, as estimated by the FLASQ model, the Game of Surface Codes model (GoSC), and the heuristic Constructive Surface Code (CSC) compiler.
		The comparison is shown as a function of the number of algorithmic qubits (x axis) for square Ising models ranging in size from \(2 \times 2\) to \(8 \times 8\).
		The layouts (Compact, Intermediate, Fast, Fast+) correspond to different fixed relationships between the number of algorithmic qubits and the total number of logical qubit patches.
	}
	\label{fig:constructive_compilation_comparison}
\end{figure}

\textbf{Comparing with the FLASQ Model.}
In \Cref{fig:constructive_compilation_comparison}, we compare the runtime estimates (in logical timesteps) provided by the FLASQ model, the GoSC model, and the constructive upper bounds generated by CSC for a single Trotter step of the 2D Ising model at various system sizes. 
In these comparisons, we vary the system size (x axis), while fixing the relationships between the number of algorithmic qubits and the total number of logical qubits to one of four possible settings for all three approaches (various colors, labeled ``Compact,'' ``Intermediate,'' ``Fast,'' and ``Fast+'').
The ``Compact,'' ``Intermediate,'' and ``Fast'' labels correspond to the spatial layouts defined by the GoSC model, each one specifying a particular scaling for the total number of logical qubit patches as a function of the number of algorithmic qubits.
The ``Fast+'' label (known as the ``Compact'' layout in prior work~\cite{molavi_dependency-aware_2025}) corresponds to an alternative choice that has slightly more space than the GoSC ``Fast'' layout.\footnote{This is because we found that the density of the ``Fast'' layout is not well-suited for use with the CSC compiler in the case where we only use one patch per qubit, instead of the GoSC two-patch per two-qubit storage scheme.}
We show the scaling of the total number of logical qubit patches with the number of algorithmic qubits in \Cref{tab:litinski_layout_comparison}, and we provide more details on the methodologies used (particularly for the CSC compiler) in \Cref{app:flasq_ising_constructive_comparison_details}.

\begin{table}
	\centering
	\caption{
		A comparison of the spatial overhead for the three data block layouts described by \cite{Litinski2019-nu} in addition to a ``Fast+'' layout with slightly more space.
		\(N\) is the number of algorithmic qubits, which is the same as the system size for these comparisons.
		The formulas represent the total number of required patches for each layout.
	}
	\label{tab:litinski_layout_comparison}
	\setlength{\tabcolsep}{8pt}
	\linespread{1.0}\selectfont%
	\renewcommand{\arraystretch}{1.2}
	\begin{tabular}{lc}
		\toprule
		\textbf{Layout} & \textbf{Total Number of Patches}
		\\
		\midrule
		Compact         & \(1.5N + 3\)
		\\
		Intermediate    & \(2N + 4\)
		\\
		Fast            & \(2N + \sqrt{8N} + 1\)
		\\
		Fast+           & \(6\lceil N/2 \rceil - 3\)
		\\
		\bottomrule
	\end{tabular}
\end{table}

\Cref{fig:constructive_compilation_comparison} reveals several distinct scaling behaviors.
The GoSC model (solid lines) exhibits linear scaling across all space settings due to its inherent serialization of T gates. Thus despite promising runtimes for small system sizes, the GoSC model quickly becomes undesirable.
In contrast, both the FLASQ model (dashed lines) and the CSC compiler (dotted lines) are capable of achieving a time complexity that is independent (or nearly so, in the case of the CSC compiler) of system size. Recall that the 2D Ising model has nearest-neighbor connectivity, which should lead to scaling independent of system size if operations are properly parallelized.

We do observe that the CSC compiler suffers from the undesirable linear scaling with system size for the three most space-constrained layouts due to the difficulty of solving the highly constrained mapping and routing problem.
In fact, for these layouts, CSC yields higher depths than the GoSC approach.
We believe that this is largely due to the fact that the layouts are designed for serialized PBC and not a more general paradigm that allows for parallel operations, but we leave the detailed analysis of CSC to the upcoming \cite{Xu2026-am}.

When comparing the same layouts, the constructive upper bound provided by CSC is consistently higher than the FLASQ estimate.
This difference quantifies the practical challenge of explicit routing that FLASQ's idealized fluid ancilla assumption does not address.
It also highlights the need to incorporate additional optimizations into our CSC compiler, particularly the walking surface codes that allow for the efficient collective motion of logical qubits.
We provide more details on how this data was generated in \Cref{app:flasq_ising_constructive_comparison_details}.

\subsection{Additional details}
\label{app:flasq_ising_constructive_comparison_details}

This section documents the methodology used to generate the comparison presented in ~\Cref{app:ising_gosc_csc_comparison} and \Cref{fig:constructive_compilation_comparison}, focusing on the configuration and implementation of the heuristic CSC compiler.

\textbf{The CSC compiler framework.}
The CSC compiler is built on top of the Rust library backend of the Marol language for specifying qubit mapping and routing problems introduced in \cite{molavi_generating_2025}.
This framework provides a generic solver (MaxState) for qubit mapping and routing problems under a variety of architectural constraints.
Concretely, in our case, we use it to schedule the usage of ancilla qubits to realize operations in the circuit according to constraints such as the ones in Table~\ref{tab:gate_costs_cmr}.
Our constraints enforce that any solution yields a valid surface code compilation and we use the solver to find a schedule with the shortest total duration.

It was necessary to extend the solver in several ways to support our particular surface code architecture:
\begin{itemize}
	\item Extend the map from data qubits to physical patch locations to keep track of richer patch state.
	      For example, we must encode the orientation of the patch boundaries and whether it contains a magic state or not.
	\item Add support for gates changing the mapping state (e.g., the locations of patch boundaries).
	\item Add support for operations with different durations.
	\item Allow transition operations (e.g.
	      Move and Rotate) to be performed in parallel with gate operations (e.g. \(H\), \(CNOT\)).
\end{itemize}

\textbf{Benchmark details.}
The benchmarks utilize circuits simulating a single Trotter step of the 2D Transverse-Field Ising Model (TFIM).
In order to make the circuits compatible with our CSC compiler, we slightly modified the workflow used for the other FLASQ estimates in this work.
Specifically, we concretely realized the arbitrary rotations using the deterministic rotation synthesis strategy realized by Gridsynth~\cite{Ross2014-zf} (the ``Diagonal'' entry in \Cref{tab:rotation_synthesis_costs_appendix}).
To approximately model the more sophisticated mixed fallback approach, we set the error threshold in Gridsynth such that the average T-count per rotation is at most \(.5\) more than the number of rotations implied by the mixed fallback formula in \Cref{tab:rotation_synthesis_costs_appendix}.

Each architecture is represented by a height, width, and list of node indices (coordinates) available for mapping data qubits.
The architecture graph has grid connectivity.
We used the same data qubit locations as in the ``Compact,'' ``Intermediate,'' and ``Fast'' layouts defined in the GoSC model~\cite{Litinski2019-nu}.
For the ``Fast+'' layout, we used the ``Compact'' architecture defined in Figure 7(b) of \cite{molavi_dependency-aware_2025}.
The initial mapping of $n$ data qubits to a sorted list of node indices $p$ is simple: $i \rightarrow p[i]$ for all $0 < i < n$.
Notably, this does not necessarily lead to a layout that keeps the interactions of the data qubits local, especially in the ``Fast'' layout.
We assumed an initial patch state where the X boundaries were aligned along the top and bottom edges.

\textbf{Operations and costs.}
\Cref{tab:gate_costs_cmr} describes the costs and constraints of realizing the operations we consider in CSC.
For simplicity, we assume a fixed depth of 6 for cultivation and configure FLASQ and GoSC to use that as well for the comparison in \Cref{fig:constructive_compilation_comparison}.

We model \(T\) gates as a composite operation of cultivation followed by injection rather than allowing cultivation as a standalone operation because of the challenges in controlling the search space.
For example, if cultivation is unrestricted, it is possible to end up in a state where all the ancilla are occupied with magic states and none are available for routing.
Further optimization of the heuristic solver or its objective function is required to avoid this situation, but we leave this to a future work.
The tradeoff of our choice is that it suboptimally blocks the ancilla used for injection for the entire duration of cultivation (and vice versa).
Another simplification we make is assuming that no \(S\) corrections need to be performed.
One way of conservatively adding this is to ensure we can always perform an \(S\) gate by extending the depth of the composite \(T\) gate by 1.5.

\begin{table}[] \centering \resizebox{\textwidth}{!}{%
	\setlength{\tabcolsep}{10pt} 
		\begin{tabular}
			{@{}llp{270pt}p{230pt}@{}}
			\toprule
			Operation
			 & Depth (\(d\))
			 & Constraint
			 & Effect on Patch State
			\\
			\midrule
			\(H\)
			 & 3
			 & 1 adjacent ancilla for Rotate.
			 & None
			\\
			\(S\)
			 & 1.5
			 & 1 adjacent ancilla along X boundary.
			 & None
			\\
			\(CNOT\)
			 & 2
			 & Connected path of ancilla from X boundary of control to Z boundary of target with at least one bend.
			 & None
			\\
			\(T\)
			 & 8
			 & 2 connected ancilla: 1 adjacent along X boundary (for injection) and 1 diagonal to argument of \(T\) gate (for cultivation).
			 & None
			\\
			Move
			 & 1
			 & Connected path of ancilla from source to destination with an even number of bends.
			 & Moves patch from source to destination.
			\\
			Rotate
			 & 3
			 & 1 adjacent ancilla for Litinski-style.
			 & Swaps the top/bottom and left/right patch boundaries.
			\\
			\bottomrule
		\end{tabular}
	}
	\caption{ Costs and constraints of operations modeled in CSC.}
	\label{tab:gate_costs_cmr}
\end{table}

\begin{figure}
	\centering
	\includegraphics[width=0.9\textwidth]{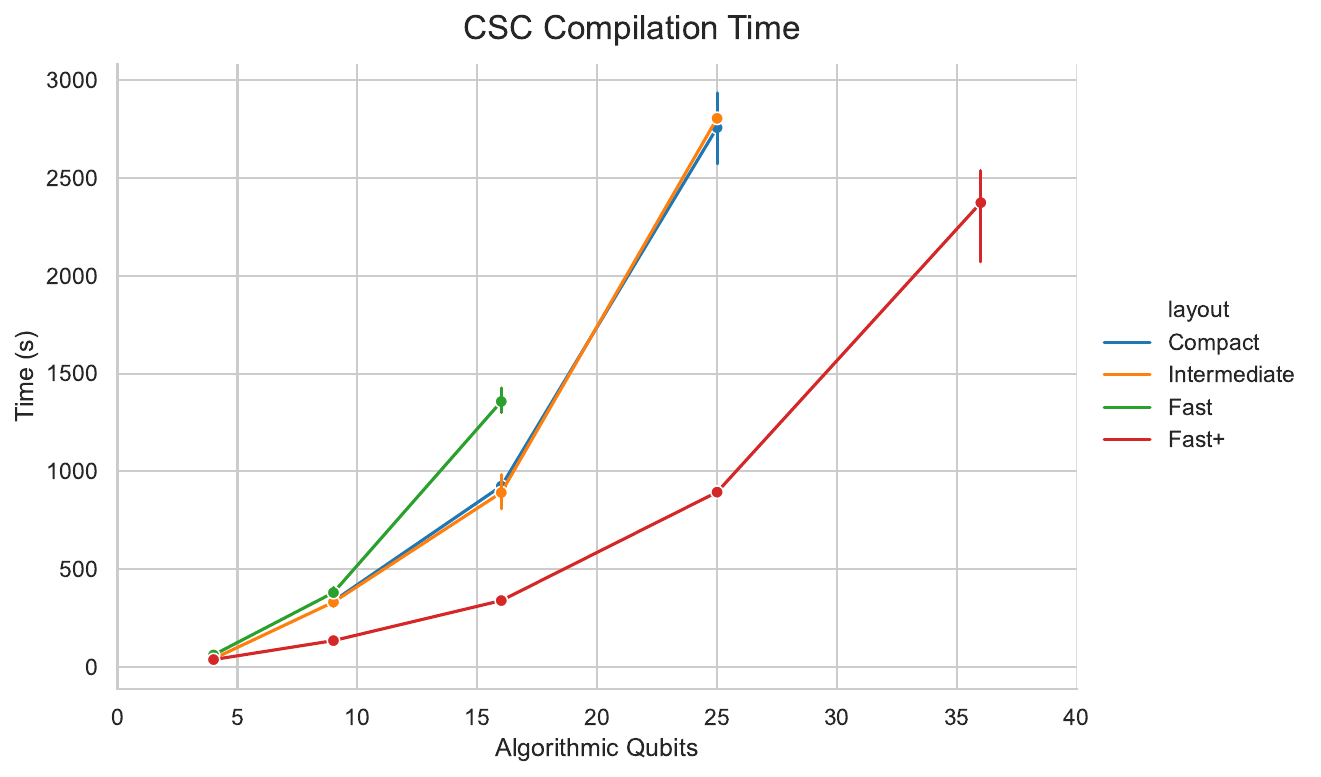}
	\caption{
		Compilation times for the CSC data in Figure~\ref{fig:constructive_compilation_comparison}.
		In comparison, GoSC and FLASQ require seconds at most.
		However, GoSC achieves this efficiency at the expense of flexibility and FLASQ does not provide a constructive solution.
	}
	\label{fig:csc_compilation_time}
\end{figure}

\textbf{Setup CSC numerical experiments.}
The data points for the CSC compiler were generated by executing the compiler with the specified circuits and architecture details, and parsing the resulting schedule to determine the total execution time (in units of logical timesteps).
We run the tool for 3 trials for each data point and report the mean and 95\% confidence interval because the algorithm is not deterministic.
Each trial was allotted 1 hour and terminated if it could not converge within this timeframe.
All experiments were run on a Large n2d instance with 64 vCPU and 120GB RAM.
The exact runtimes for the instances we were able to obtain a solution for within this amount of time are shown in \Cref{fig:csc_compilation_time}.
As expected, the growing difficulty of the problem at larger system sizes or under more constraints leads to quickly increasing runtimes.

\textbf{GoSC Methodology.}
We use the \verb|LitinskiTransformation| pass in Qiskit v2.2.0~\cite{javadi-abhari_quantum_2024} to transform the circuit into the PBC representation.
Then we compute the depth of each circuit by following the rules for each layout in \cite{Litinski2019-nu}.
Because the Qiskit transformation does not consolidate the $n$ Clifford measurements at the end, we upper bound the cost using the maximum depth (for each layout).
We use the same depth for the ``Fast+'' layout as the ``Fast'' layout because the model does not have a meaningful way of using the extra space.

The depths we compute for GoSC optimistically assume the computation time is driven by magic state consumption. That is, to achieve the depth computed, we would need a constant amount of extra space along the perimeter of the architecture to ensure a magic state is always available when needed.

\section{More details on the Hamming weight phasing case study}
\label{app:hwp_details}

This appendix contains additional details and numerical data to support the Hamming weight phasing case study presented in \Cref{sec:hwp_example}.

\subsection{Qubit placement strategy}

To generate the upper bound (pessimistic estimate) to the FLASQ cost for the HWP analysis in \Cref{sec:hwp_example}, we require a concrete 2D layout of the qubits. 
The FLASQ cost model's distance-dependent costs (e.g., for CNOT or Toffoli gates within the HWP arithmetic circuits) are a function of the Manhattan distance between the involved qubits. 
Since the analyzed implementation of HWP does not natively provide an optimized 2D layout, we use a simple naive placement strategy to lay the circuit out on a grid and determine these distances.

We divide a two-dimensional grid into two regions, one for the data qubits (the \(N\) qubits being rotated) and one for the algorithmic ancilla qubits required by the HWP algorithm.
Within these regions, qubits are placed sequentially following a zig-zag pattern. For the analysis presented in \Cref{sec:hwp_example}, we assume a fixed width of 10 columns for this placement pattern. 
The data qubits are placed in rows with non-negative indices.
For example, the data qubits fill the first row from left to right (0, 0) to (0, 9), then wrap to the next row, filling it from right to left (1, 9) to (1, 0), and so on.
The algorithmic ancilla qubits are placed in rows with negative indices (in the order they are first used in the HWP implementation).
Specifically, the ancilla qubits follow the same zig-zag pattern starting at (-1, 0).

This deterministic, albeit unoptimized, layout ensures that all qubits involved in the HWP computation have well-defined coordinates and that the data and ancilla qubits are grouped close to each other.
This provides the necessary basis for calculating the Manhattan distances used in the FLASQ volume estimates, thereby enabling the calculation of the pessimistic bound shown in \Cref{fig:HWP_big_plot} (and the additional figures presented below).

\subsection{Varying the assumptions}

In this section, we present two variations of the analysis we performed in \Cref{fig:HWP_big_plot} using slightly different assumptions.
Both variations highlight the robustness of the conclusions we presented in \Cref{sec:hwp_example}.

\begin{figure}
	\centering
	\includegraphics[width=.75\textwidth]{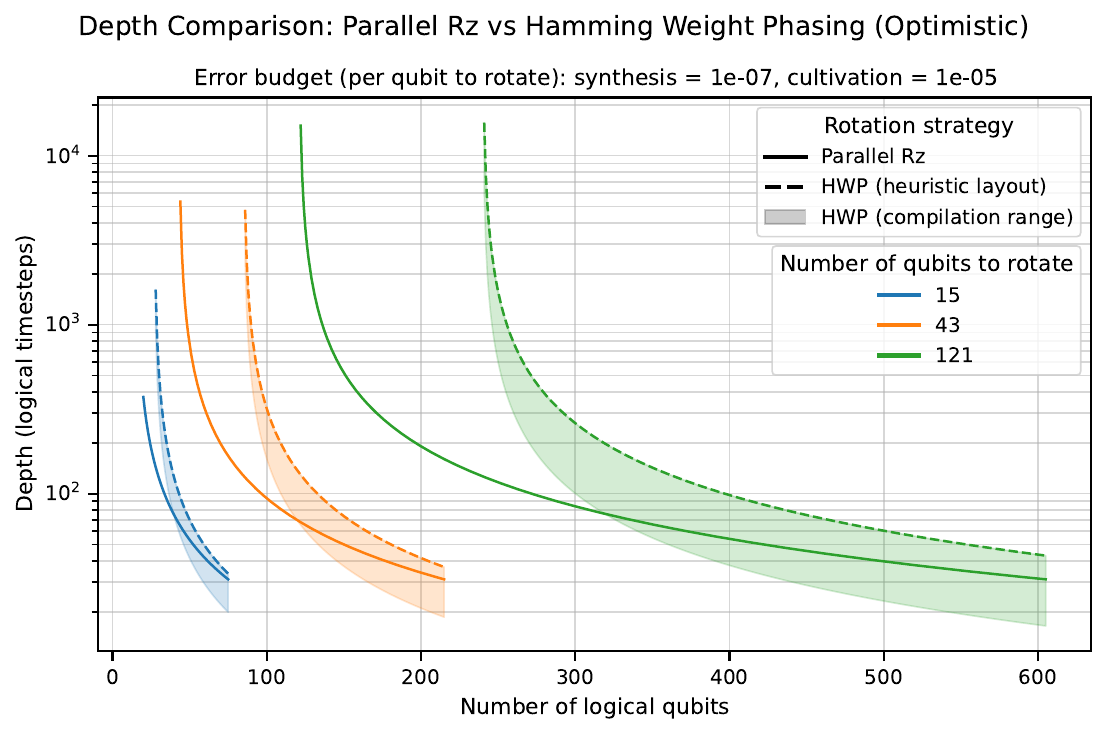}
	\caption{
		The number of logical timesteps required to perform \(N\) parallel \(R_Z\) rotations, targeting an error from rotation synthesis of \(N \times 10^{-7}\) and an error from magic state cultivation of \(N \times 10^{-5}\).
		This plot uses the optimistic FLASQ estimates of \Cref{tab:gate_costs_large} rather than the default conservative ones used in \Cref{fig:HWP_big_plot} Solid lines represent the naive Parallel Rz approach.
		For Hamming weight phasing (HWP), the shaded region represents a range of potential performance bounded by using a heuristic to lay out the qubits on a 2D grid before applying the FLASQ model (dashed upper line) or neglecting the distance-dependent components of the cost (bottom edge of the shaded region).
		All curves begin only when enough total logical qubits are available and terminate when the total number of logical qubits is \(5N\).
	}
	\label{fig:HWP_optimistic_plot}
\end{figure}

In \Cref{fig:HWP_optimistic_plot}, we perform the same procedure using the optimistic estimates of the ancilla volume for individual gates (see \Cref{app:flasq_volume_details}) instead of the default conservative ones.
While the absolute depths are lower, the qualitative results remain the same: HWP performance is highly sensitive to compilation, and the heuristic layout is generally outperformed by the Parallel Rz strategy in this regime.
We actually find that the potential benefit of HWP under these assumptions is slightly smaller than under the conservative assumptions.
The reduction in cost for the allocation of ancilla space and routing is outweighed by the reduction we also assume for the cost of T state cultivation and rotation synthesis.

\begin{figure}
	\centering
	\includegraphics[width=.75\textwidth]{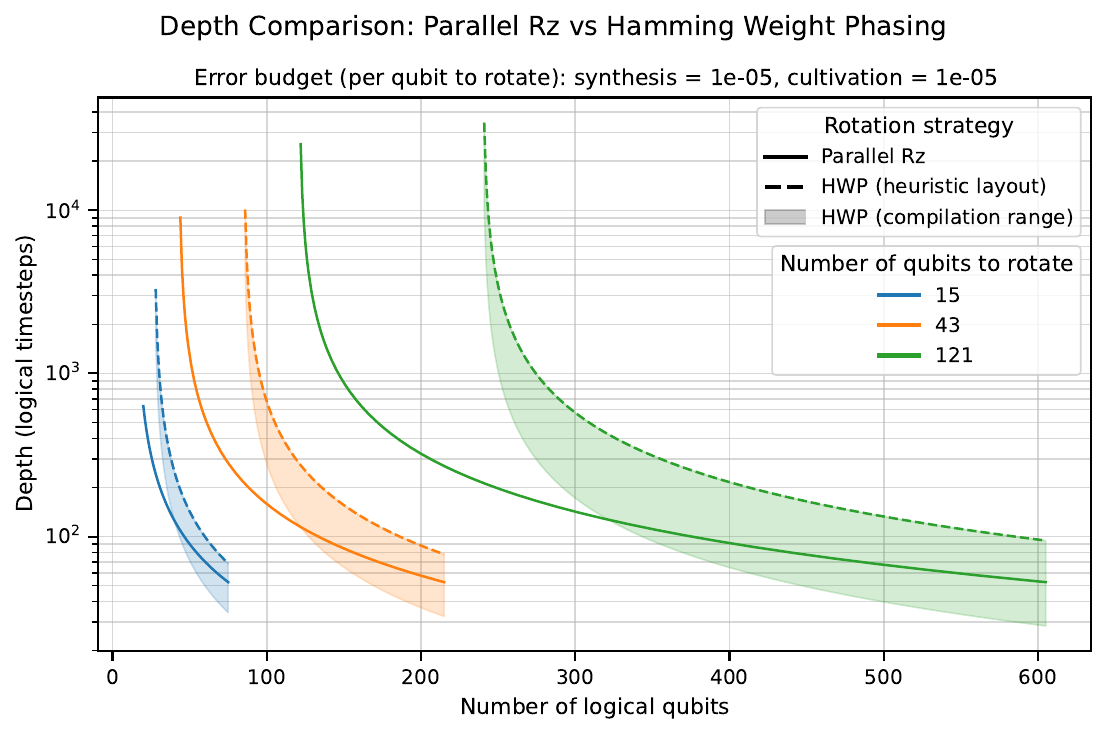}
	\caption{
		The number of logical timesteps required to perform \(N\) parallel \(R_Z\) rotations, targeting an error from rotation synthesis of \(N \times 10^{-5}\) and an error from magic state cultivation of \(N \times 10^{-5}\).
		Solid lines represent the naive Parallel Rz approach.
		For Hamming weight phasing (HWP), the shaded region represents a range of potential performance bounded by using a heuristic to lay out the qubits on a 2D grid before applying the FLASQ model (dashed upper line) or neglecting the distance-dependent components of the cost (bottom edge of the shaded region).
		All curves begin only when enough total logical qubits are available and terminate when the total number of logical qubits is \(5N\).
	}
	\label{fig:HWP_conservative_plot_high_rotation_error}
\end{figure}

\Cref{fig:HWP_conservative_plot_high_rotation_error} varies the assumptions of \Cref{sec:hwp_example} and \Cref{fig:HWP_big_plot} along another axes.
In this figure, we present an analysis that uses the default conservative parameters for the FLASQ model but we allow for an increased rotation synthesis error of \(10^{-5}\) per rotation rather than \(10^{-7}\).
This parameter regime might be relevant if we were to allow for \(\mathcal{O}\left( 1 \right)\) overall error from rotation synthesis across the execution of an early fault-tolerant algorithm with tens or hundreds of thousands of rotations.
Recent work indicates that this level of error could be accounted for using error mitigation~\cite{Koczor2024-mb}.
Provided that this error budget is acceptable, we find it will be even more challenging to benefit from HWP: for the implementation we consider the potential advantage is less than a factor of \(2\) under these conditions even when distance-dependent costs are neglected.

\end{document}